\documentclass{aa}
\usepackage{times,graphicx}
\usepackage{natbib}		% \citet et \citep 
\begin{document}

%-----------------------------------------------
%-----------------------------------------------

\title{Oscillatory motions observed in eruptive filaments}

\author{ K.~Bocchialini~(1), F.~Baudin~(1), S.~Koutchmy~(2), G.~Pouget~(1), J.~Solomon~(1)}

\institute{  (1) Institut d'Astrophysique Spatiale, B\^at. 121, UMR 8617, CNRS/Univ. Paris-Sud, F-91405 Orsay, France \\ 
             (2) Institut d'Astrophysique de Paris, UMR 7095, CNRS/UPMC, 98 bis Boulevard Arago, F-75014 Paris, France}

\titlerunning{Oscillatory motions observed in eruptive filaments}
\authorrunning{K. Bocchialini et al.}

\date{Received XXXX / XXXX }

\abstract{\\

Context: The origin of the variable component of the solar wind is of great intrinsic interest for heliophysics and space-weather, e.g.
the initiation of coronal mass ejections, and the problem of mass loss of all stars. It is also related to the physics of coronal neutral sheets and streamers, 
occurring above lines of magnetic polarity reversal.  Filaments and prominences correspond to the cool coronal component of these regions.\\

Aims: We examine the dynamical behaviour of these structures where reconnection and dissipation of magnetic energy in the turbulent plasma are occurring. 
The link between the observed oscillatory motions and the eruption occurrence is investigated in detail for two different events.\\

Method: Two filaments are analysed using two different datasets:
time series of spectra using a transition region line (He~I at 584.33 ${\rm \AA}$) and a coronal line 
(Mg~X at 609.79 ${\rm \AA}$) measured with CDS on-board SOHO, observed on May 30, 2003, 
and time series of intensity and velocity images from the NSO/Dunn Solar Telescope in the H$\alpha$ line 
on September 18, 1994 for the other.
The oscillatory content is investigated using Fourier transform and wavelet analysis and is compared to different models.\\

Results: In both filaments, oscillations are clearly observed, in intensity and velocity in the He~I and Mg~X lines, in velocity in H$\alpha$, 
with similar periods from a few minutes up to 80 minutes, with a main range from 20 to 30 minutes, cotemporal with eruptions. 
Both filaments exhibit vertical oscillating motions. 
For the filament observed in the UV (He I and Mg X lines), we provide evidence of damped velocity 
oscillations, and for the filament observed in the visible (H$\alpha$ line), we provide evidence that parts of the filament are oscillating, while the 
filament is moving over the solar surface, before its disappearance.
}
 
\maketitle

%-----------------------------------------------
%-----------------------------------------------
\section{Introduction}

Prominences (observed above the limb) and filaments (observed on the solar disk) are complex structures of 
cool, dense plasma embedded in the 2 orders of magnitude less dense corona, which is also 2 orders of magnitude hotter.
They are maintained in dynamical equilibrium against gravity by the magnetic field, and they can channel waves 
propagating outward from the low chromosphere. Filaments and prominences form in both active and quiet-Sun regions. 
Filaments and prominences that have been stable for weeks can suddenly erupt (e.g. \cite{Oliver_09}). 
Some of the erupting filaments, heated and becoming visible in 
coronal emission lines, do not produce a CME: the filament reforms after cooling due to radiative losses.
Such an example was reported by \citet{Filippov_02} who discuss the possible origin of the heating related to the dynamical
behaviour of the filament, including oscillations.\\

As pointed out by \citet{Isobe_Tripathi_06} and \citet{Tripathi_09}, oscillations of erupting filaments provide an alternative tool for probing the onset 
mechanisms of eruptions. SOHO/SUMER observations of a prominence oscillation were discussed in the context of the physics of CMEs \citep{Chen_08}. 
It was suggested that repetitive reconnections between emerging flux and the pre-existing magnetic field are the trigger mechanism of Doppler velocity 
oscillations leading to the eruption. Periods near 20 minutes, lasting 4 hours before the eruption, were detected.
However, much of the literature deals only with quiescent filaments or prominences. The oscillations are 
classified in two groups based on their amplitude: large amplitudes of several tens 
of km/s  \citep{Ramsey_66}, and small amplitudes of a few km/s or less \citep[e.g.][]{Terradas_02, Oliver_02, Engvold_08}.
Large amplitude oscillations in filaments, typically 20 km/s or more, are rare events \citep{Tripathi_09}; they are due to flares or to a local 
emergence of magnetic flux near the filament, while small amplitude oscillations do not 
affect the whole structure at a time but are of a local nature; intermediate values are also reported \citep[e.g.][]{Bocchialini_01} 
with velocities up to around 13 km/s.\\

The oscillations in quiescent prominences are classified into three categories traditionally measured from their Doppler signal: 
short-period (less than 5 minutes), intermediate periods (in the range of 6 to 20 minutes) detected in individual threads of prominences,
and long-period (between 40 to 90 minutes) detected over the whole prominence or filament \citep{Molowny_97, Oliver_02, Engvold_08, Mackay_10}. 
\citet{Ning_09} report threads in prominences exhibiting vertical and horizontal motions from Hinode/SOT observations close to the H$\alpha$ line 
center at 6563+0.076 ${\rm \AA}$; the predominant period of 
the small-amplitude velocity oscillations is about 5 minutes. \citet{Berger_10} recently reported small-scale turbulent 
upflows in quiescent prominences observed with Hinode/SOT, in the Ca II H line at 3969 ${\rm \AA}$ and the H$\alpha$ line at 6563 ${\rm \AA}$. 
The recurrence time of these upflows is roughly 5-8 minutes and the location remains active 
on timescales from tens of minutes up to several hours. \citet{Isobe_Tripathi_06} and \citet{Isobe_07} report the oscillatory 
motion of a polar crown filament, observed at 195 ${\rm {\rm \AA}}$ at a 12 minute cadence and in H$\alpha$, with a long and constant period of 
about 2 hours: 5 km/s amplitude oscillations, measured in the plane of the sky, are detected in 195-${\rm {\rm \AA}}$ images, and occured during 
the slow rise of the filament, which erupted after 3 periods. \citet{Pinter_08} observed the same filament using the same EUV dataset and wavelet spectra to
extract intensities, which show that the filament oscillates as a rigid body, with a period of about 2.5-2.6 hours, which is constant along the filament. 
All the above analysis do not reveal any signature of the eruption. Recent detection of long and ultra-long period oscillations of up to several 
hours have been reported:  
\citet{Foullon_04} report the first detection of 8-27 hour period in oscillatory intensity variations of a quiescent filament observed at 195 ${\rm {\rm \AA}}$. 
The dominant 
period is 12.1 hours. \citet{Pouget_06} report 5-6 hour period oscillations detected in the Doppler velocity of a quiescent filament, in He~I at 
584.33 ${\rm {\rm \AA}}$. \citet{Foullon_09} show that the long period detected in a quiescent filament, using 195 ${\rm {\rm \AA}}$ images, increases prior 
to its eruption.\\

Small-amplitude oscillations have been extensively modelled over the past 20 years in parallel with the development of coronal seismology 
\citep{Ballester_06}, which aims to interpret the oscillations detected in the corona and its structures in terms of MHD waves as a tool
to diagnose those structures. However, most of these MHD models of prominence oscillations deal with stable filaments and cannot 
predict their eruption. 
\citet{Vrsnak_90a} developed a simple model of cylindrical prominences to study their eruptive instability and provided a scenario for the eruption 
onset in terms of mass loss, increase of the electric current caused by slow reconnection below the filament, and twisting motions in the 
footpoints. Indeed, \citet{Koutchmy_08} observed such behaviour in a prominence just before it erupted and gave rise to a CME, compatible 
with the standard model \citep{Priest_00}, which invoke resistive instabilities. 
\citet{Filippov_08} proposed a scenario where the slow rising of a filament is interpreted as an increase of the non-potentiality of the prominence
magnetic field inside the filament before its eruption, when compared to the potential field in the surrounding corona, computed from the observed 
surface fields.\\

Several authors \citep[e.g.][]{Wiehr_89, Molowny_98,Terradas_02, Vrsnak_07, Ning_09} discussed damped oscillations in quiescent filaments, 
disappearing after a few periods, i.e. quickly damped by one or several mechanisms \citep{Oliver_09}. \citet{Arregui_10} reviewed the theoretical 
damping mechanisms in quiescent prominences as thermal effects, through non-adiabatic processes, mass flows, resonant damping and partial ionization effects.
\citet{Vrsnak_90b} reported damped post-eruptive oscillations in an active prominence, however
clear observations of prominences at limb during their eruption are extremely rare \citep{Berger_10}. We focus here on the analysis of two active 
filaments, observed during their eruption, from two sets of observations performed at high cadence: one with the Coronal Diagnostic Spectrometer (CDS) 
\citep{CDS_95} on-board SOHO \citep{Fleck_95} 
and one with the NSO/Dunn Solar Telescope (DST). These data sets are complemented by observations from 
the extreme UV imager EIT on-board SOHO and the soft X-ray telescope (SXT) on-board Yohkoh. The two data sets are thus quite different: one consists of 
spectroscopic observations in the UV (SOHO/CDS), the other of narrow band filtergrams taken in the H$\alpha$ line (DST).

However in both cases, wavelet analysis of the observed signals in the two filaments leads to evidence of intermediate-period, small-amplitude 
oscillations (less than 20 km/s) apparently cotemporal with a total or partial eruption of the filaments. 
Moreover, the CDS observations in He~I at 584.33 ${\rm \AA}$ and in Mg~X at 609.79 ${\rm \AA}$ reveal intensity oscillations in phase in both lines, 
preceded by velocity oscillations in phase in both lines. Both data sets reveal a vertical oscillatory motion inside the filaments: 
the CDS observations in He~I at 584.33 ${\rm \AA}$ exhibit a damped oscillation. The DST 2D imaging observations in the H$\alpha$ wings 
reveal that parts of the filament are moving vertically in opposite directions, while the filament is moving over the solar surface, before its 
disappearance; no damping is measured for this event. In addition, a flare and a loop eruption are observed in Yohkoh/SXT.

\section{Damped oscillatory motions in an eruptive filament, observed simultaneously in He~I and Mg~X}
\subsection{CDS data set properties}

During the 11th MEDOC campaign we obtained long observations of a filament with 
CDS. 
The observations were carried out on May 30, 2003, 
between 6:40 UT and 13:51 UT, in He~I at 584.33 ${\rm \AA}$ and Mg~X at 609.79 ${\rm \AA}$. 
Table \ref{tabcdsfile} summarizes the observational parameters of the sequence.\\ 

\begin{table}%[htbp]
\caption{CDS observational parameters.}
    \label{tabcdsfile}
\centering
\begin{tabular}{c c}
    \hline
    Observed Lines &  He~I (584.33 ${\rm \AA}$) \\  
                   & Mg~X (609.79 ${\rm \AA}$) \\
    Slit dimensions (x $\times$ y) & 2\arcsec{}$\times$240\arcsec{} \\   
    Spatial resolution along $y$ &  1.68\arcsec{} \\
    Average temporal resolution $\Delta t$ &  24.6 s \\
    Duration &  7 h 10 \\
    Beginning (UT) & 6:40 \\
    End (UT)& 13:51 \\
    \hline
\end{tabular}

\end{table}

At the beginning of the observations, the slit centre was located on 
the solar disc at $x$ = 690 arcsec and $y$ = 418 arcsec. During the observations, the slit's $x$ position was regularly shifted, in order to 
compensate for solar rotation in order to follow the same part of the filament; at the end of 
the observation the slit was at $x$ = 721 arcsec. The data have three dimensions: time, wavelength and spatial position on the solar disc. At each time $t_i$, 
a spectrum within a constant wavelength range $\Delta\lambda$ (centred on either the He~I or Mg~X line) was obtained for each $y_j$ position along the 
slit, which is 141 pixels high (240 arcsec). The usual corrections were applied to these data: flat fielding, cosmic ray removal, and correction of some 
geometric instrumental effects. The method applied to identify the signal due to the filament is described by \citet{Pouget_06}: for each time
$t$, the pixels ($t,y$) that belong to the filament were those whose intensity is below a threshold 
chosen such that the filament width remained constant in time.\\
To compute the Doppler velocity, we used a reference profile, averaged over the whole field of view and over
time: the velocity studied below is consequently a relative quantity, however the spectral analysis is not affected. For each pixel, 
the line profile is fitted to a Gaussian profile (plus a component taking into account the asymmetric shape of the profiles obtained by CDS 
after the 1998 SOHO recovery, see \citet{Thompson_99} allowing the extraction of its central wavelength\footnote{The routine used is FIT$\_$SPEC, available 
in the SolarSoftWare Library of IDL}). However, it should be noted that it is very difficult to get accurate absolute line-shifts measured 
by CDS \citep{Pouget_06}. The typical uncertainty of the relative velocities is 6 km/s, due to the uncertainty of the pixel position on 
the CCD in the spectral dimension.\\

\begin{figure*}
   \centering

   \includegraphics[width=9cm]{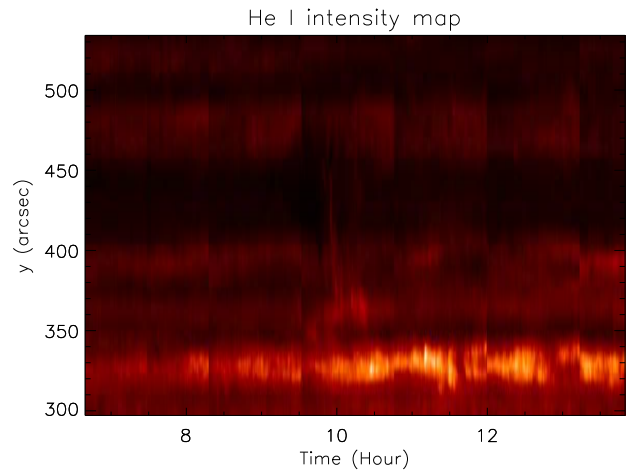}
   \caption{He I intensity map, as a function of time and along the slit.  The filament is located between $y$ = 409\,arcsec and $y$ = 459\,arcsec.}
              \label{series_intensity_file}
\end{figure*}

\begin{figure*}
   \centering
   \includegraphics[width=9cm]{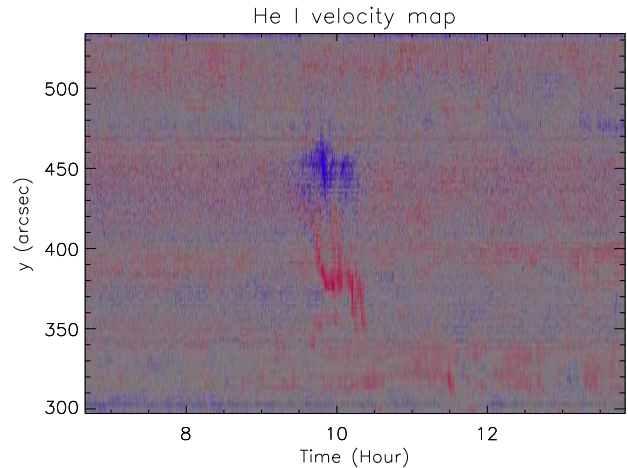}
   \caption{He I velocity map, as a function of time and along the slit.  The filament is located between $y$ = 409 and  459\,arcsec.}
              \label{series_velocity_file}
\end{figure*}

Figures \ref{series_intensity_file} and \ref{series_velocity_file} show 
the intensity averaged over the waveband centred on He~I and the associated velocity series: the filament is visible in intensity, as 
the dark zone between slit positions $y$ = 409\,arcsec and 459\,arcsec. The eruption occured between 9:00 and 11:00 UT: a strong blue shift is 
clearly visible in the filament, while a strong red shift is visible between the positions $y$ = 350\,arcsec and $y$ = 400\,arcsec, below the position 
of the filament (Figure \ref{series_velocity_file}). At the same time, an intensity enhancement around position $y$ = 334\,arcsec is clearly visible 
(Figure \ref{series_intensity_file}). \\ 

\begin{figure*}
   \centering
   \includegraphics[width=8cm, angle=90]{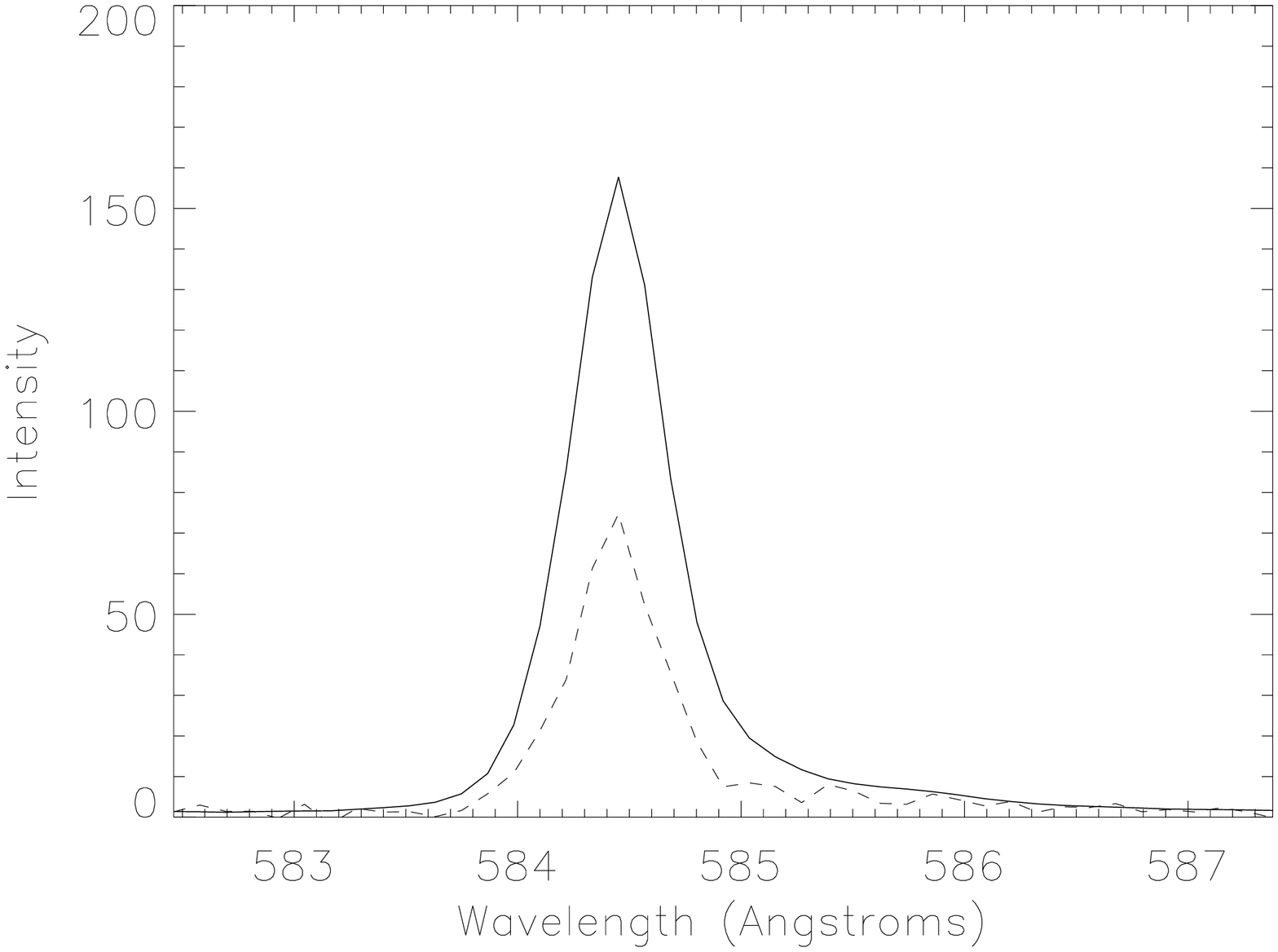}
   \caption{Mean He~I profiles (arbitrary units): solid line, average over the $y$ dimension and over the whole temporal sequence; 
    dashed line, average over the filament and the whole temporal sequence.}
              \label{profilsmoyensHeI}
\end{figure*}

\begin{figure*}
   \centering
   \includegraphics[width=8cm, angle=90]{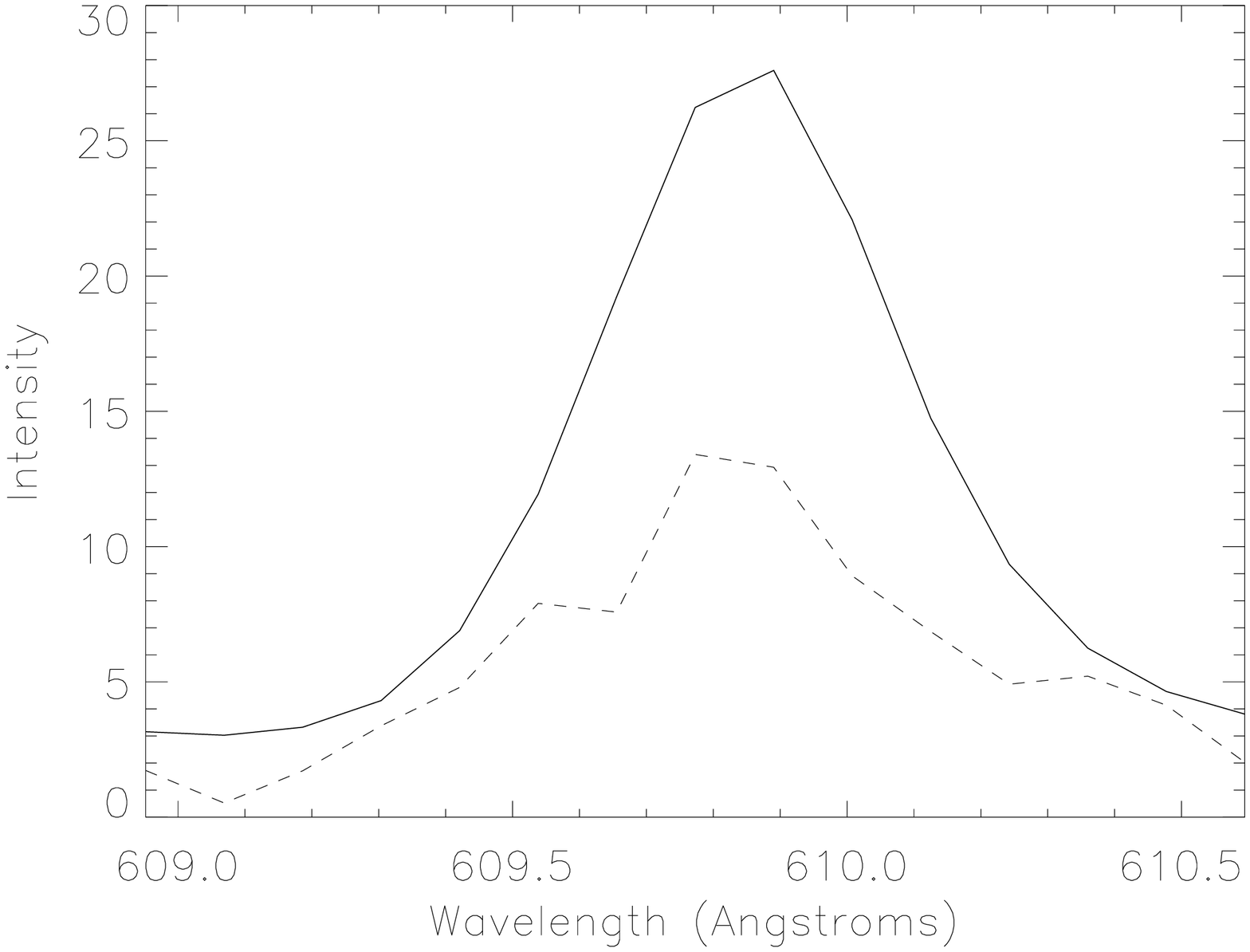}
   \caption{Mean Mg~X profiles (arbitrary units): solid line, average over the $y$ dimension and over the whole temporal sequence; 
    dashed line, average over the filament and the whole temporal sequence.}
              \label{profilsmoyensMg10}
\end{figure*}

Figure \ref{profilsmoyensHeI} and Figure \ref{profilsmoyensMg10} show, respectively, the He~I and Mg~X profiles, averaged 
over the $y$ dimension and the whole temporal sequence, which are compared respectively to the He~I and Mg~X profiles, averaged 
over the filament, between the positions $y$ = 409\,arcsec and $y$ = 459\,arcsec (and the whole temporal sequence). 
The profiles are in emission, but the filament profile is clearly less intense by more than a factor of two at its maximum than the average profile 
along the slit, for both lines. The optically thick He~I line is formed at a temperature $T\sim$20000\,K, as mentionned by \citet{Regnier_01}, and 
it is sensitive to the motions of the prominence, while the coronal Mg~X line is formed at $T\sim$1.2\,MK.


\subsection{Overview of the observed area}

\begin{figure*}
   \centering
   \includegraphics[width=17cm, angle=90]{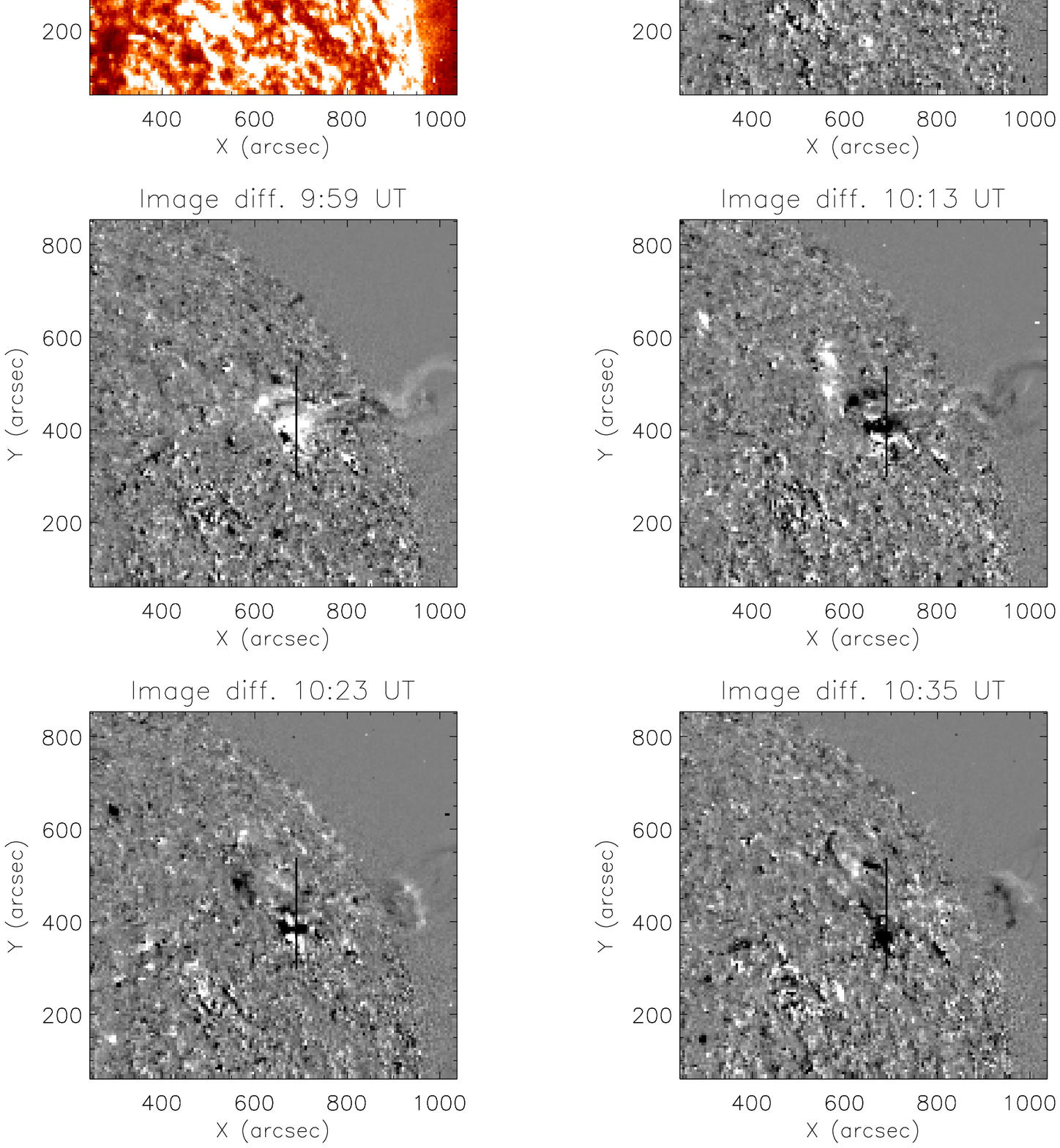}
   \caption{Top left: EIT image of the area of the filament studied, in 304 ${\rm \AA}$ obtained at 9:35 UT, on May 30, 2003. Top right, middle 
and bottom rows: EIT difference image sequence of the filament in 304 ${\rm \AA}$ between 9:47 and 10:35 UT (from left to right, top to bottom), 
obtained at a near-12 minute cadence. 
The filament eruption begins at 9:35 UT.  The CDS slit position is indicated by a vertical black line}
              \label{EITcontext}
\end{figure*}

\begin{figure*}
   \centering
   \includegraphics[width=12cm, angle=90]{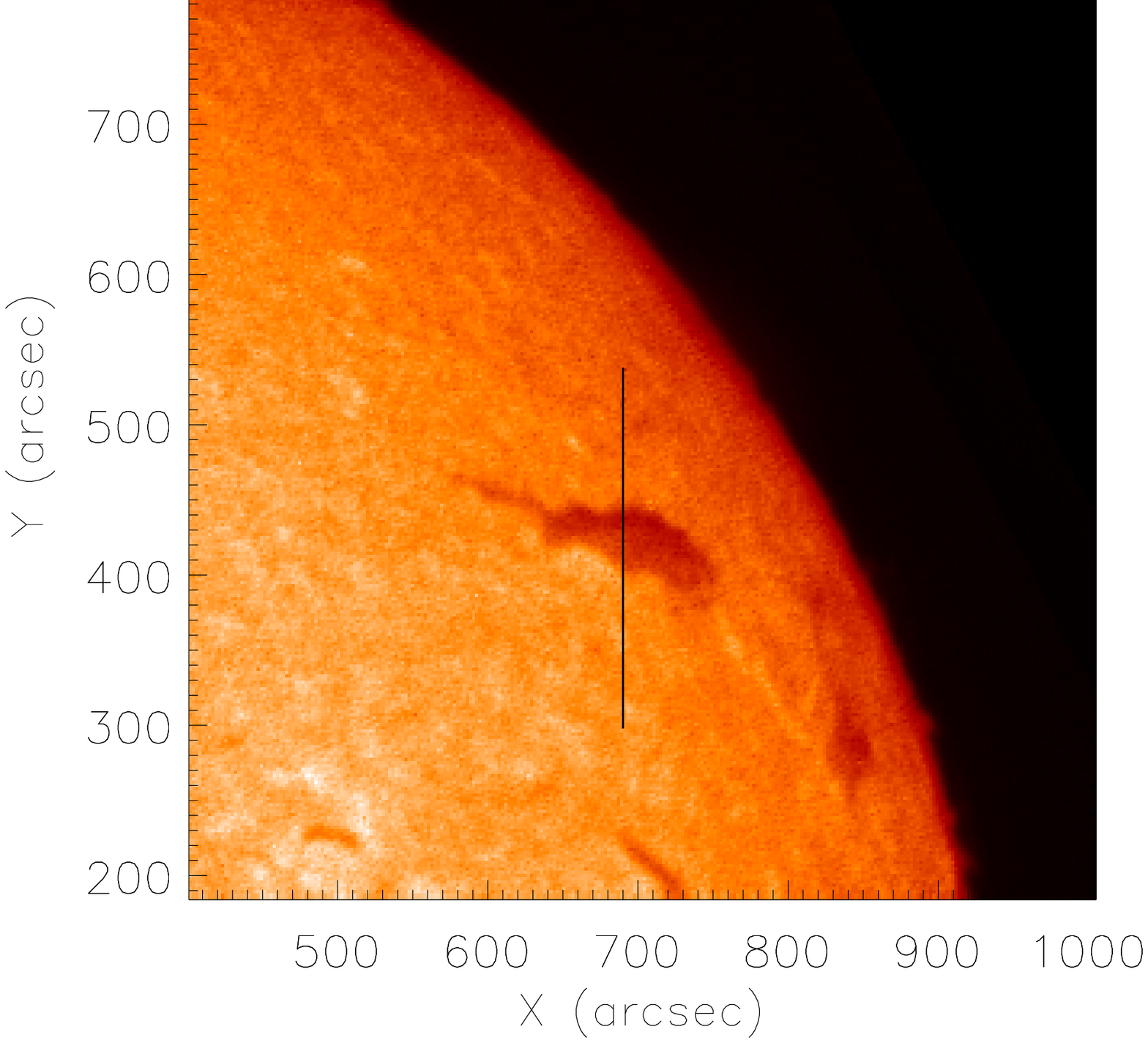}
   \caption{Kiepenheuer Institute image of the area of the filament studied, in H$\alpha$, obtained at 8:07 UT on May 30, 2003. 
The CDS slit position is indicated by a vertical black line.}
              \label{Halphacontext}
\end{figure*}

The EIT \citep{EIT_95} instrument provided the 
contextual field of view. Figure \ref{EITcontext} shows (top left) an EIT image of the area around the filament studied in He~II at 304 ${\rm \AA}$ 
obtained at 9:35 UT. The CDS slit position is indicated by a vertical black line on the disc. Figure \ref{Halphacontext} shows an H$\alpha$ image 
obtained at 8:07 UT at the Kiepenheuer Institute. The CDS slit position is indicated by a vertical black line.

The EIT differences between two successive images in 304 ${\rm \AA}$ (Figure \ref{EITcontext}) reveal that a part of the 
filament lifts off from 9:35 to 10:35~UT (co-temporal with the CDS observations) and is seen as a prominence on the west limb, 
while the part on the disc does not erupt. 
The prominence disappears into the corona but is not associated with a visible Coronal Mass Ejection, as confirmed by 
SOHO/LASCO observations. The event is comparable to the so-called confined filament eruption studied by \citet{Filippov_02}.\\
The CDS slit cuts across the part of the filament that does not erupt. The velocities detected in this part of the filament could be 
the signature of an excitation due to the erupting part of the filament.\\
Previously, a flare was observed with EIT in the same region at 6:30~UT.  The source of the destabilization of the filament could be a wave 
resulting from this flare.\\
The sequence of MDI magnetograms \citep{MDI_95} obtained between 8:03 UT and 12:47 UT does not exhibit any particular signature that could 
be linked to the eruption.

\subsection{Spectral analysis of the signal in the filament}
The Doppler velocity and the intensity time series are derived in the filament for both lines, He~I and Mg~X, by 
averaging the 30 profiles between  the positions $y$ = 409\,arcsec and 459\,arcsec, for each temporal exposure. 
In the following, as the signals show strong variations in time, we used a wavelet analysis to explore their frequency content.

\begin{figure*}
   \centering
\includegraphics[width=6cm]{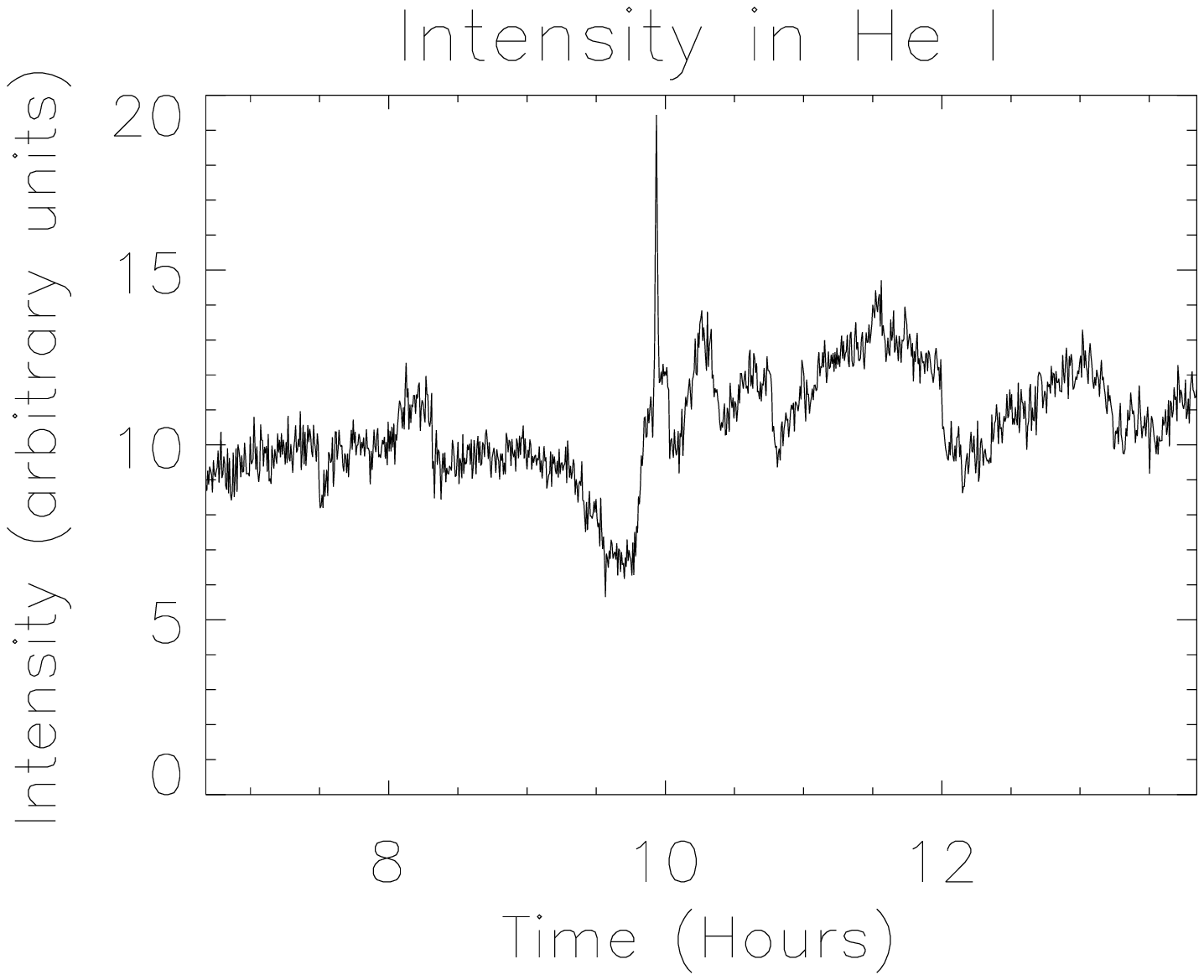}
\includegraphics[width=6cm]{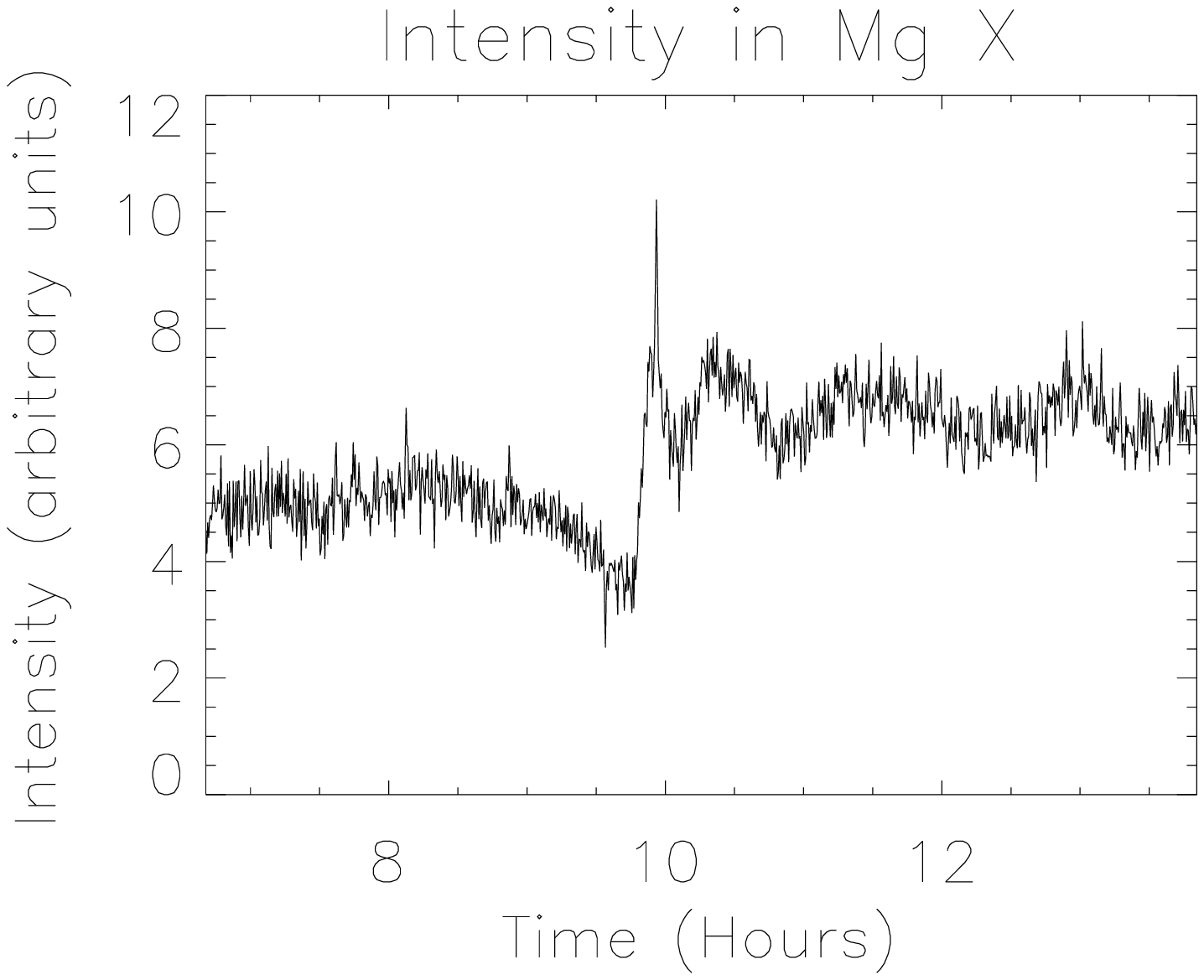}\\
\includegraphics[width=4.8cm, angle=90]{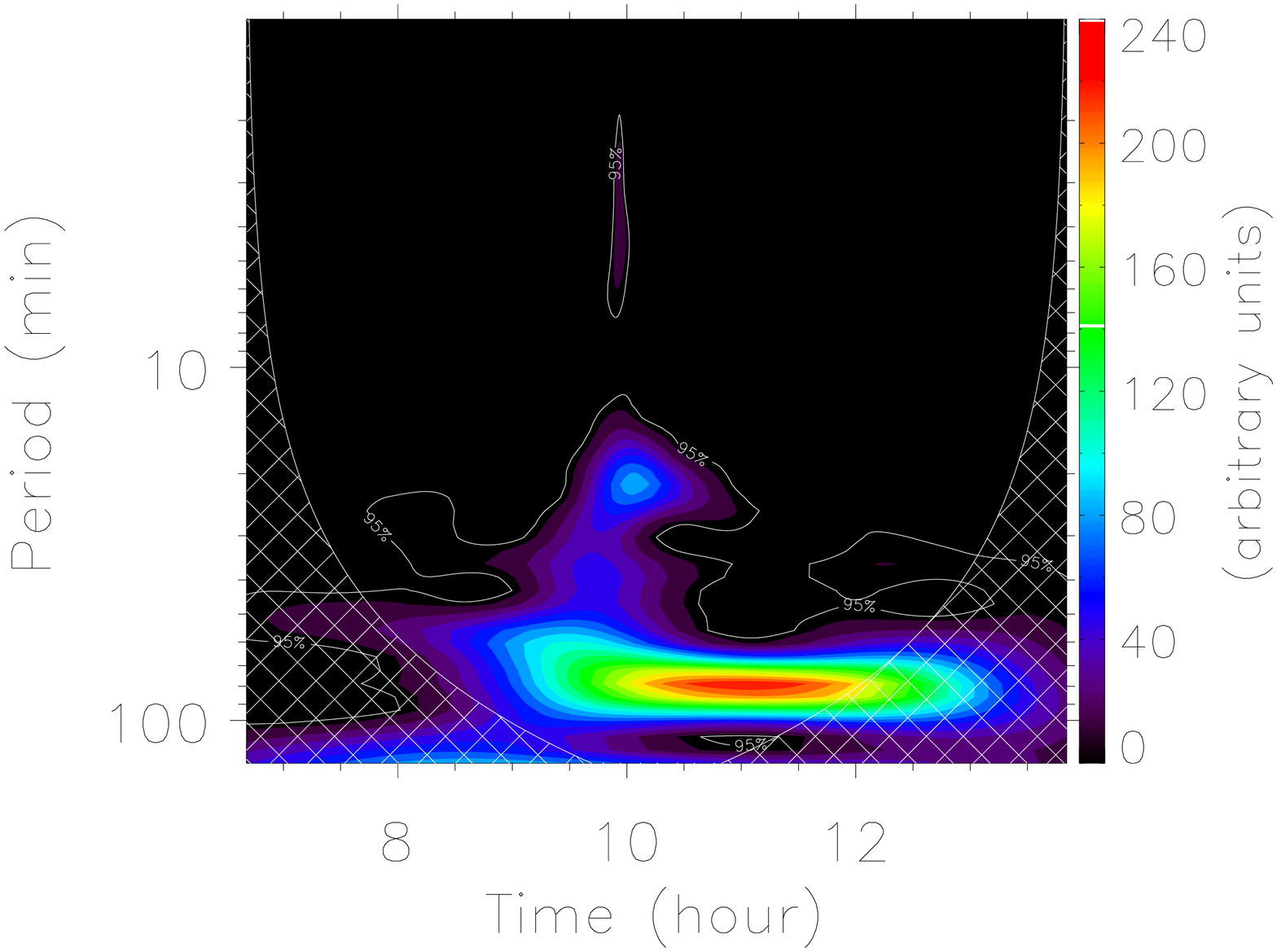}
\includegraphics[width=4.8cm, angle=90]{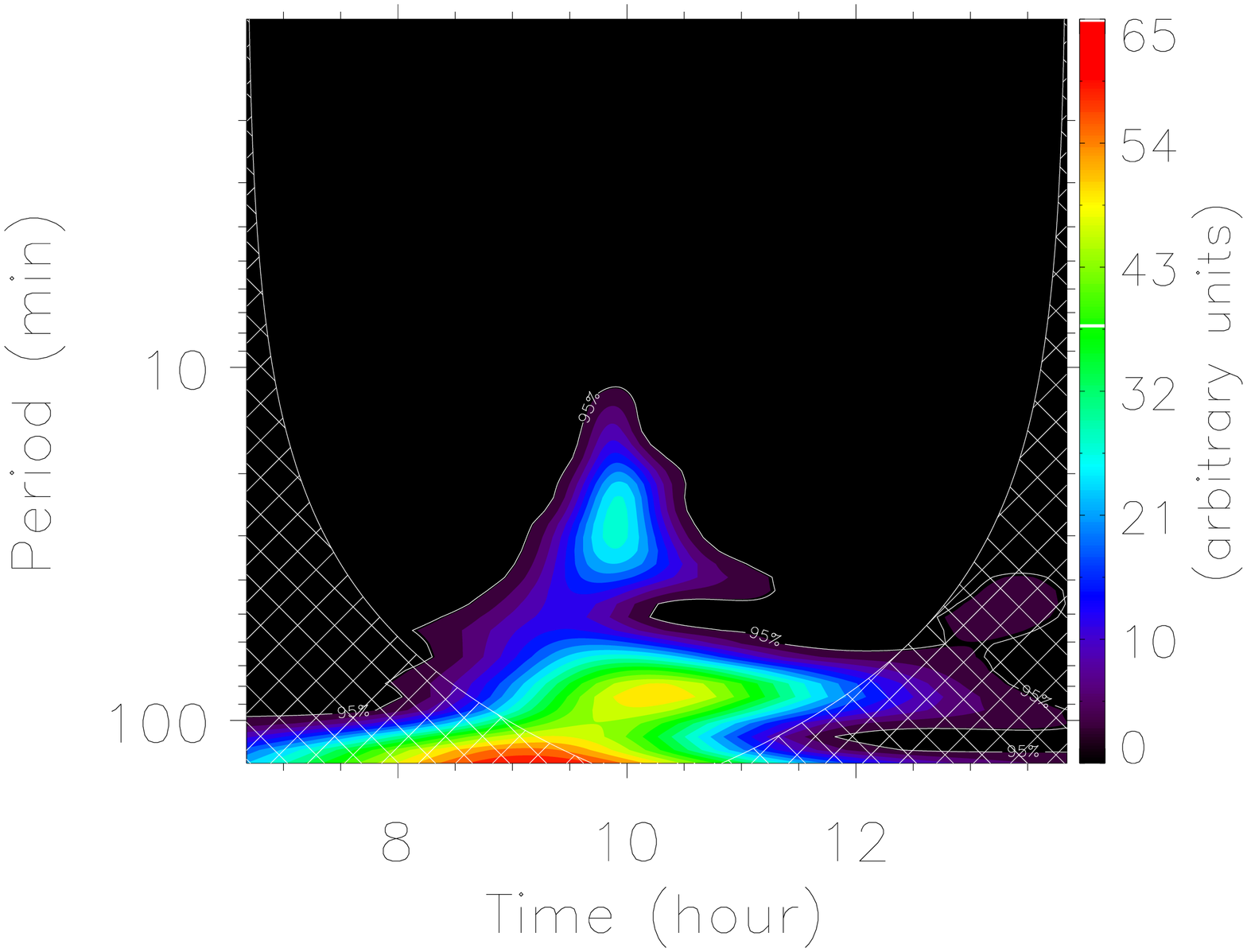}\\
   \caption{Intensity in the filament (top) as a function of time
   in He~I (left) and in Mg~X (right), and the corresponding wavelet analysis (bottom); significant power is within the 95$\%$
confidence level contour and outside the hatched cones of influence.}
              \label{he1mg10ifile}
\end{figure*}

Results in intensity:\\
Figure \ref{he1mg10ifile} shows the intensity variations and the corresponding wavelet analysis, for the He~I and Mg~X lines. 
We used the code written by \citet{Torrence_98}\footnote{available at http://paos.colorado.edu/research/wavelets}.
In both lines, the intensity begins to decrease slowly before the eruption, and then increases impulsively 
at the beginning of the eruption, around 9:55 UT, simultaneously in He~I and Mg~X. 
The intensity in He~I oscillates with a period around 22 minutes, from 9:00 UT to 11:00 UT.
The intensity oscillates also with a period around 80 minutes, from 9:00 UT to 13:00 UT. 
In Mg~X, the intensity oscillates with a period around 30 minutes, from 9:00 UT to 11:00 UT and also oscillates with a period around 80 minutes 
from 9:00 UT to 11:00 UT.

\begin{figure*}
   \centering
\includegraphics[width=6cm]{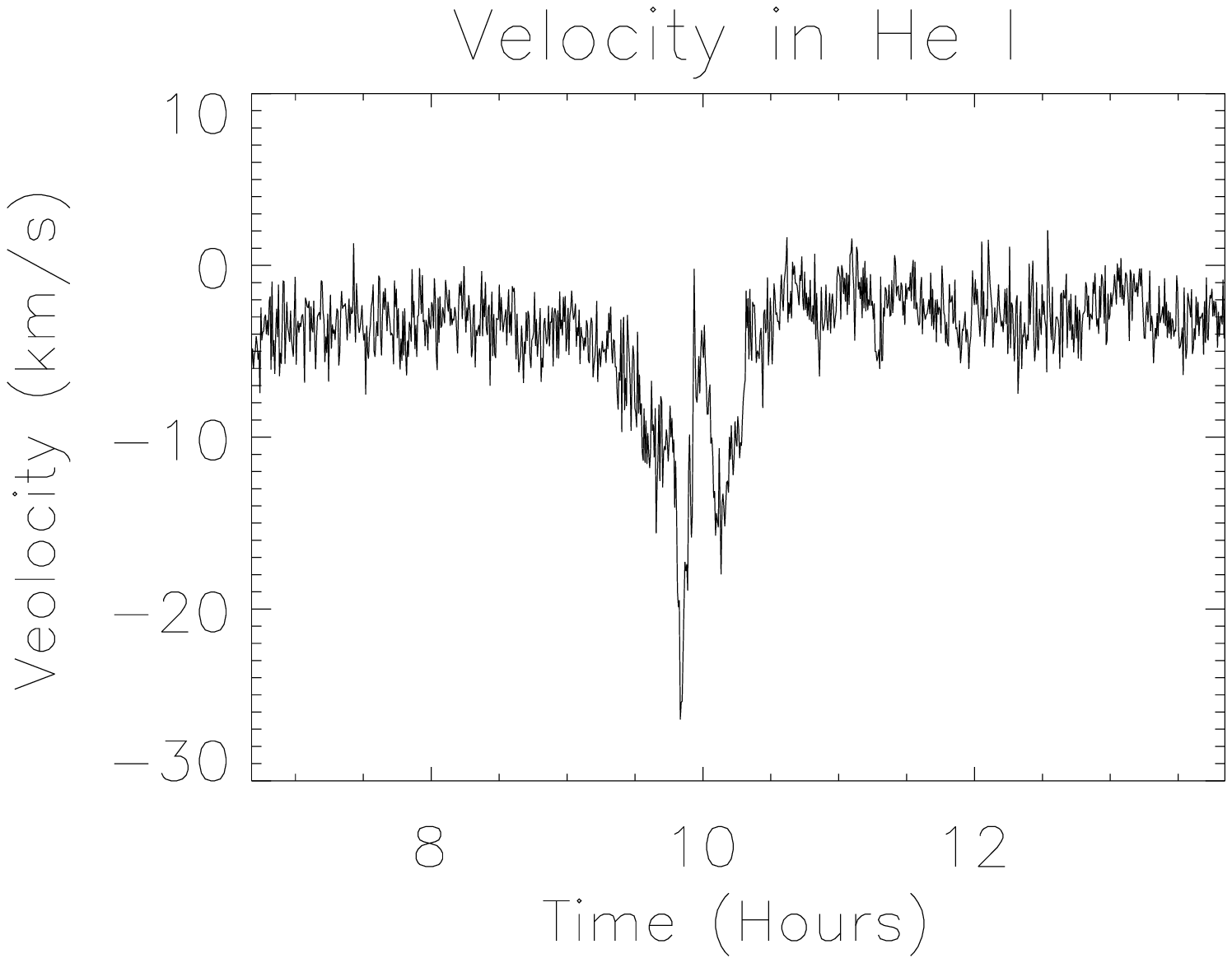}
\includegraphics[width=6cm]{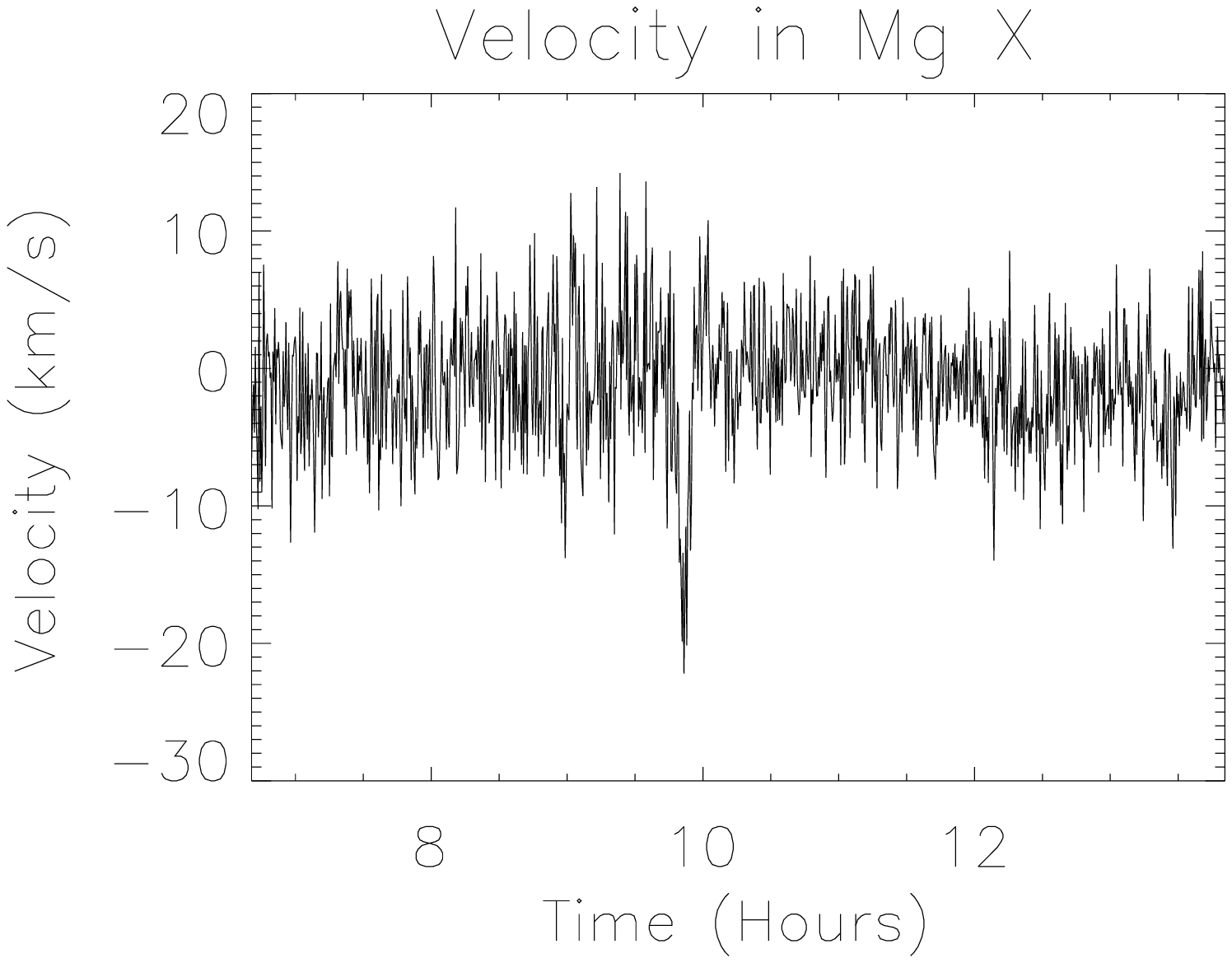}\\
\includegraphics[width=4.8cm, angle=90]{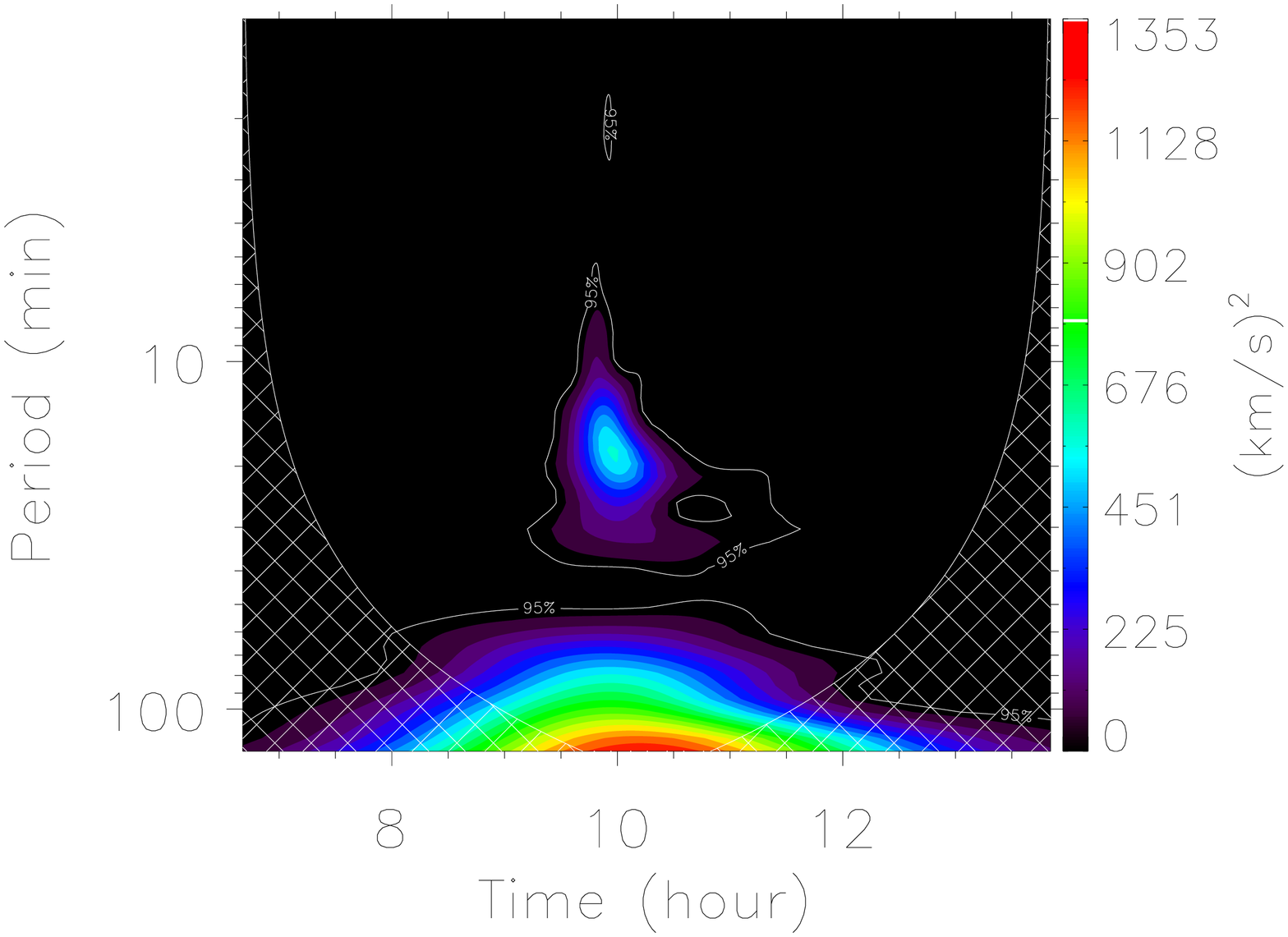}
\includegraphics[width=4.8cm, angle=90]{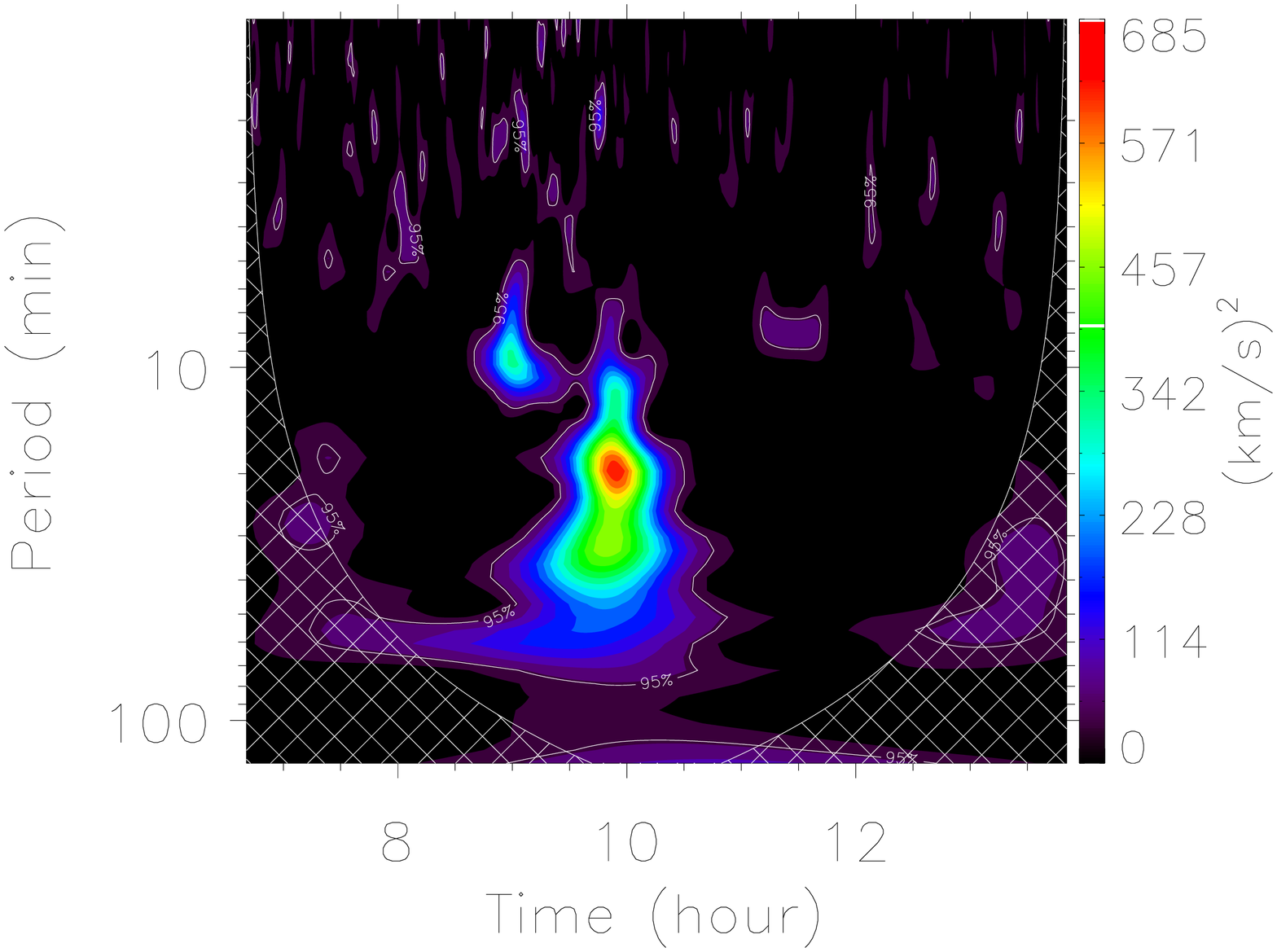}\\
   \caption{Velocity in the filament (top), as a function of time, in He~I (left) and in Mg~X (right), and the corresponding wavelet analysis (bottom); 
significant power is within the 95$\%$ confidence level contour and outside the hatched cones of influence.}
              \label{he1mg10vfile}
\end{figure*}

Results in velocity:\\
Figure \ref{he1mg10vfile} shows the velocity and the corresponding wavelet analysis, in He~I and Mg~X. 
In both lines, a strong upward impulse is detected simultaneously with the eruption, around 10:00 UT. The wavelet analysis confirms 
that the period is about 20 minutes in both lines. Mg~X also exhibits 10 minute period oscillations. The oscillations last roughly one hour 
and are quickly damped.

\begin{figure*}
  \centering
  \includegraphics[width=8cm, angle=90]{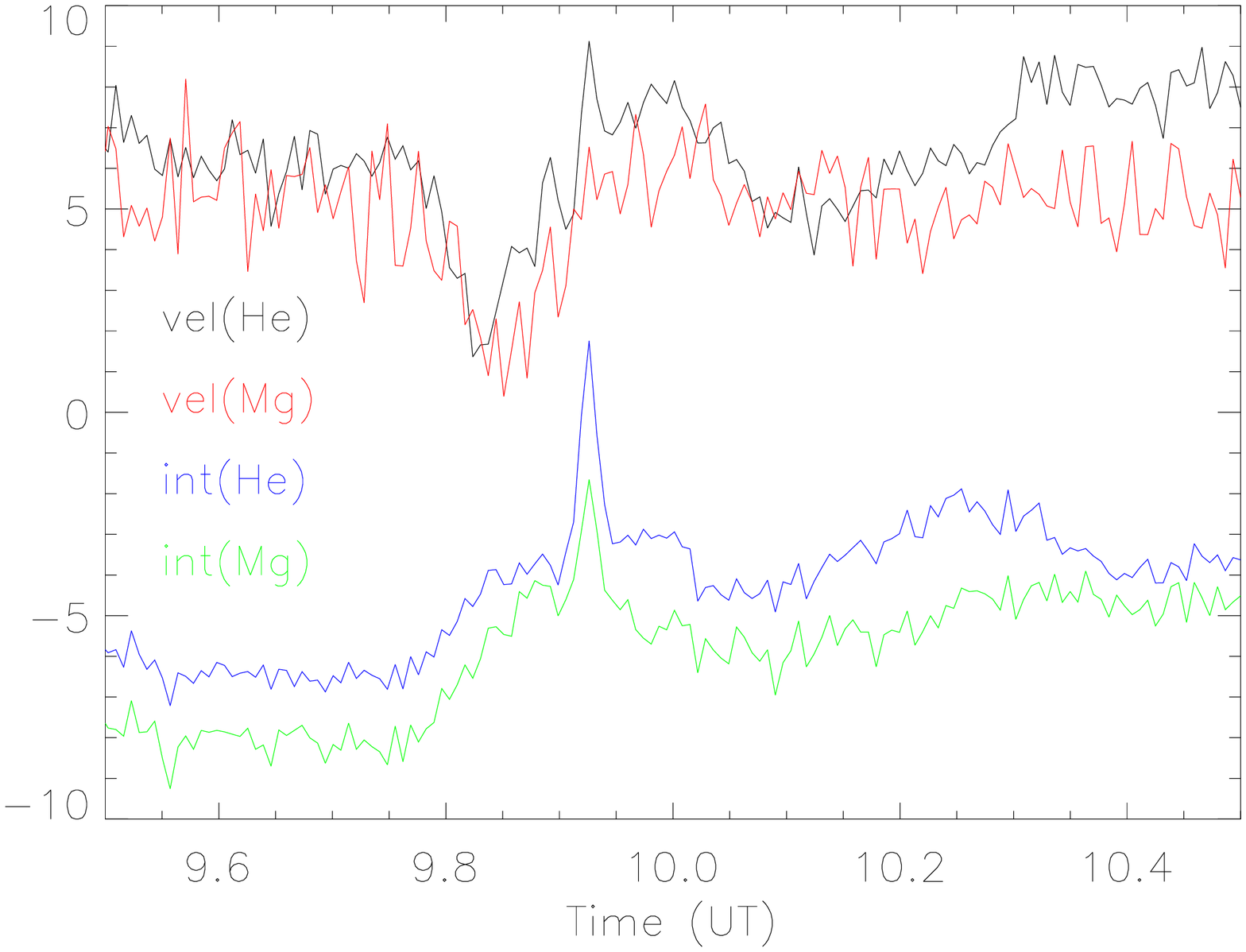}
  \caption{Comparison of the signals in the filament between 9:00 UT and 11:00 UT, in intensity (He~I in blue and Mg~X in green) and in velocity 
 (He~I in black and Mg~X in red).}
              \label{dephasage}
\end{figure*}

Figure \ref{dephasage} displays a close-up around the eruption of the intensity and velocity in both lines simultaneously 
(in arbitrary units, for easier 
temporal comparison between the signals).  The velocity oscillations in Mg~X are in phase with the velocity oscillations in He~I, 
suggesting that the hot corona in which the filament channel is embedded responds coherently to the oscillating filament.
The intensity signals in both lines are in phase, showing simultaneous variations as well.
Moreover, the peak of intensity seen in both lines appears after the beginning of the oscillations detected in velocity,
which indicates that the dynamical effects precede the thermal effects.\\

\begin{figure*}
   \centering

\includegraphics[width=10cm, angle=90]{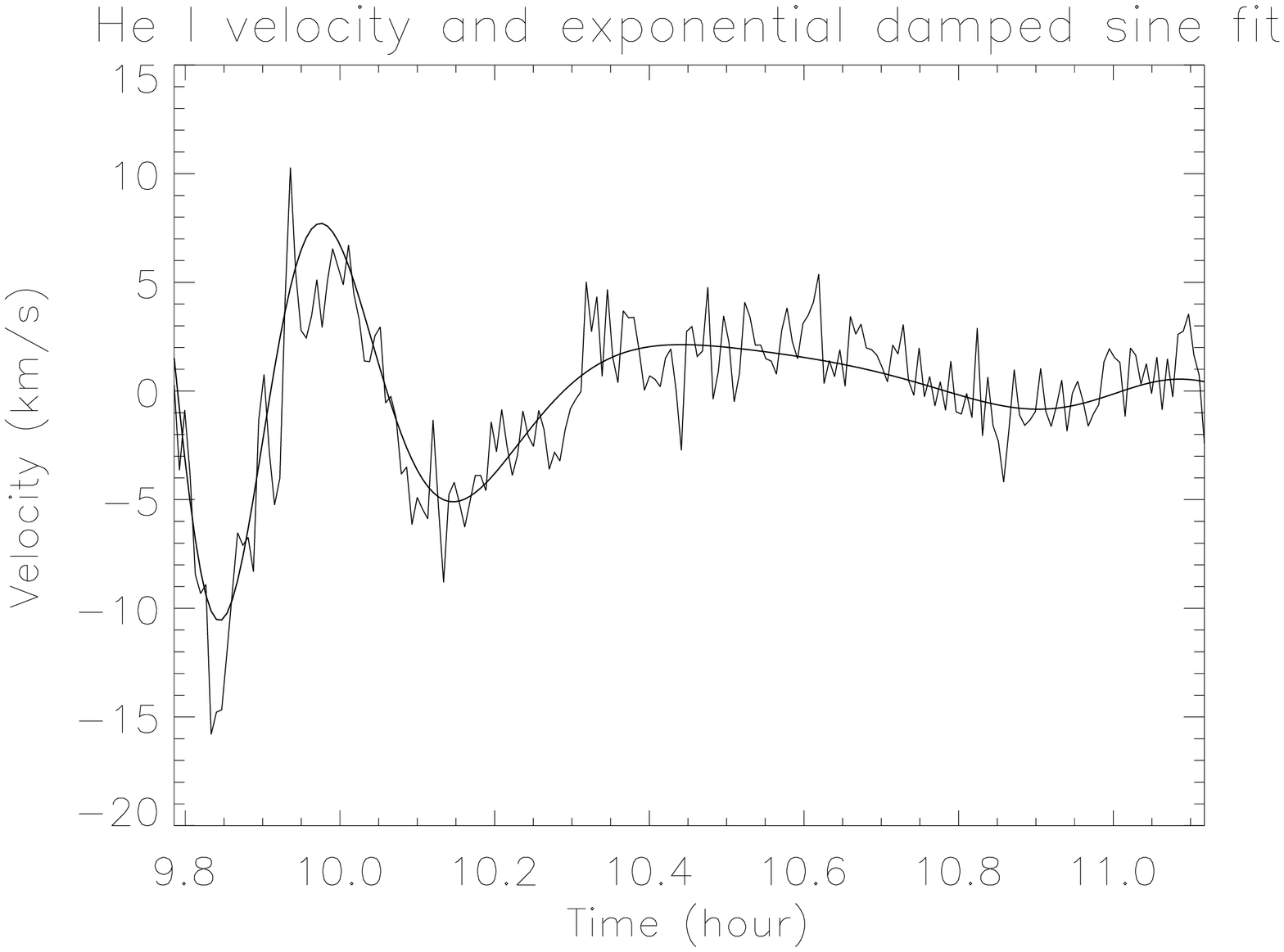}
   \caption{Residual velocity in the filament in He~I during the eruption, and its fit. In order to emphasize and fit the low period signal, 
the long period signal has been removed.}
              \label{fit-sinus-amortie}
\end{figure*}

A more refined analysis of the velocity shows that the signal in He~I can be separated into two parts: one related to the oscillation 
itself with an apparent period 
around 22 minutes seen with the wavelet analysis on Figure \ref{he1mg10vfile} and another with a period longer than 80 minutes. 
The residual velocity (after the subtraction of the long-period trend 
using a 50 minute boxcar running average) is displayed on Figure \ref{fit-sinus-amortie}
and corresponds to an oscillation that can be fitted by an exponentially damped sine function, with a frequency linearly decreasing with time. 
The velocity can then be expressed as followed: $v(t)=A\sin(2\pi(b-a(t - t_0))(t - t_0)+\phi)e^{-(t - t_0)/\tau}$, with $A$ = 12.2 km/s, 
$a= 2.34 10^{-4} \, \mathrm{mHz/s}$, $b = 1.26 \, \mathrm{mHz}$, $\phi=3.02$ rad, $\tau=25.08$ minutes and $t_0$ = 9.78 hours, with $\tau$/T=1.9 (and $T = 1/b$ 
the initial period of the oscillations). The pseudo-period becomes roughly 50 minutes, one hour after the velocity maximum.\\

\subsection{Summary of the analysis}
The analysis of the CDS observations performed on an eruptive filament in the He~I and Mg~X lines exhibit strong intensity and velocity
oscillations during the eruption. Periods range from 20 to 30 minutes in intensity, periods longer than 80 minutes are also detected.
For the velocity measured in He~I, the amplitude of the main 20 minute period oscillations is 12.2 km/s, with a damping time of 25 minutes.
The EIT 304 ${\rm \AA}$ images show that the filament erupts during the CDS observations.\\

The Fig.\ref{dephasage} compares the signals in intensity and velocity. First, one can note that the oscillation in velocity starts at the same moment than a 
long period signal in intensity but that a sudden brightening in intensity occurs well after the oscillations started. It could suggest that a thermal event 
occurs after the oscillatory behaviour started. More precisely, just before the brightening, between 9.83 UT and 9.95 UT, one can distinguish 
short period oscillations of a few minutes in the He~I velocity signal. 
These oscillations can be seen on the wavelet diagrams Fig.\,\ref{he1mg10ifile} and \ref{he1mg10vfile}, both in intensity and velocity, in He~I.

In the next section, we analyse the similar period oscillations detected in another eruptive filament, measured from the ground, in 
the wings of the H$\alpha$ line.

\section{Oscillation and rotational motions detected in an eruptive filament, observed in the wings of H$\alpha$}
\subsection{DST dataset properties}
\begin{figure*}
   \centering
   \includegraphics[width=5cm]{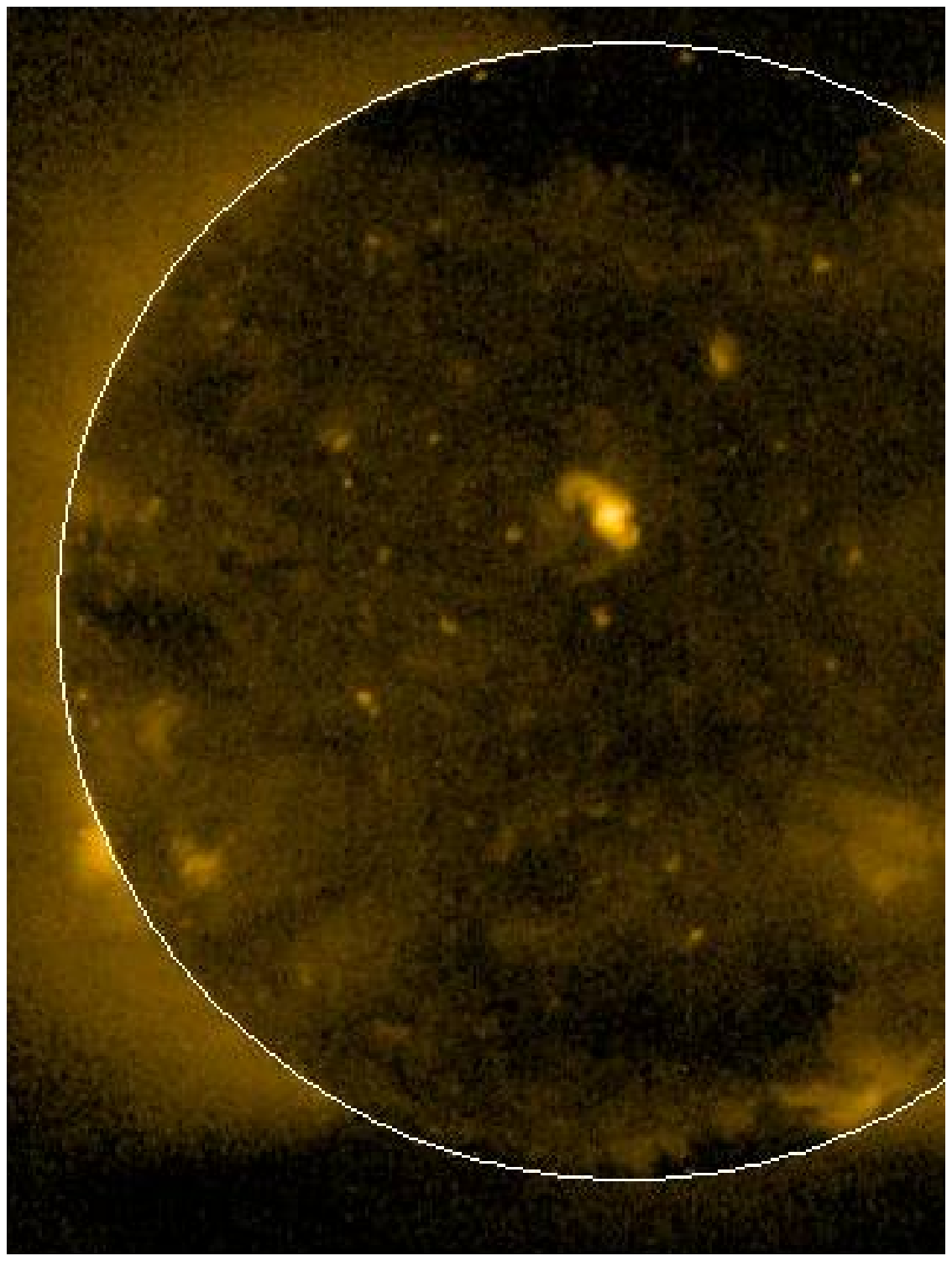}
   \includegraphics[width=5cm]{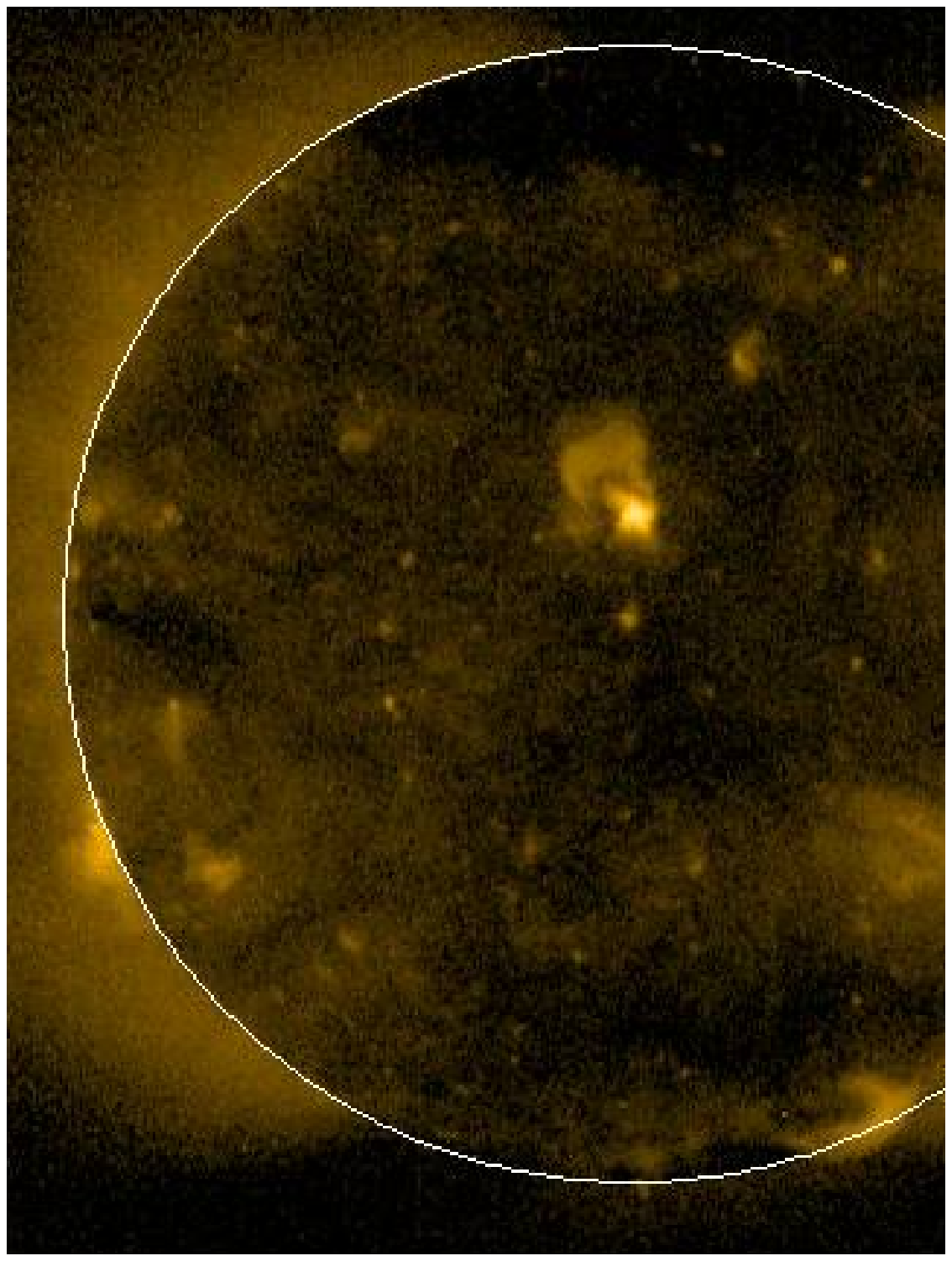}
   \caption{Yohkoh/SXT images taken at 15:32 UT (left) and 19:18 UT (right) on September 18, 1994, showing the observed region and its eruption, near 
the disk centre.}
              \label{yohkoh}
\end{figure*}

In the following, we analyse ground-based observations of a similar event observed in NOAA AR 7779 on September 18, 1994, 
in the blue and red wings of H$\alpha$ line at $\lambda_{\pm} = 6562.8 \pm 0.75 {\rm \AA}$, from 15:32 UT to 17:22 UT, using a Universal 
Birefringent Filter UBF, operated in the 0.24 ${\rm \AA}$ Full Width at Half Maximum (FWHM) mode.
This event occurred very near the solar disc center, while the event studied previously was located near the limb.
The square images are 350\,arcsec wide and centred on N 17.3, E 1.3. Each set of filtergrams was obtained at a cadence of 
1 minute, with a 0.34 arcsec/pixel spatial resolution, during 110 minutes at the NSO/Sacramento Peak Dunn Solar Telescope
\citep{Baudin_97, Baudin_98}.\\
 The photospheric context of the region, including a discussion of dynamical effects, was given in a preliminary paper by \citet{Baudin_97}.
 Following this observation sequence, a large soft X-ray (SXR) eruption was recorded by the Yohkoh Soft X-Ray Telescope (YOHKOH/SXT)
(Figure \ref{yohkoh}).\\

\begin{figure*}
   \centering
   \includegraphics[width=8cm,angle=90]{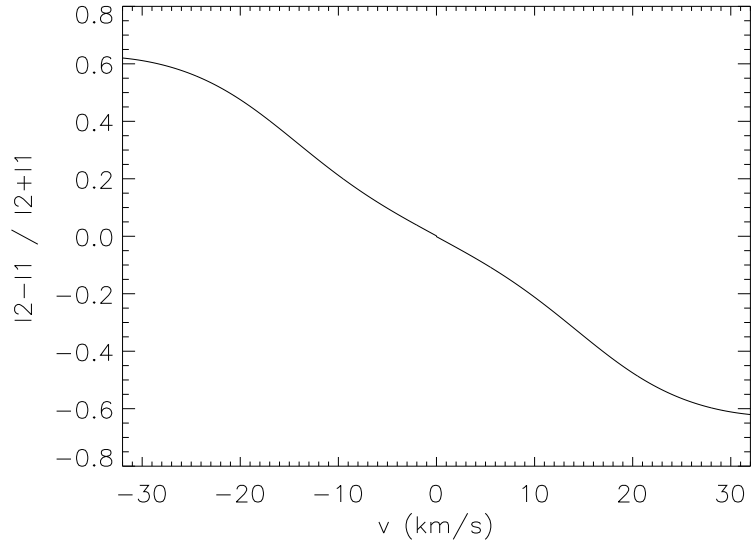}
   \caption{Calibration curve $\frac{I_{2}-I_{1}}{I_{2}+I_{1}}=f(v)$ for NSP/DST observations.}
              \label{etalonnage-i-v}
\end{figure*}

\citet{Abramenko_96} obtained vector magnetograms of NOAA AR\,7779 and determined
the helicity and corresponding electric currents. Although AR\,7779 was located close to the equator (N17E02), the negative
current helicity found is typical for a northern hemisphere event; the value of the current helicity could satisfy the energy requirements for the filament 
eruption.\\
Our analysis is now focused on the study, in intensity and Doppler velocity, of the eruption of the filament centred in the f.o.v.. 
The wing intensity is the sum of the signals obtained in the red and blue wings. The Doppler velocity is 
obtained using the following method \citep{Malherbe_97, Georgakilas_01}:\\
an H$\alpha$ reference center-disk profile \citep[]{Delbouille_73}\footnote{available at the BASS2000 database: {\tt http://bass2000.bagn.obs-mip.fr/}}
is convolved with a Gaussian profile with a 0.24 ${\rm \AA}$ Full Width at Half Maximum, corresponding to the width of 
the filter used for the observations. 
This profile is shifted by several steps $\Delta \lambda = \frac{FWHM}{2}$; for each shifted profile, the ratio
$\frac{I_{2} - I_{1}}{I_{2} + I_{1}}$ is computed, where $I_{2}$ is the intensity in the red wing, and $I_{1}$ is the intensity in the blue wing, at
$\lambda_{\pm} = 6562.8 \pm 0.75 {\rm \AA}$. A calibration curve $\frac{I_{2} - I_{1}}{I_{2} + I_{1}}= f(v)$ is obtained assuming that the shifting is equal to
 $\frac{v}{c} \lambda_{0}$, with $\lambda_{0} = 6562.8 {\rm \AA}$.
The curve is displayed on Figure \ref{etalonnage-i-v} assuming no differential changes 
of the line profile during the event.  We note that this assumption is model-independent but may lead to an underestimation of the deduced velocities 
\citep{Labrosse_10}. The velocities are readily deduced from this calibration curve, in agreement with the calibration of \citet{Georgakilas_01} who used 
the same UBF mode.

%\clearpage

\begin{figure*}
   \centering
   \includegraphics[width=4cm,angle=90]{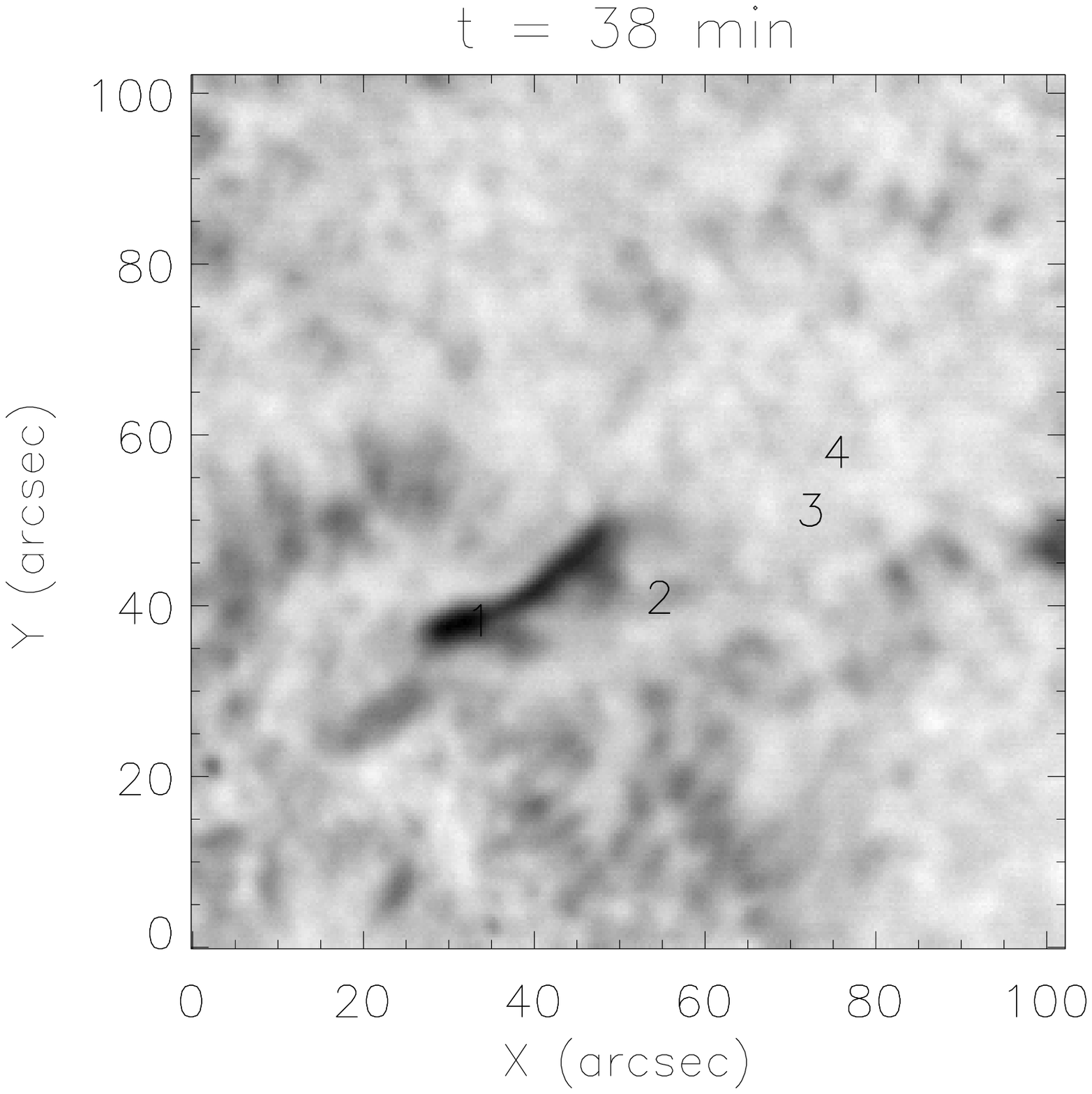} 
   \includegraphics[width=4cm,angle=90]{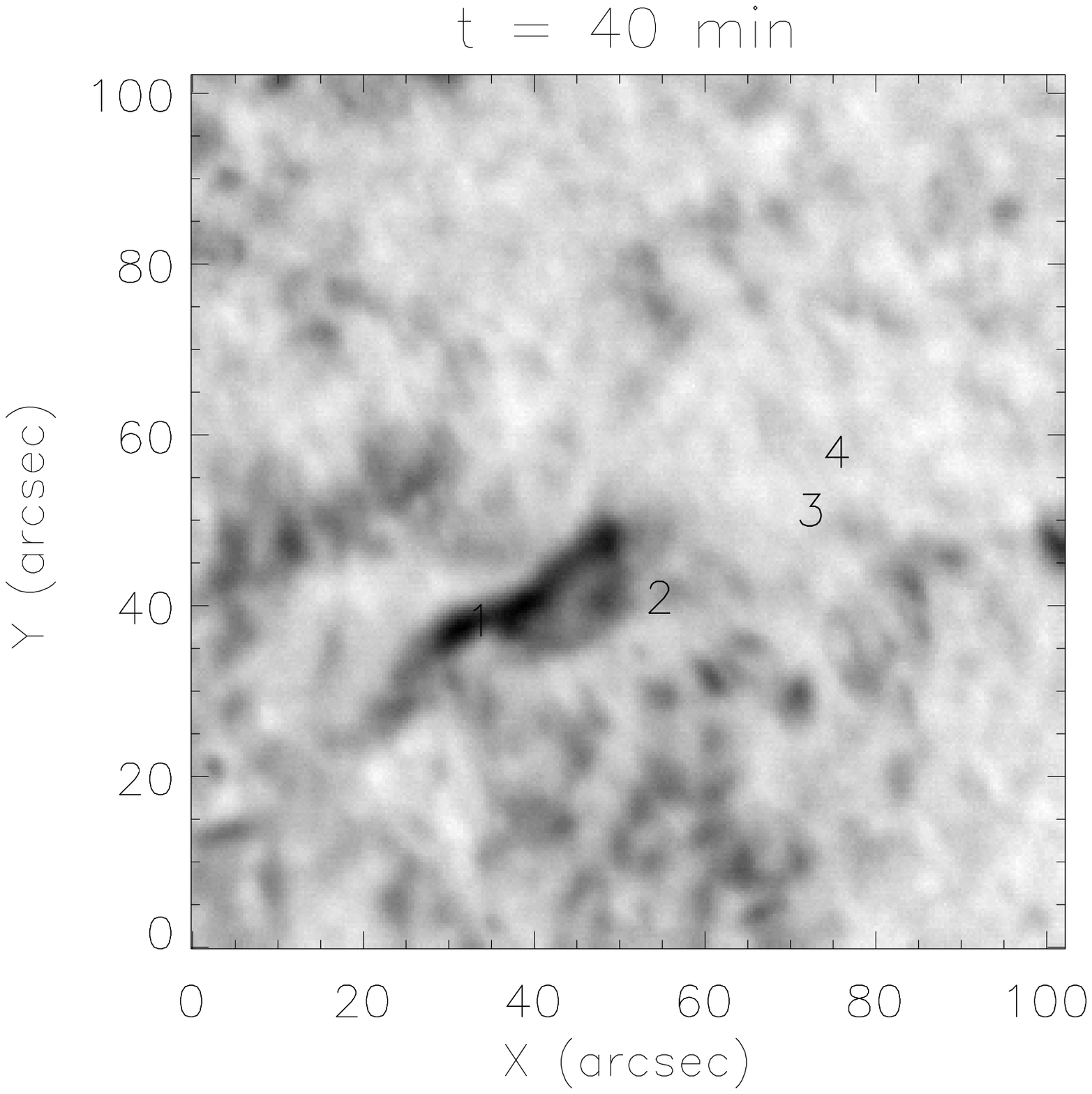} 
   \includegraphics[width=4cm,angle=90]{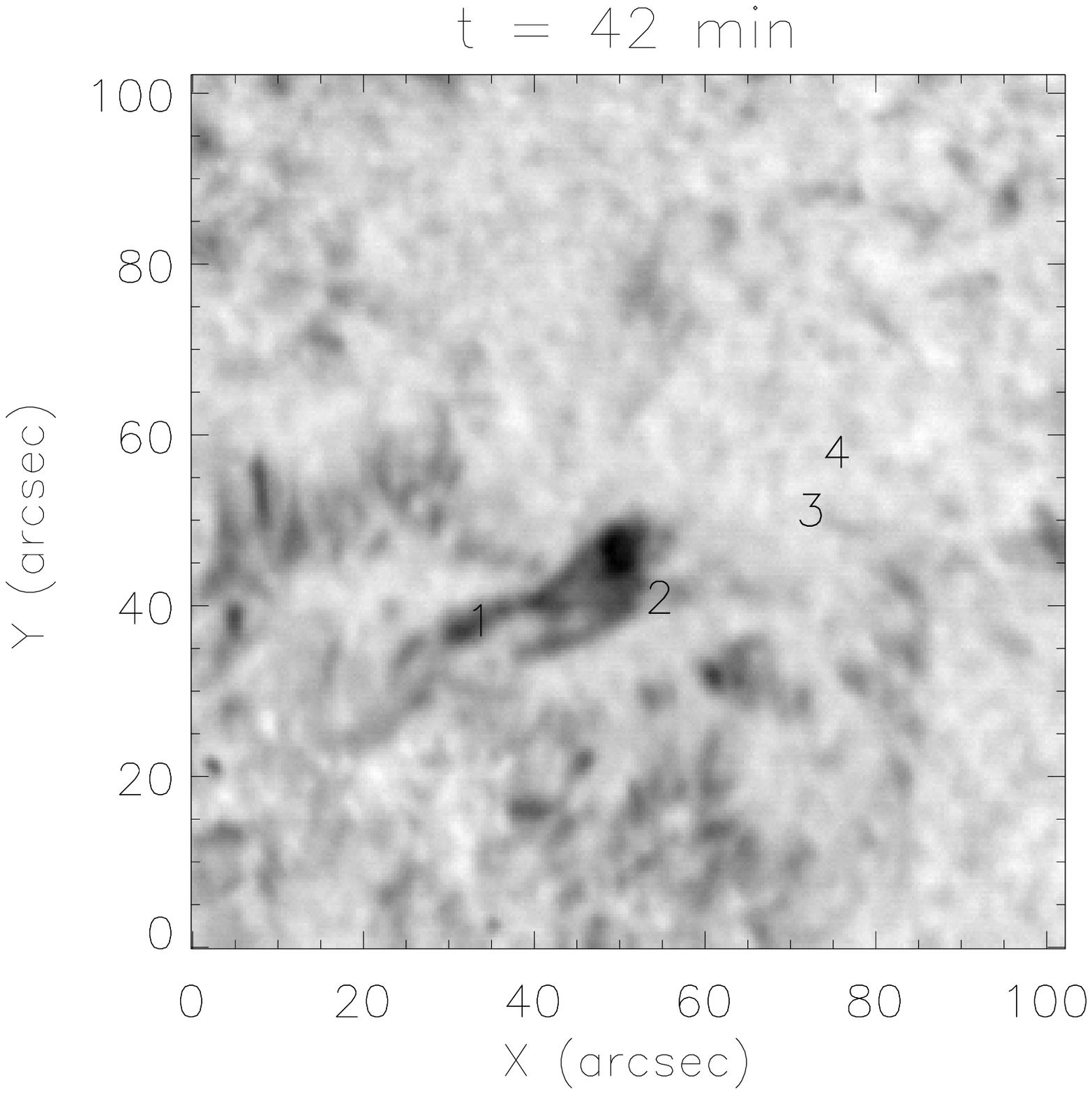} 
   \includegraphics[width=4cm,angle=90]{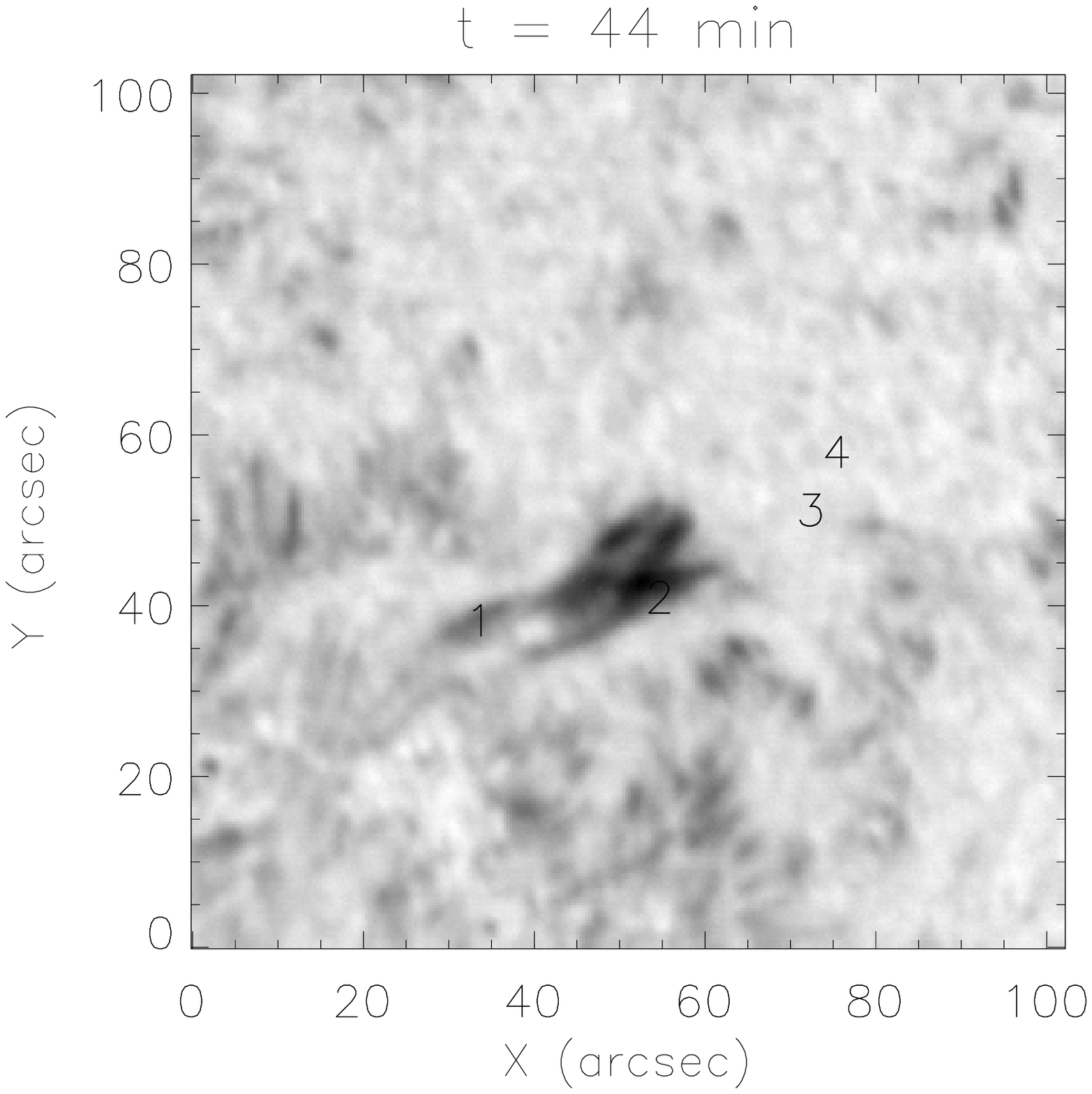} 
   \includegraphics[width=4cm,angle=90]{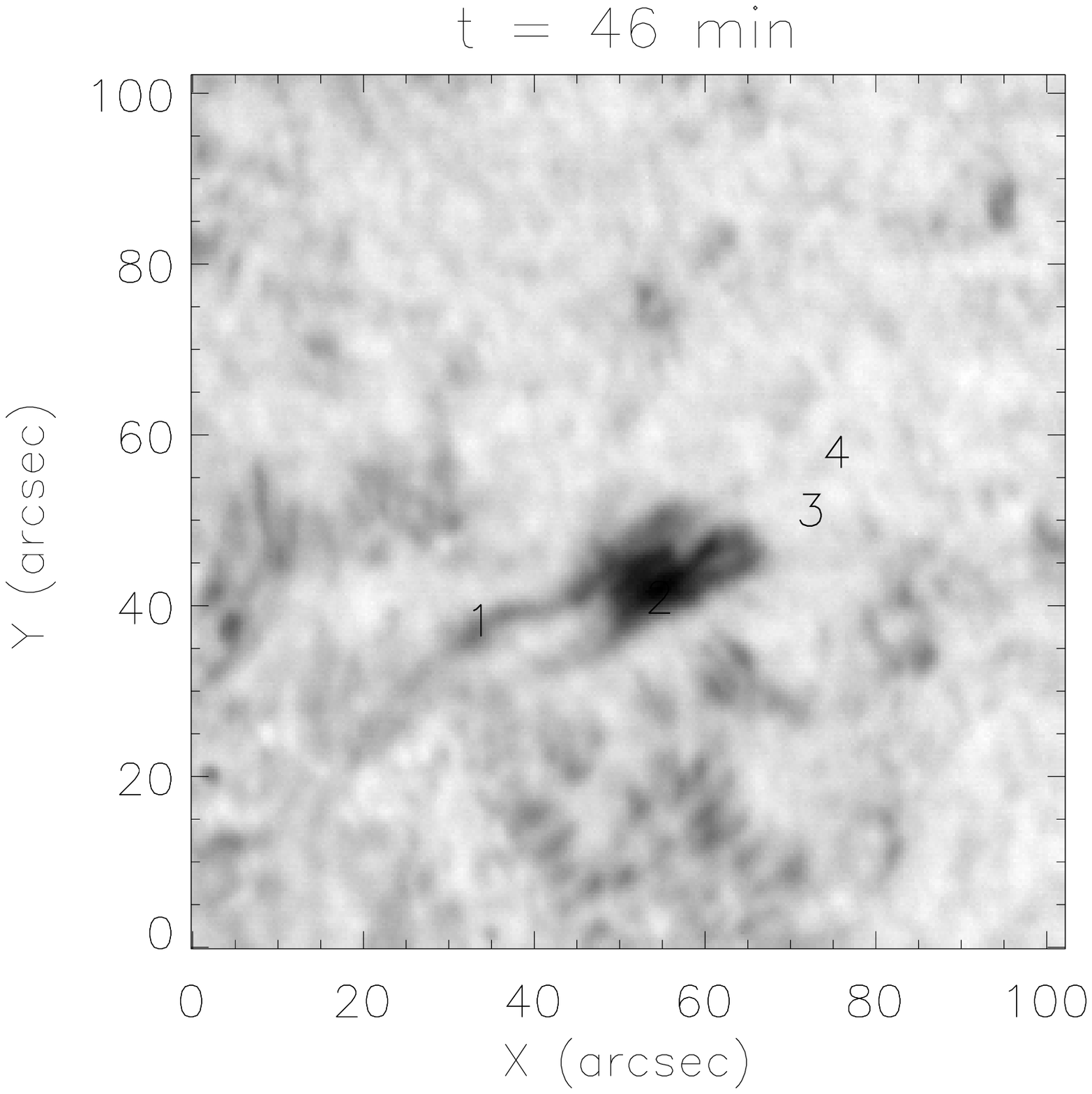} 
   \includegraphics[width=4cm,angle=90]{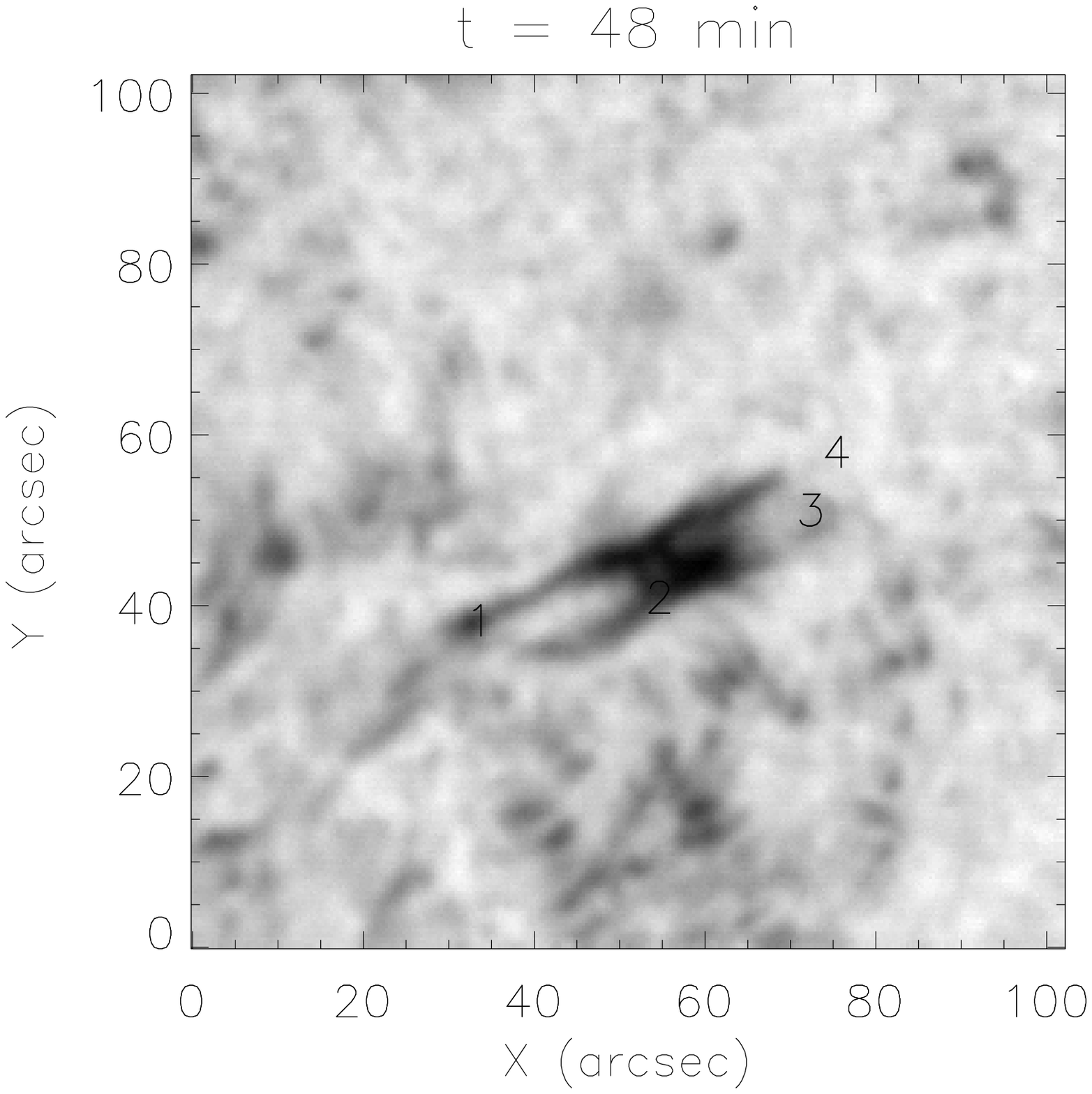} 
   \includegraphics[width=4cm,angle=90]{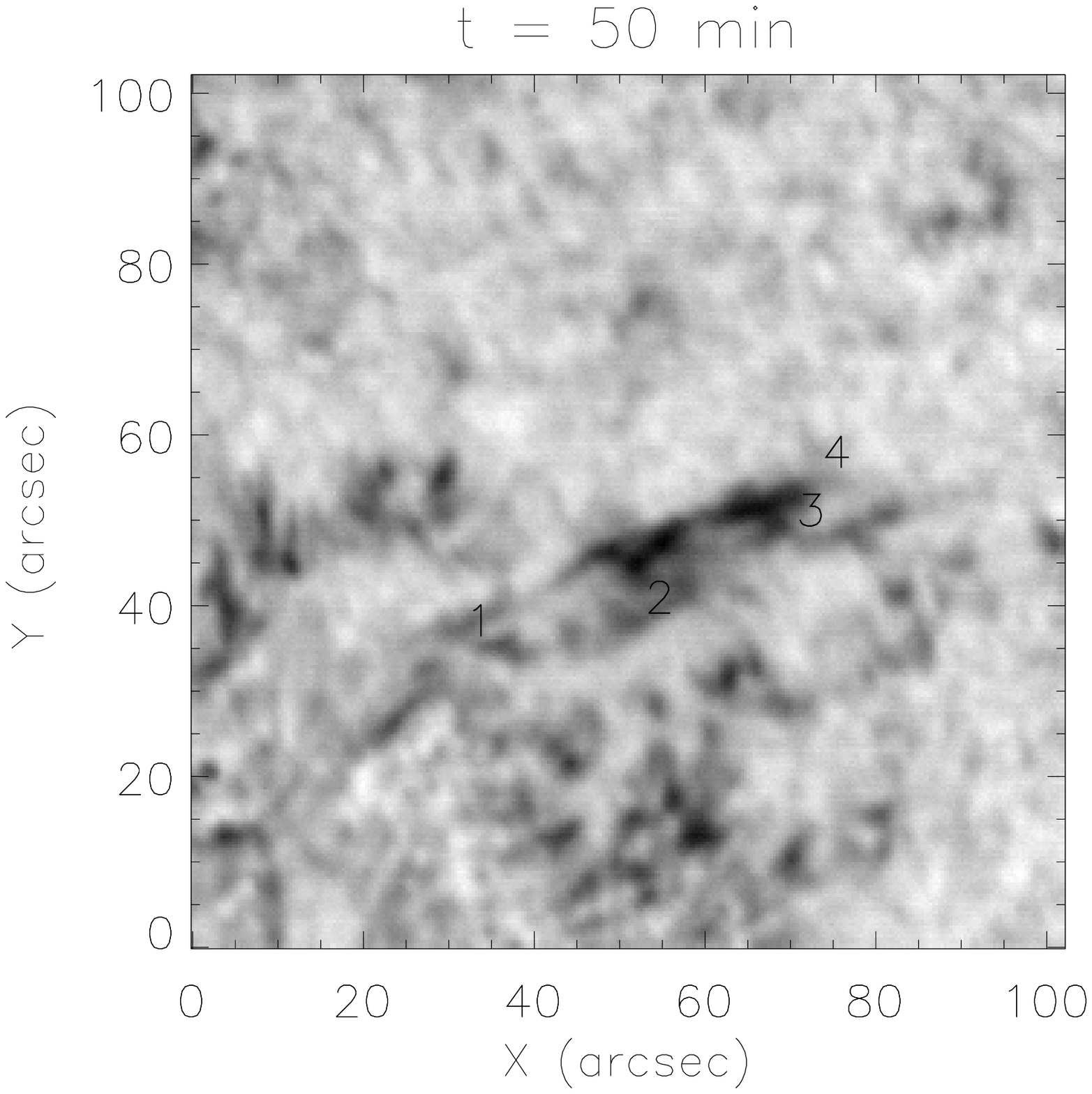} 
   \includegraphics[width=4cm,angle=90]{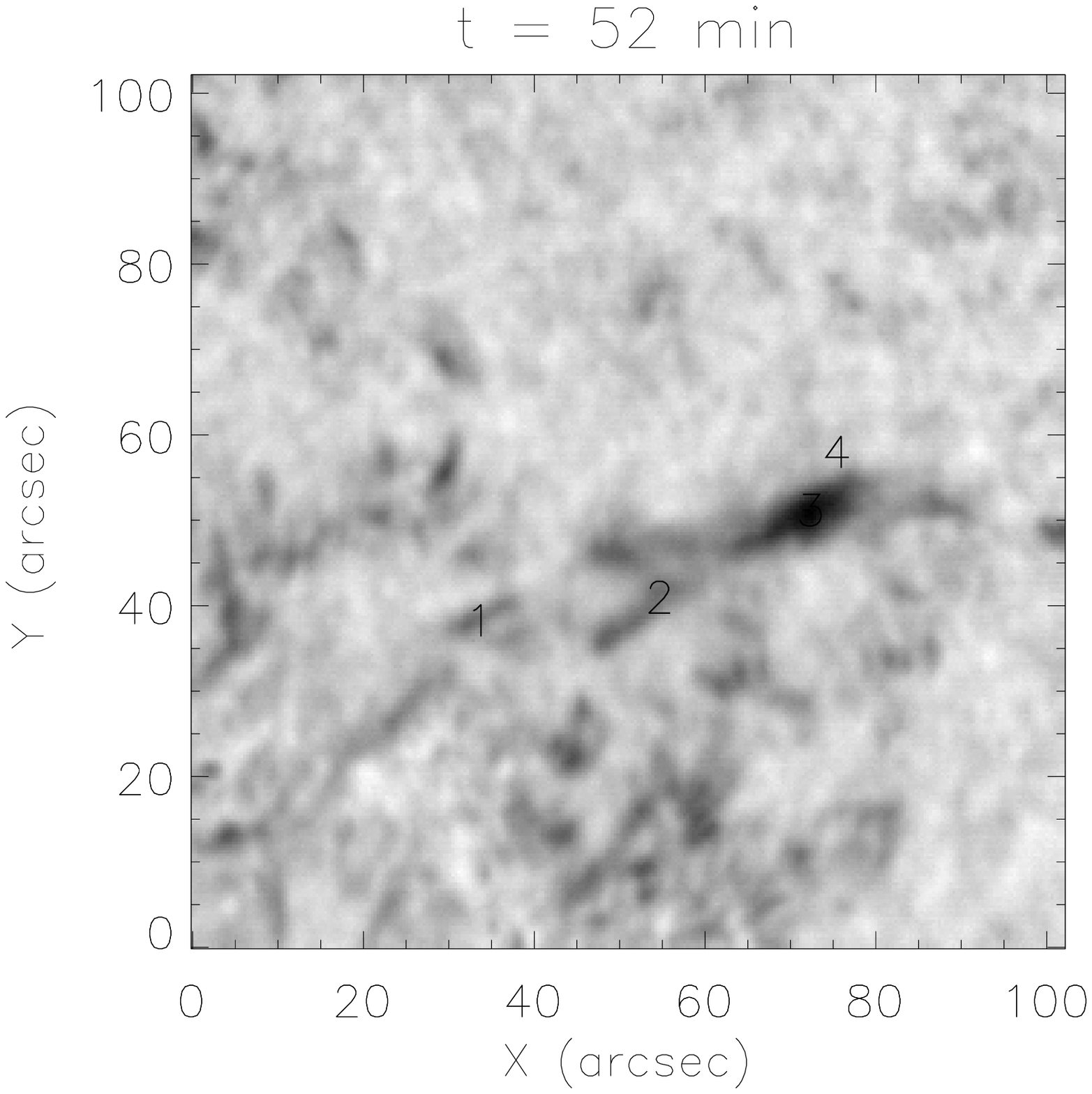}
   \includegraphics[width=4cm,angle=90]{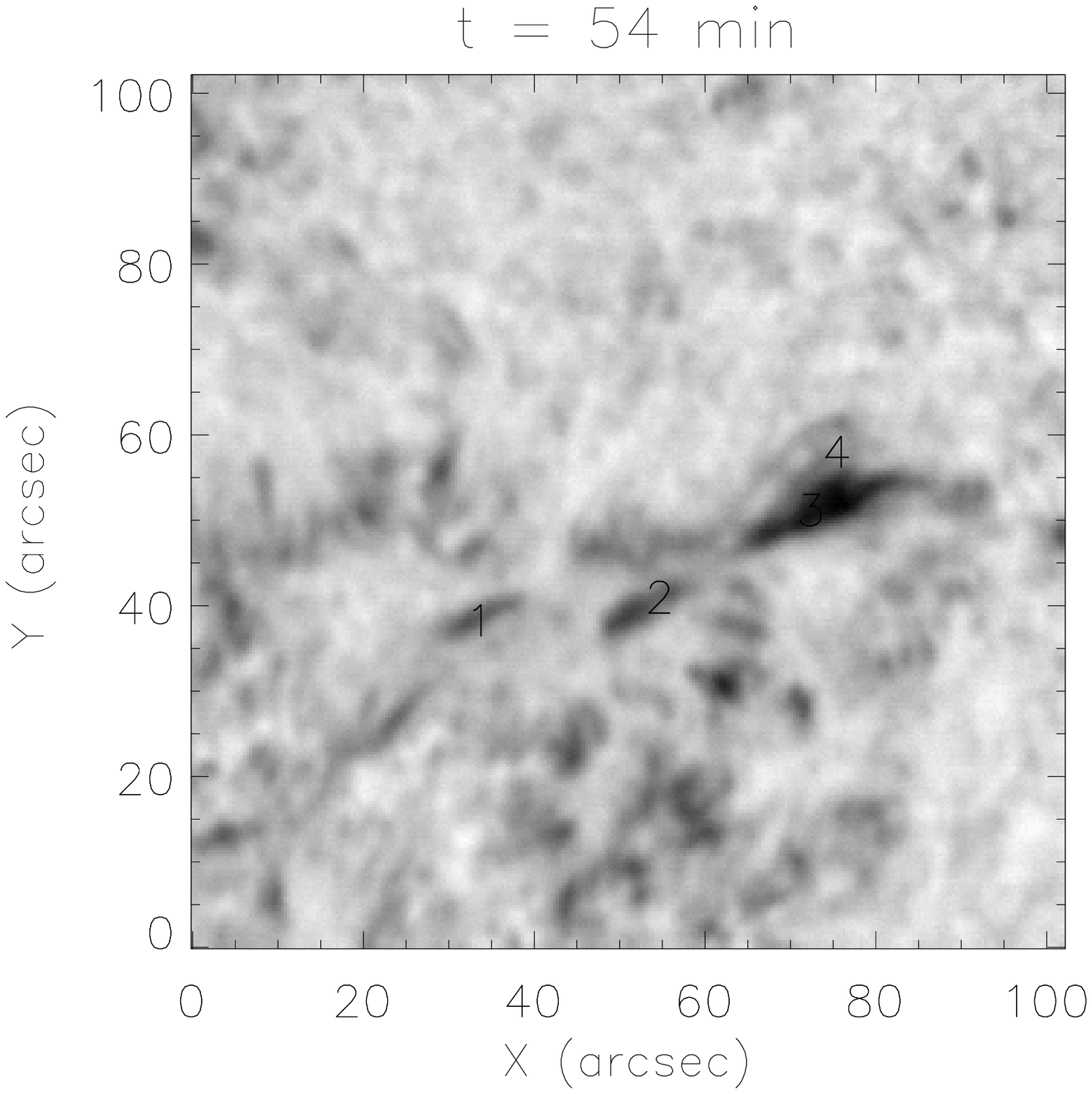} 

  \caption{Part of the intensity sequence in summed H$\alpha$ wings from  $t$=38 min to  
  $t$=54 min, after the beginning of the observations.}
              \label{seqi1}
\end{figure*}

\begin{figure*}
   \centering
   \includegraphics[width=4cm,angle=90]{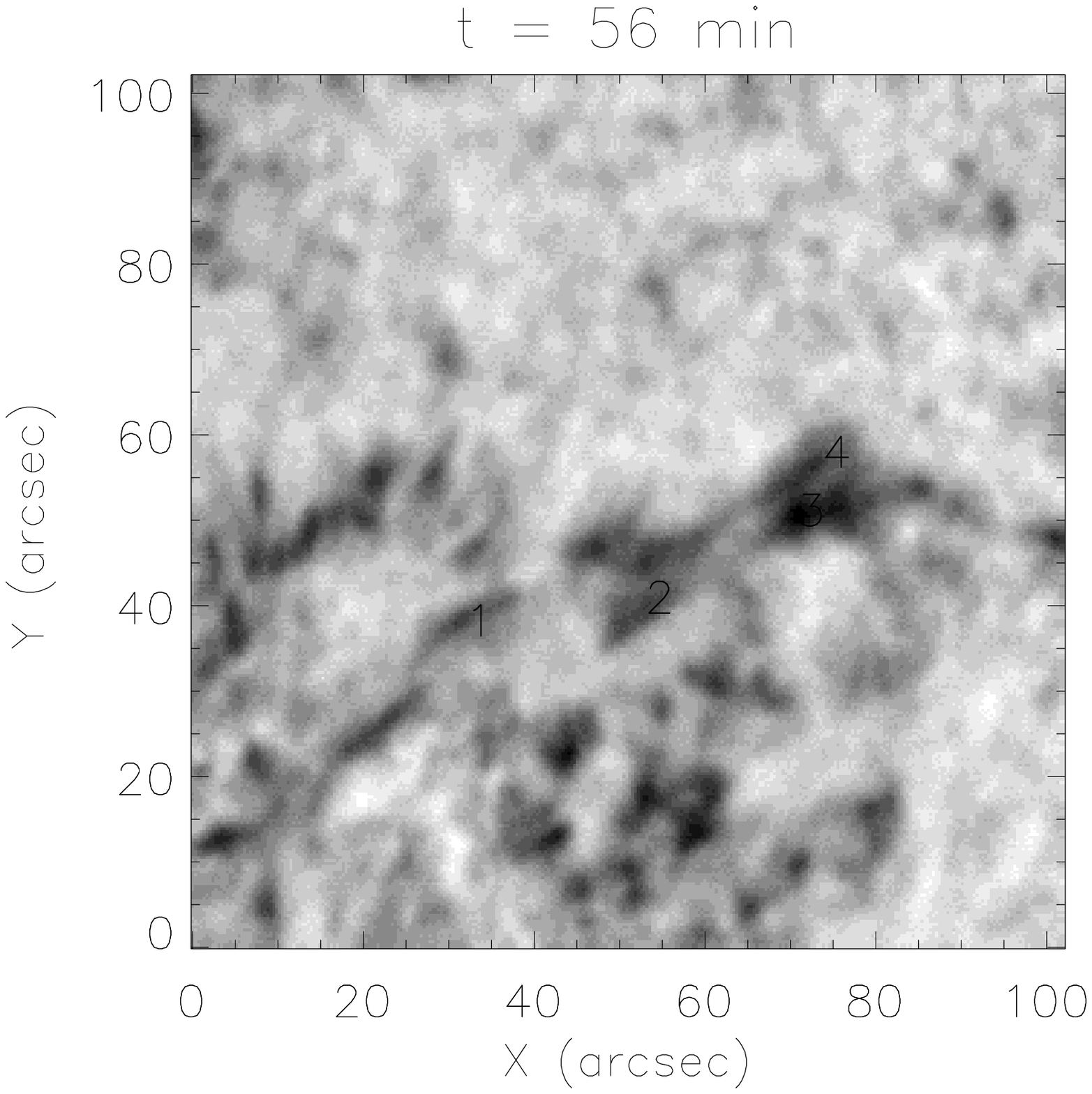}
   \includegraphics[width=4cm,angle=90]{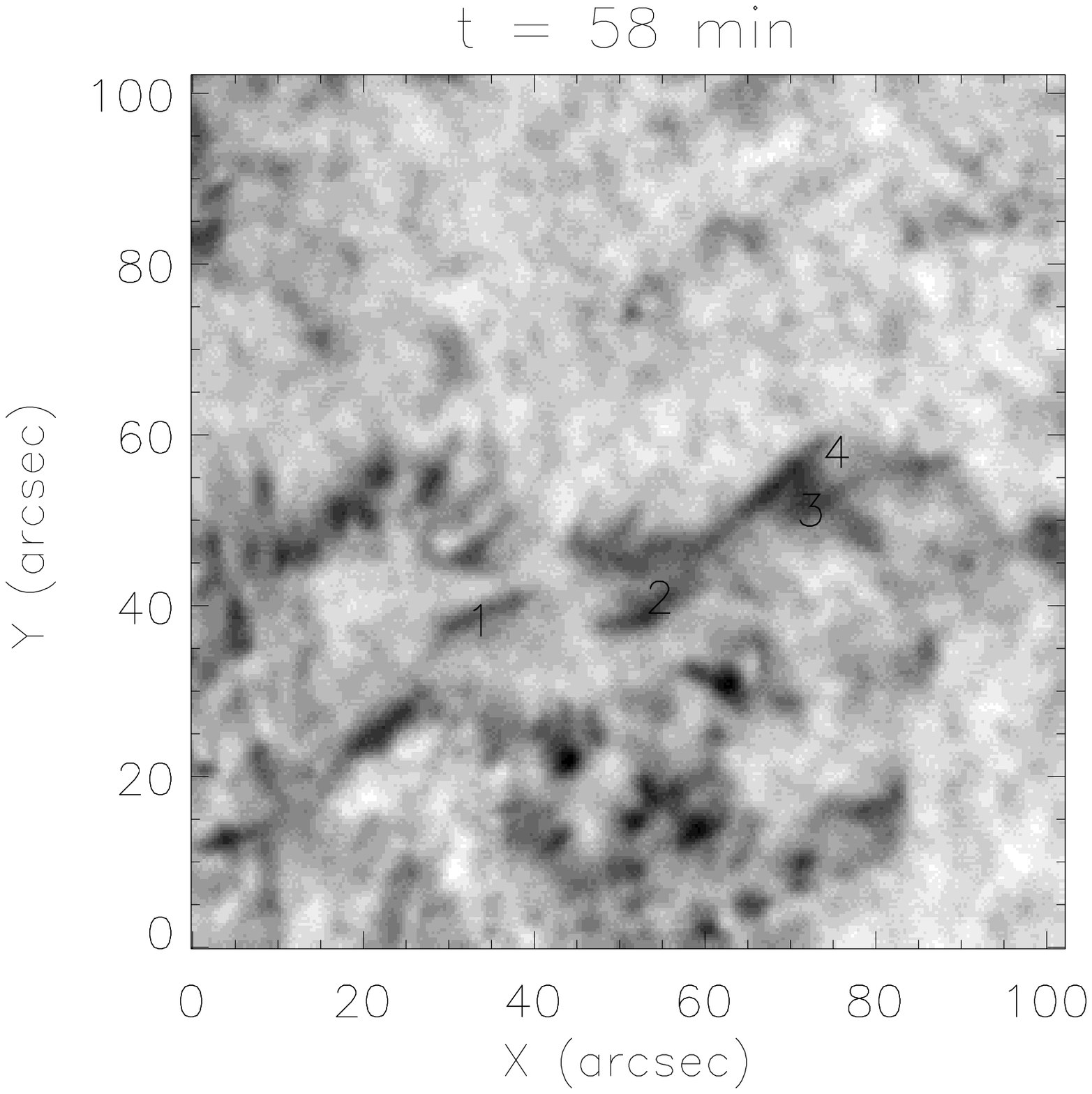}
   \includegraphics[width=4cm,angle=90]{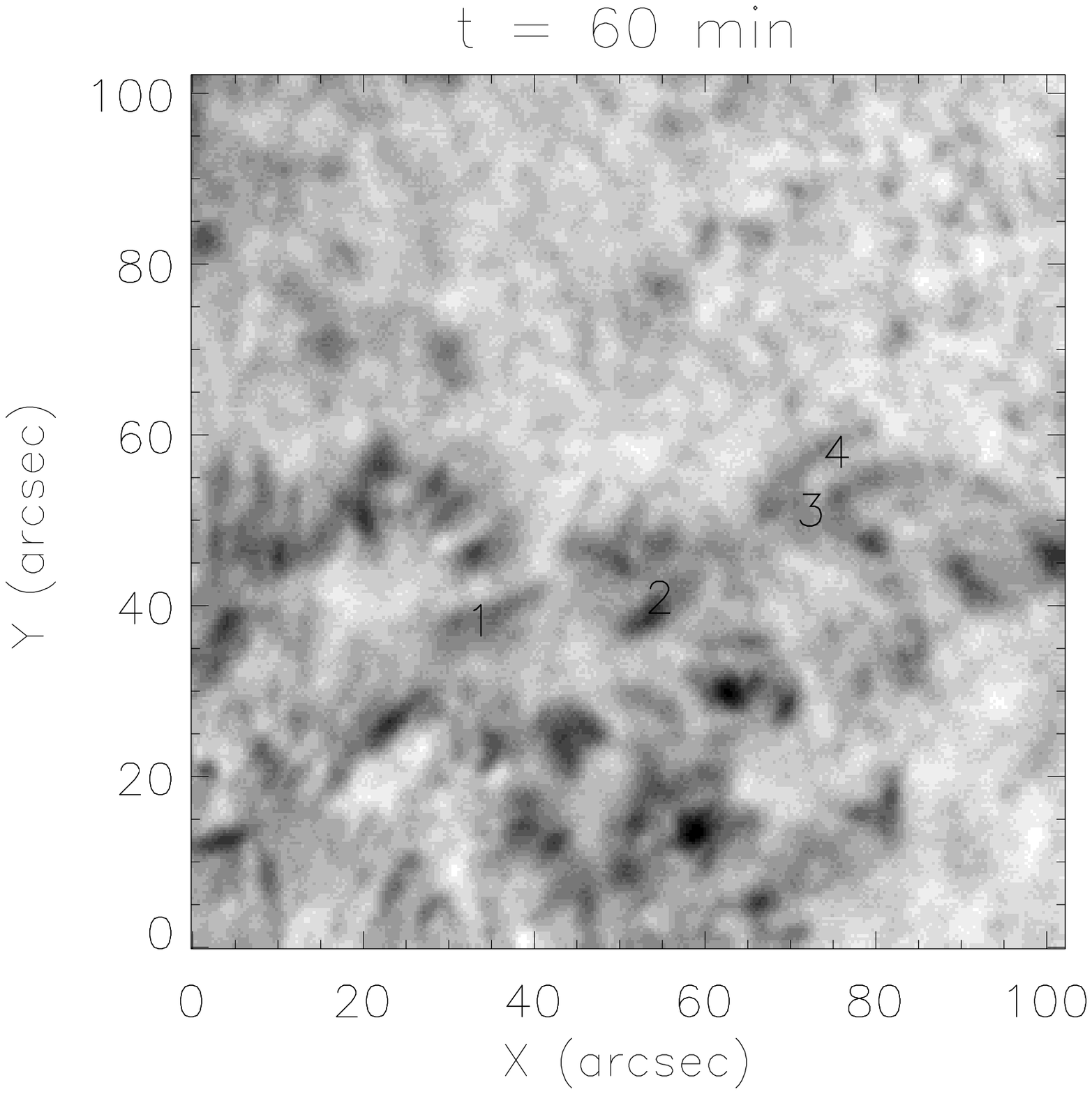}
   \includegraphics[width=4cm,angle=90]{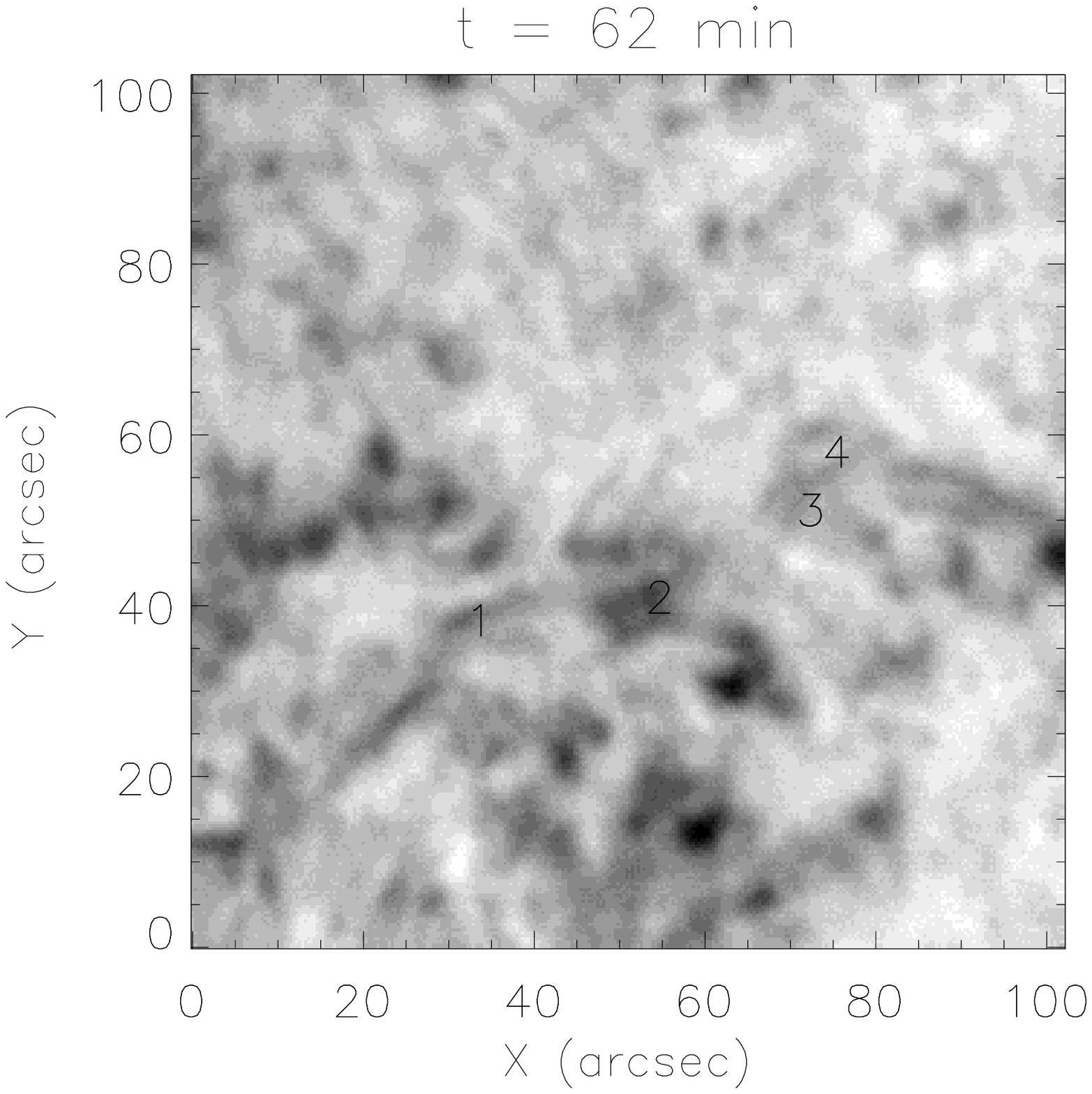}
   \includegraphics[width=4cm,angle=90]{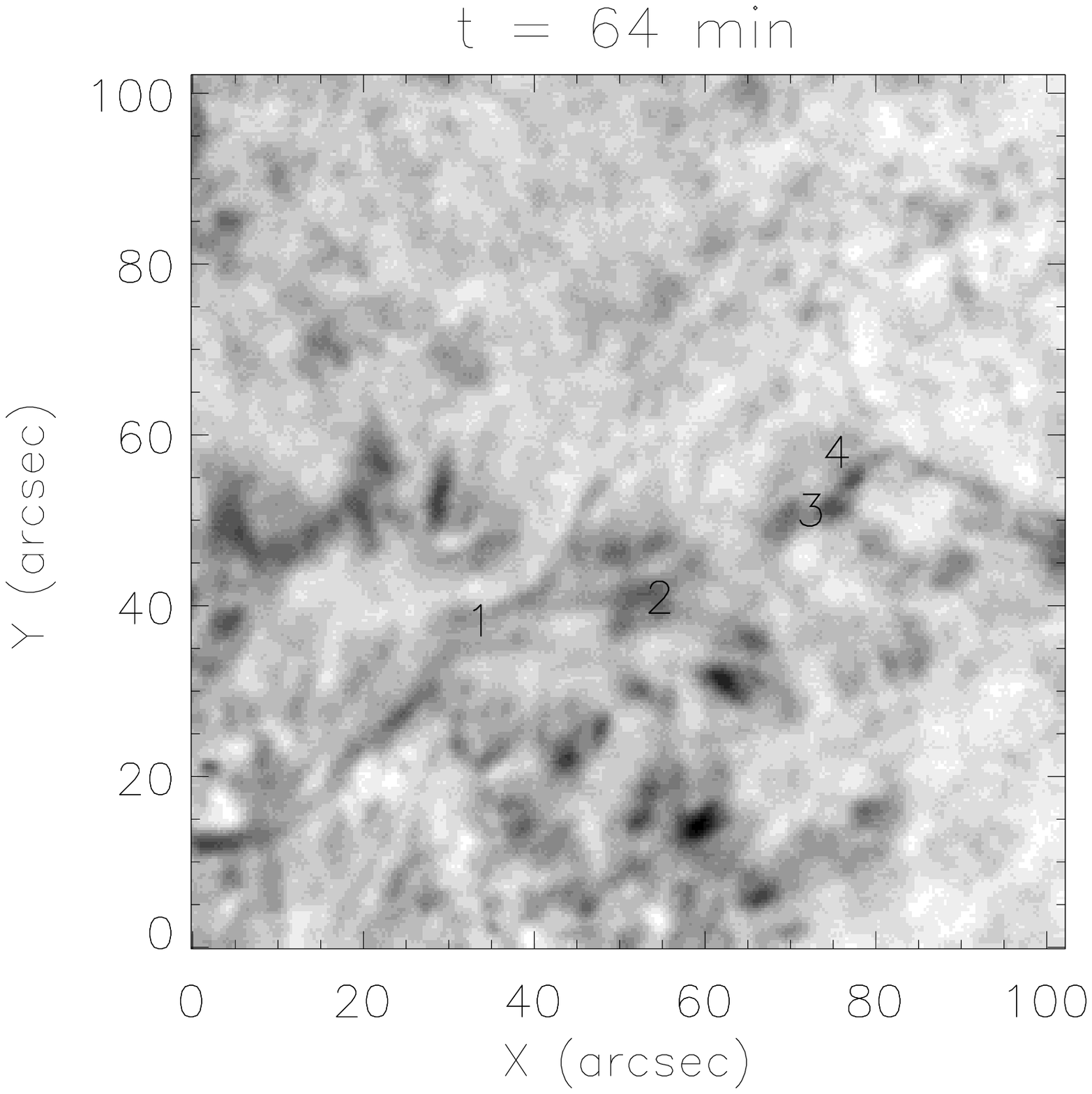}
   \includegraphics[width=4cm,angle=90]{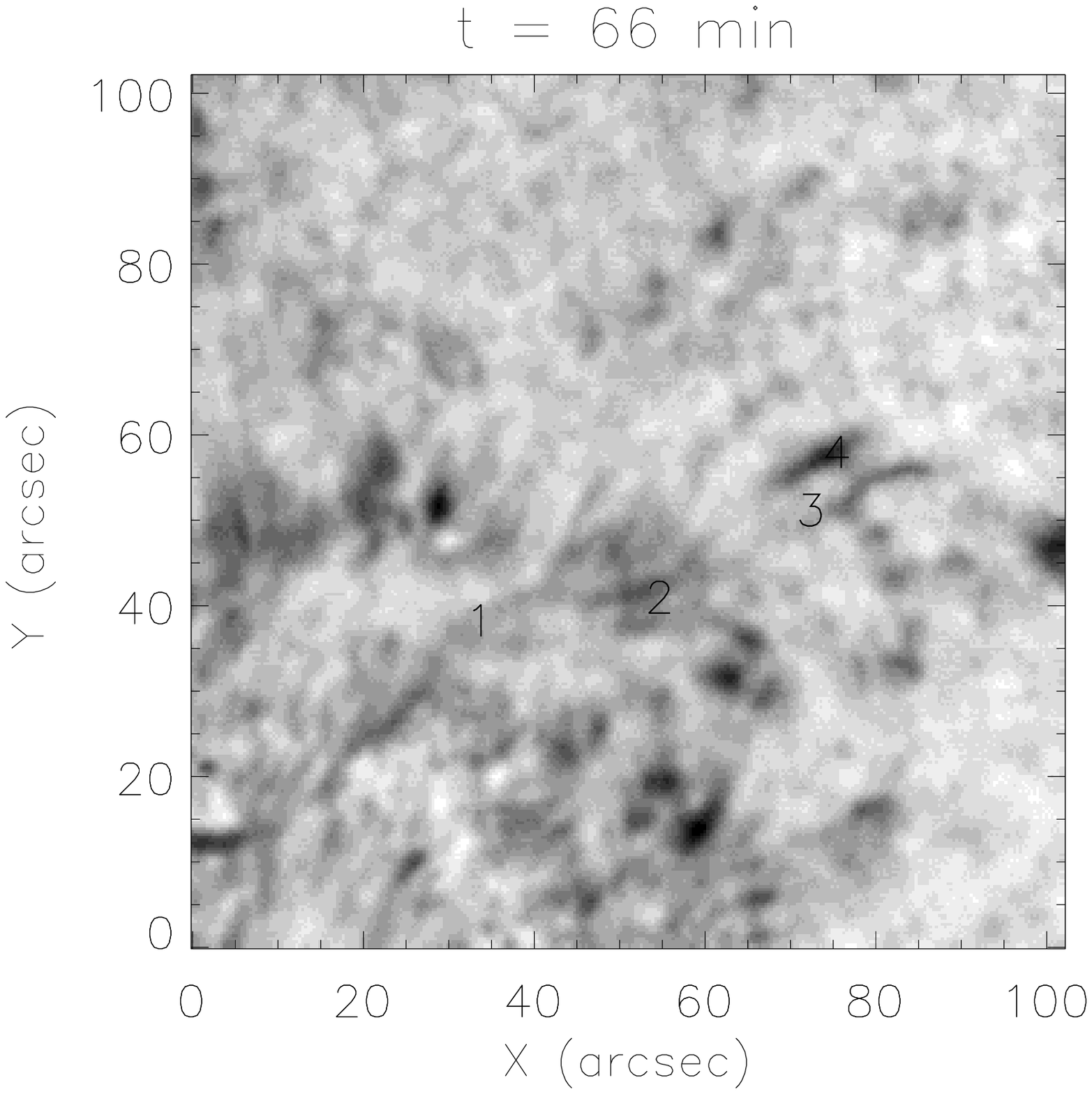}
   \includegraphics[width=4cm,angle=90]{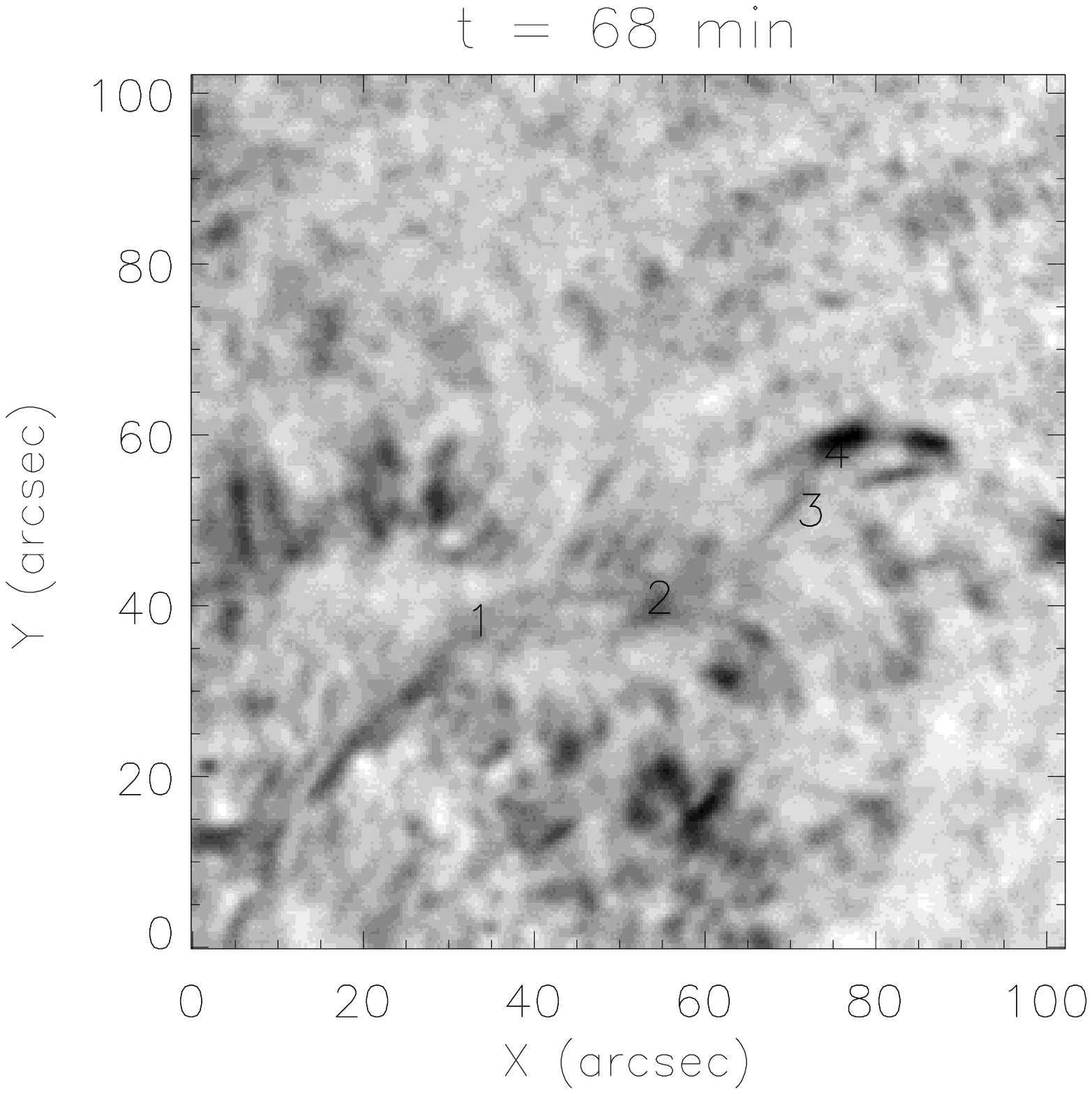}
   \includegraphics[width=4cm,angle=90]{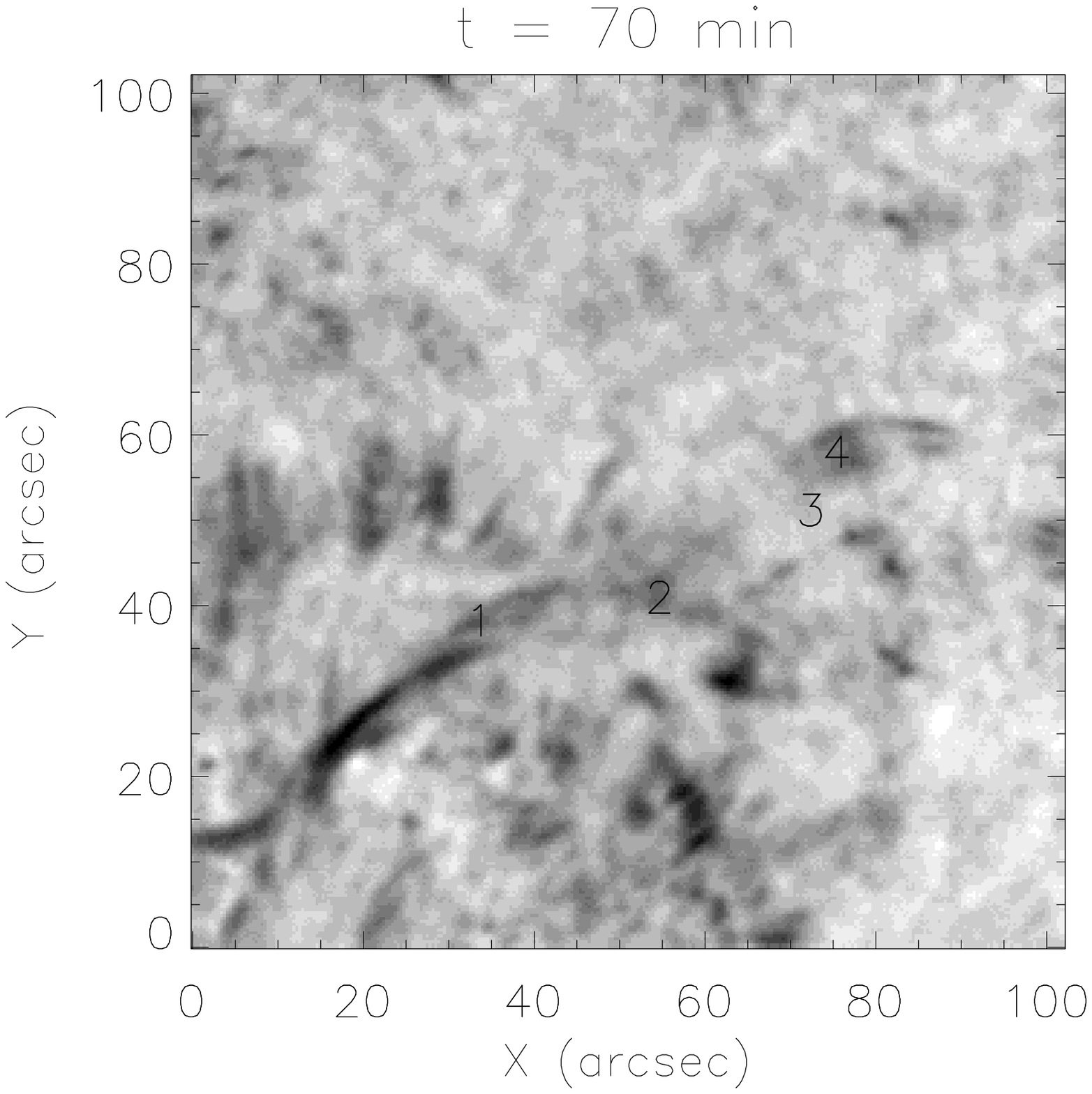}
   \includegraphics[width=4cm,angle=90]{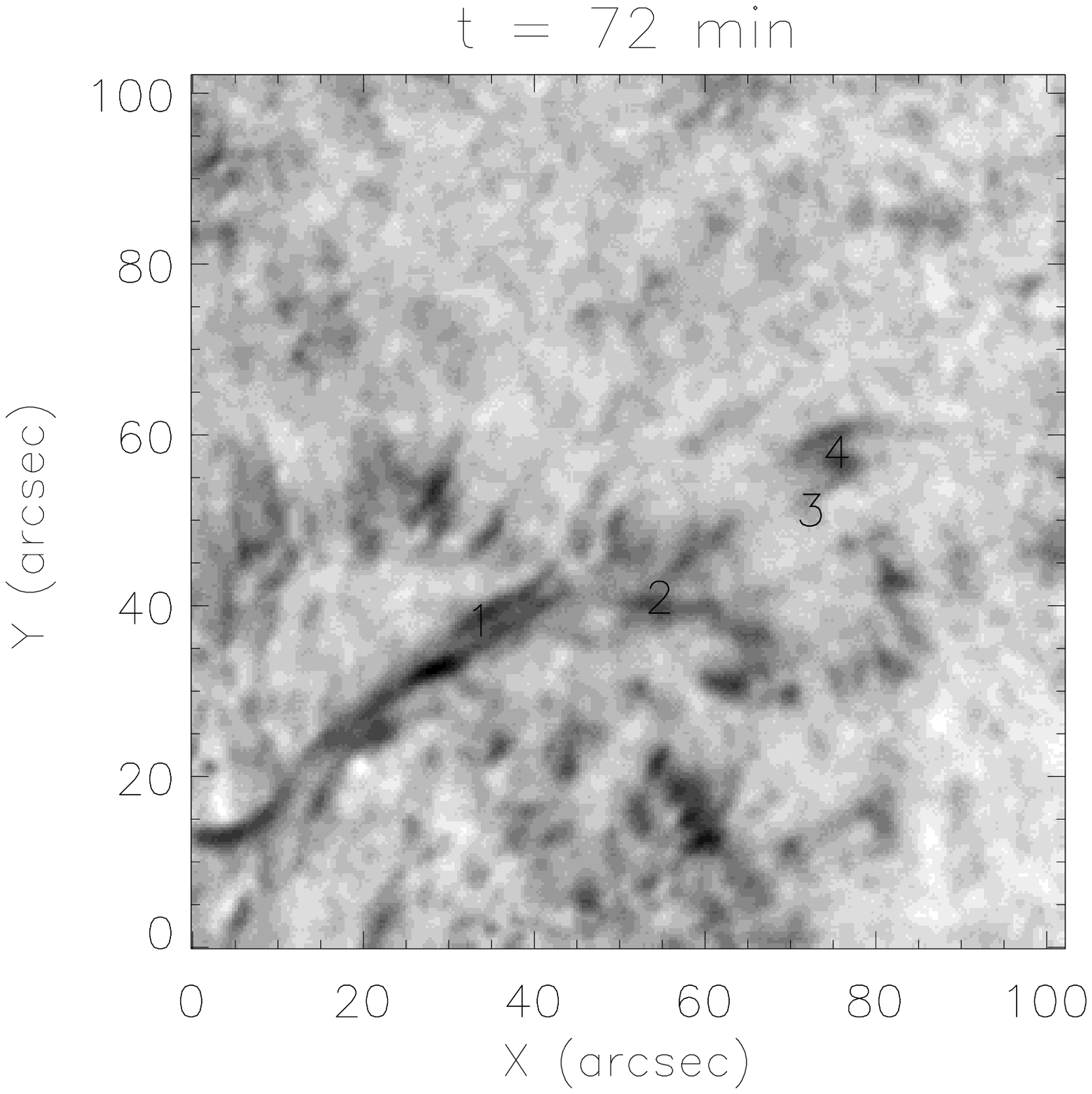}
   \caption{Part of the intensity sequence in summed H$\alpha$ wings from  $t$=56 min to  
$t$=72 min, after the beginning of the observations.}
              \label{seqi2}
\end{figure*}

\begin{figure*}
   \centering
   \includegraphics[width=4cm,angle=90]{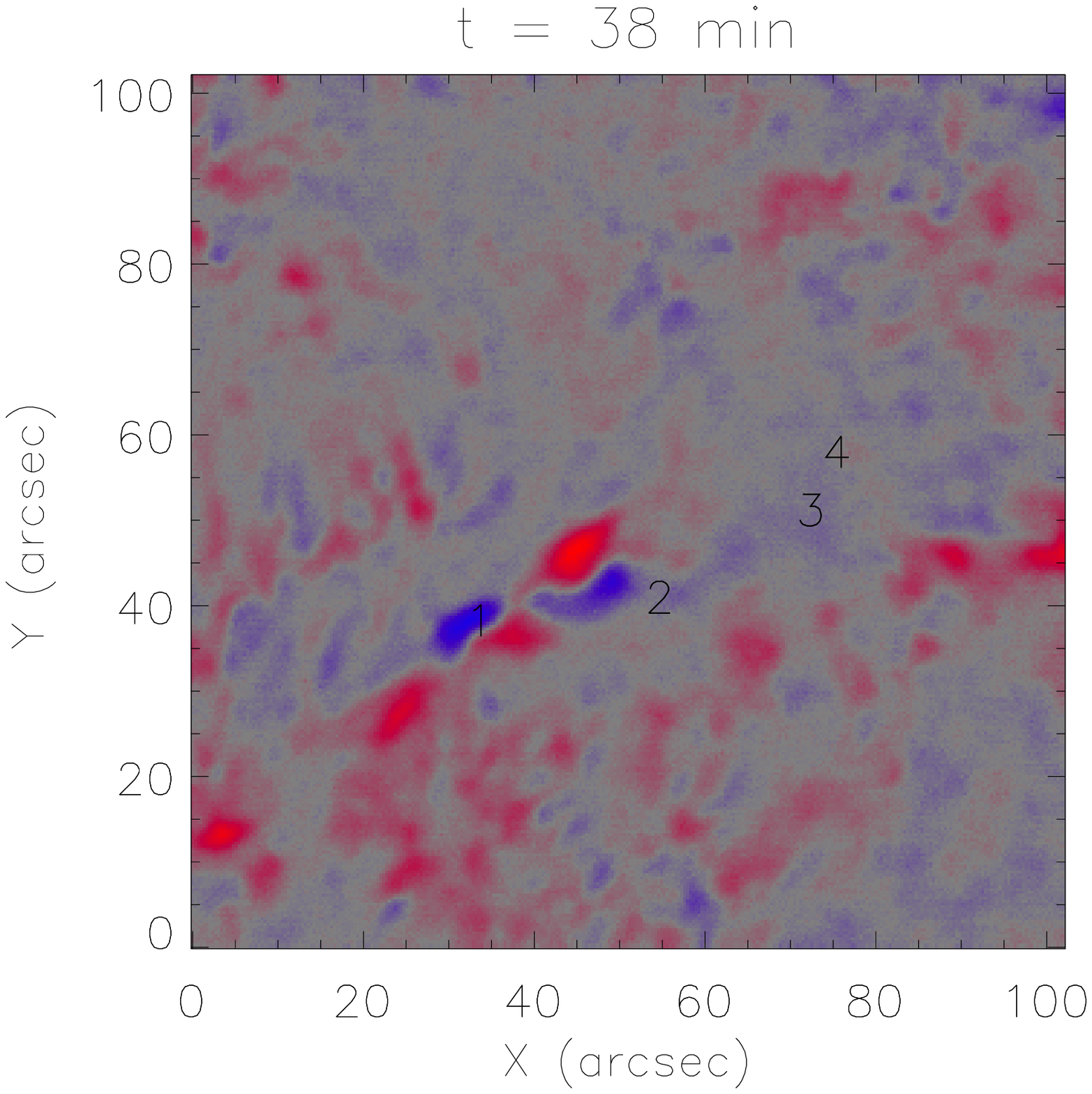}
   \includegraphics[width=4cm,angle=90]{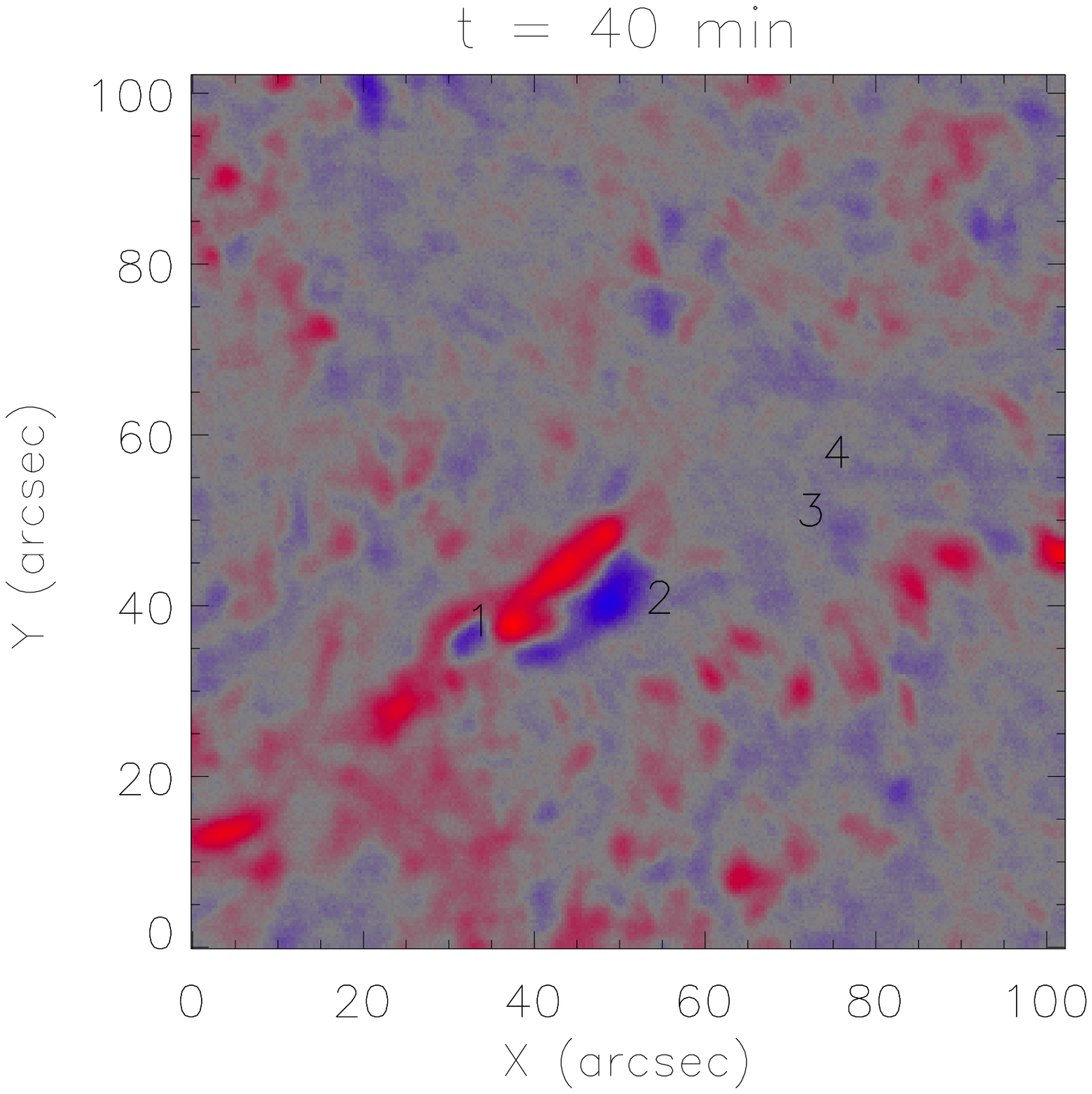}
   \includegraphics[width=4cm,angle=90]{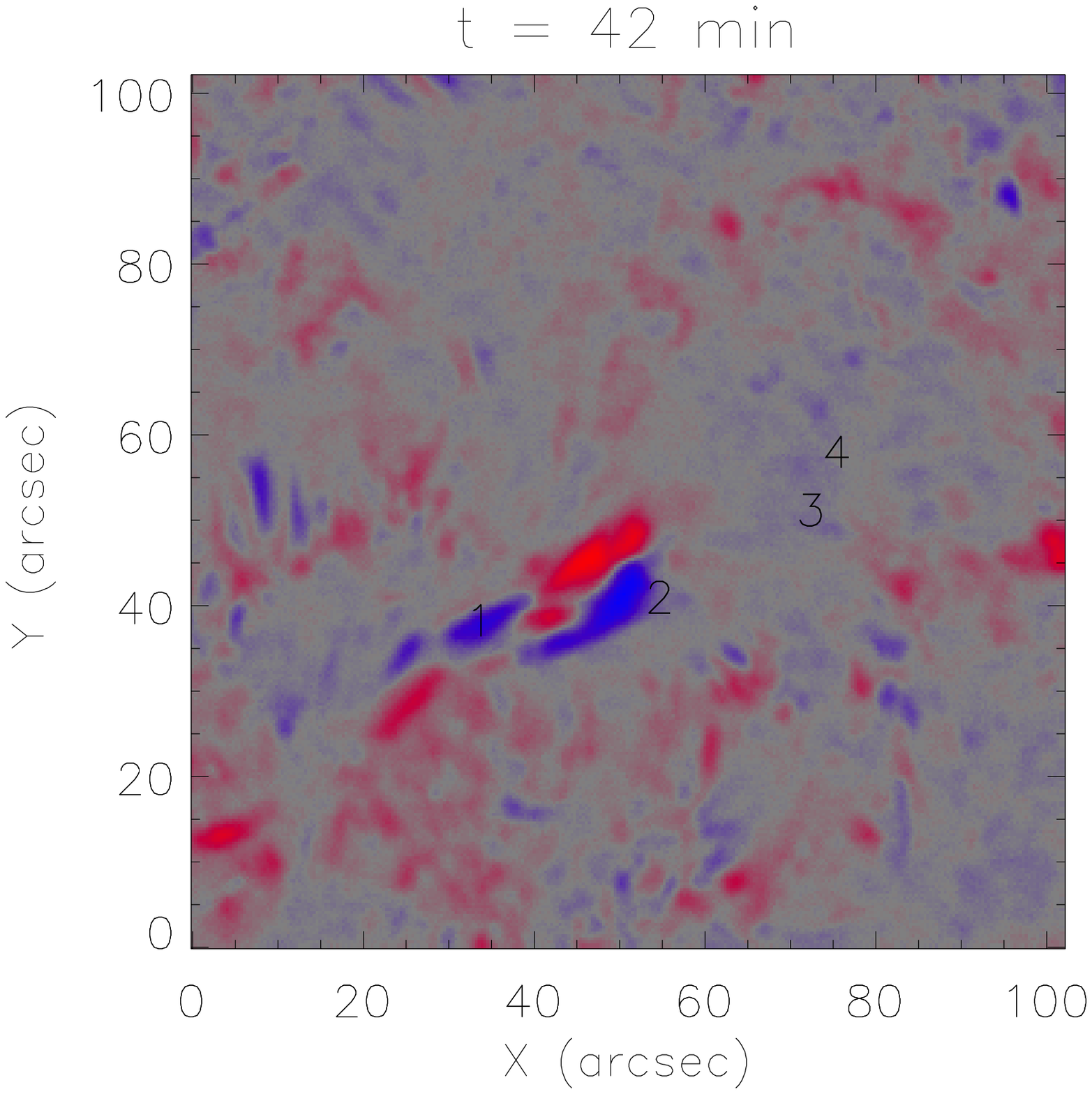}
   \includegraphics[width=4cm,angle=90]{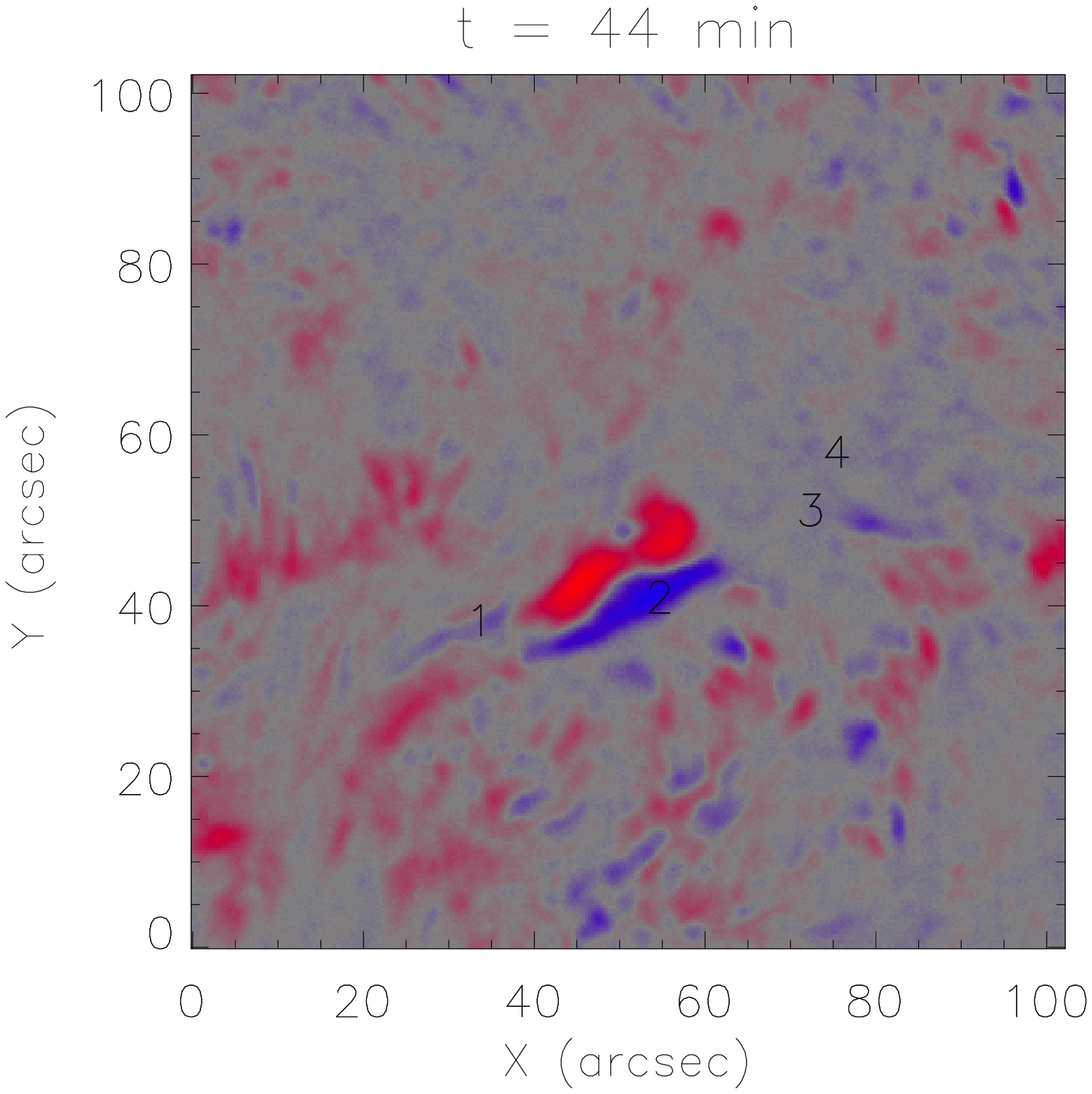}
   \includegraphics[width=4cm,angle=90]{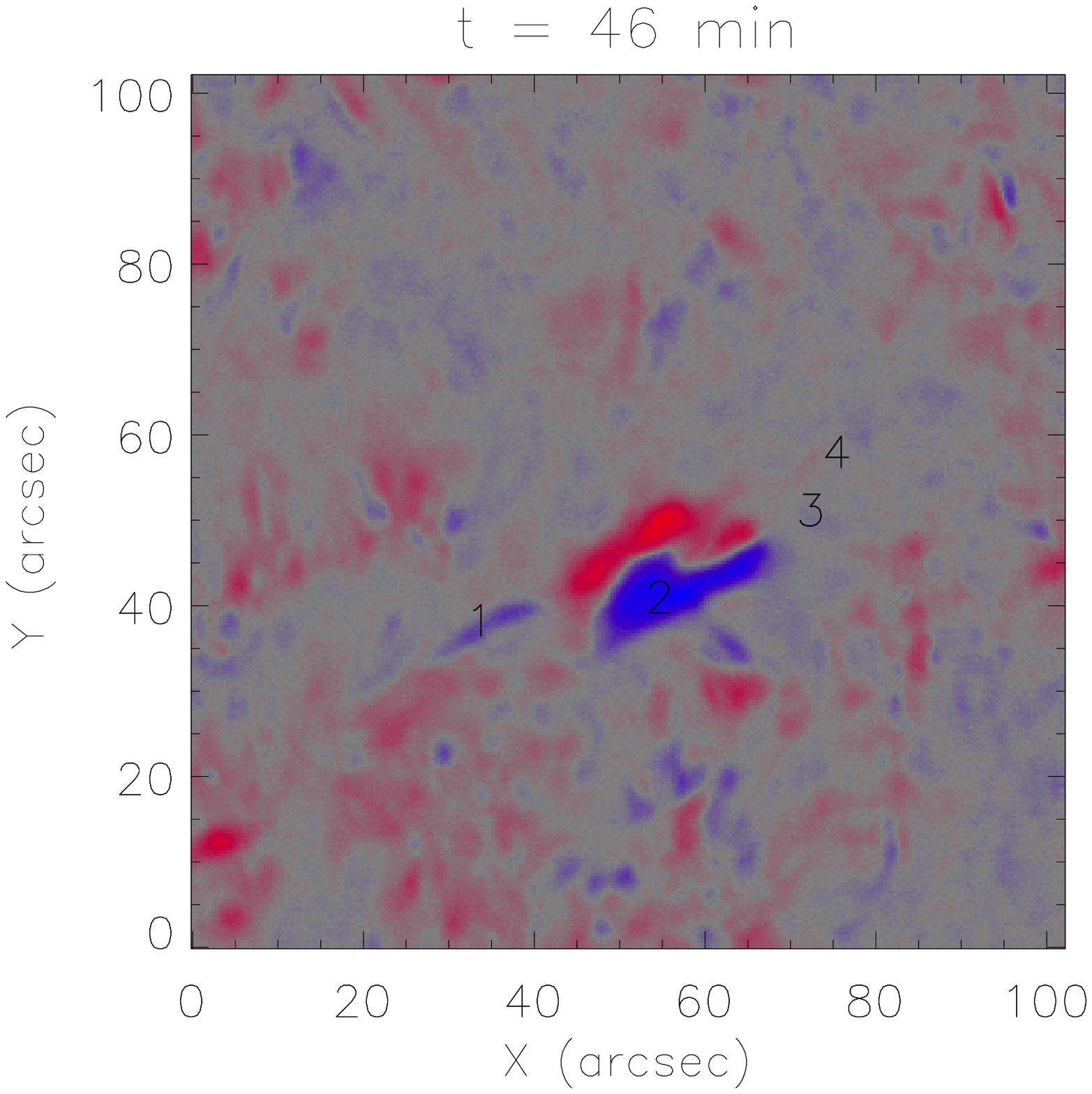}
   \includegraphics[width=4cm,angle=90]{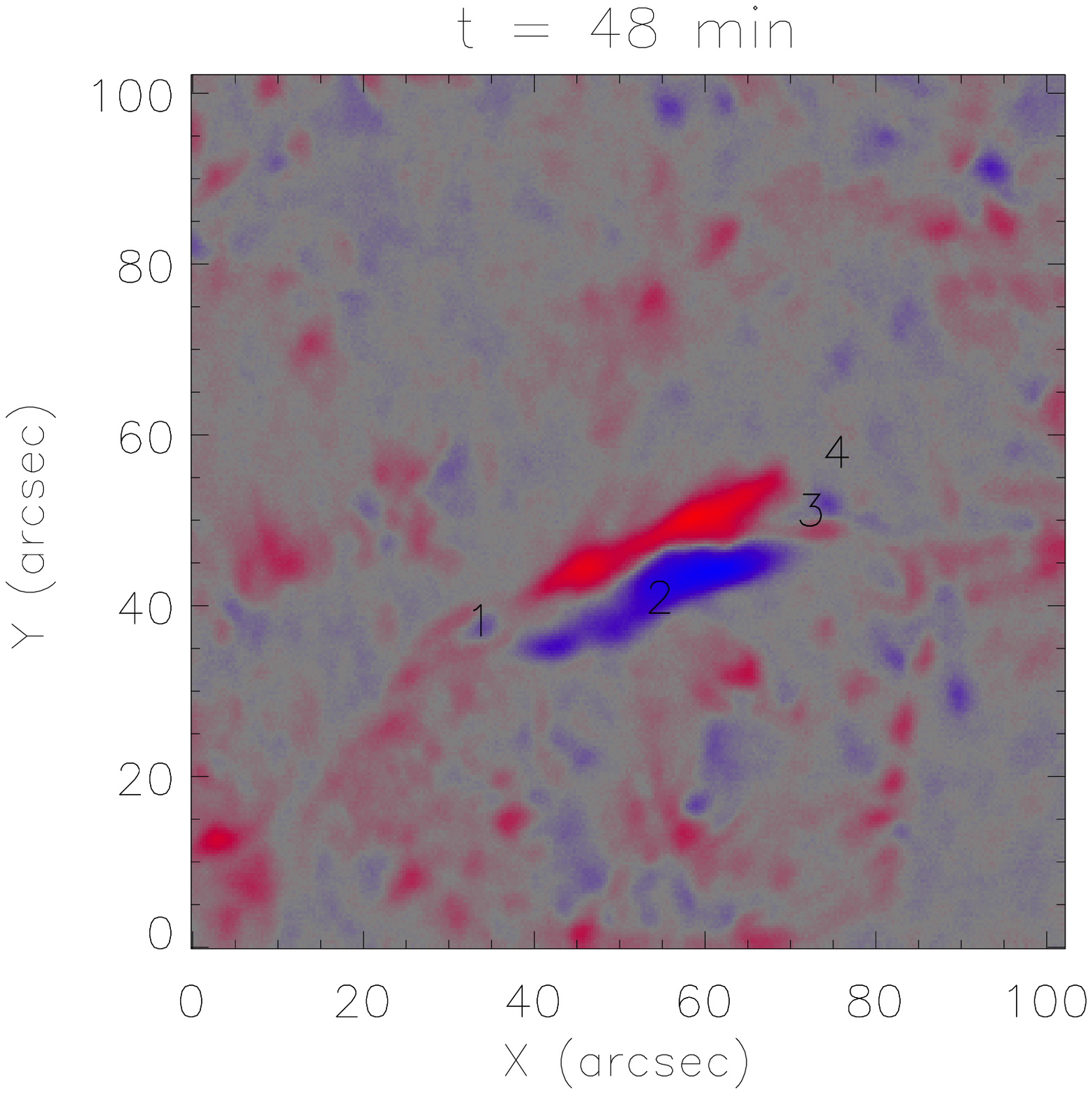}
   \includegraphics[width=4cm,angle=90]{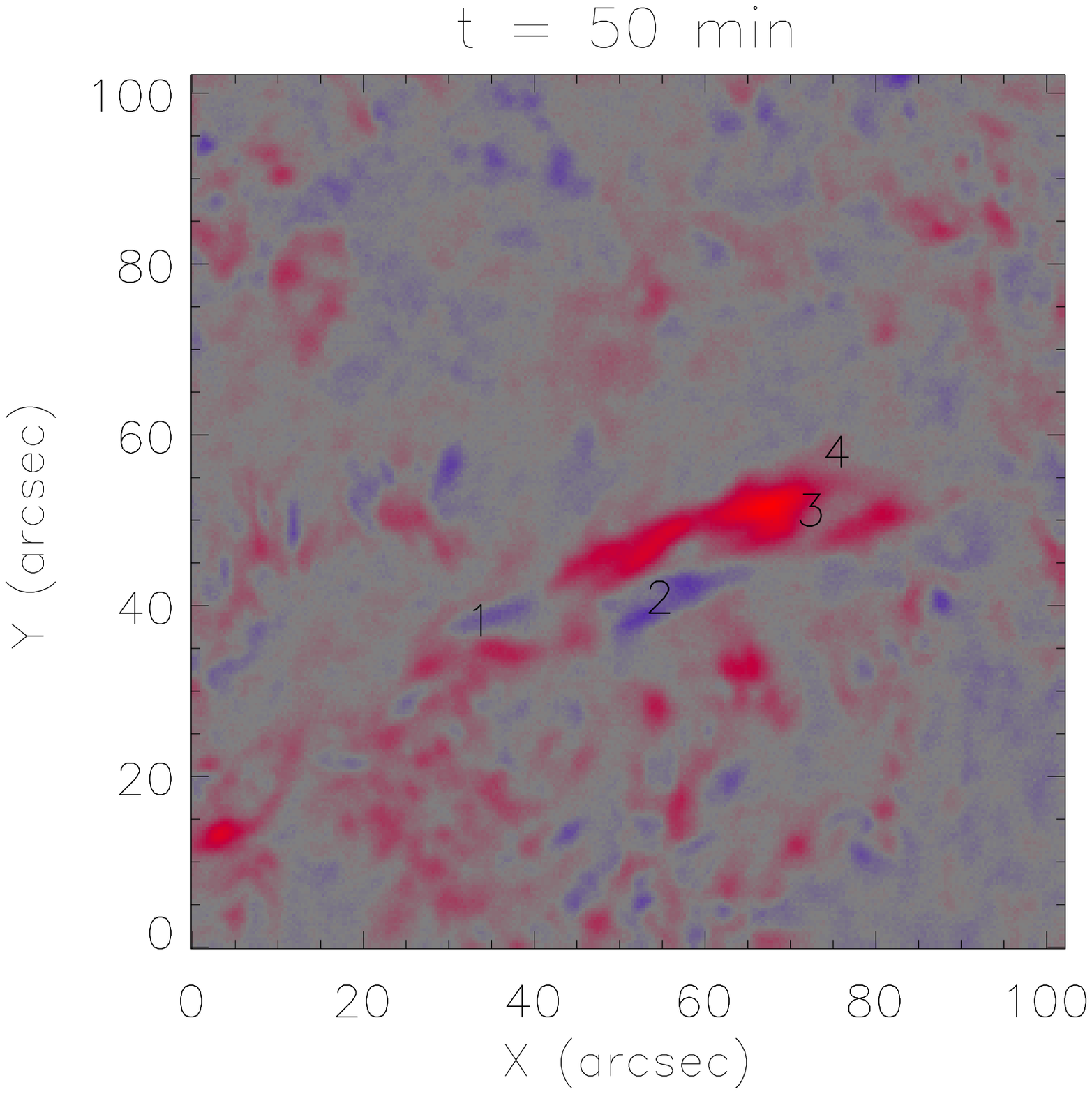}
   \includegraphics[width=4cm,angle=90]{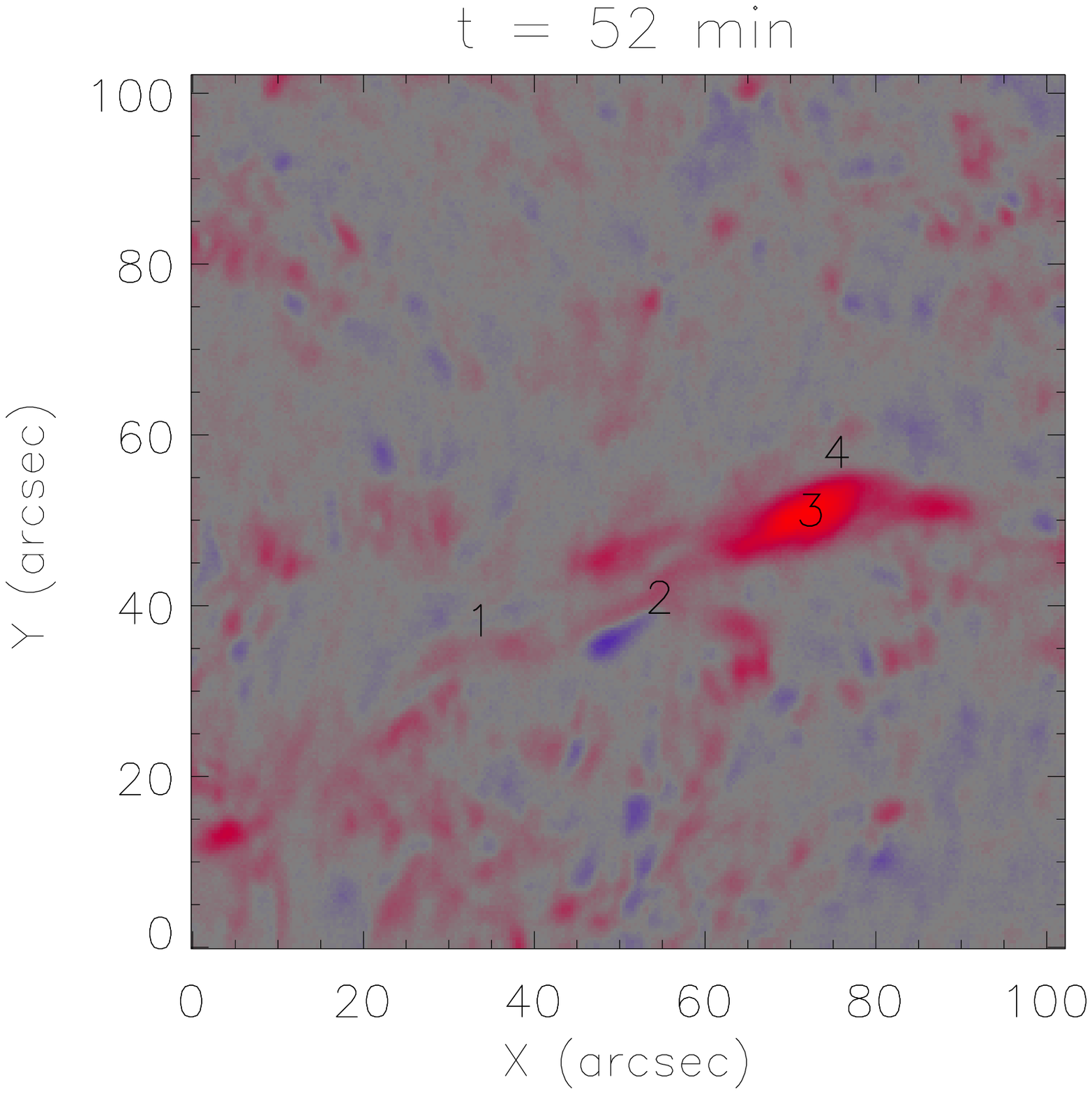}
   \includegraphics[width=4cm,angle=90]{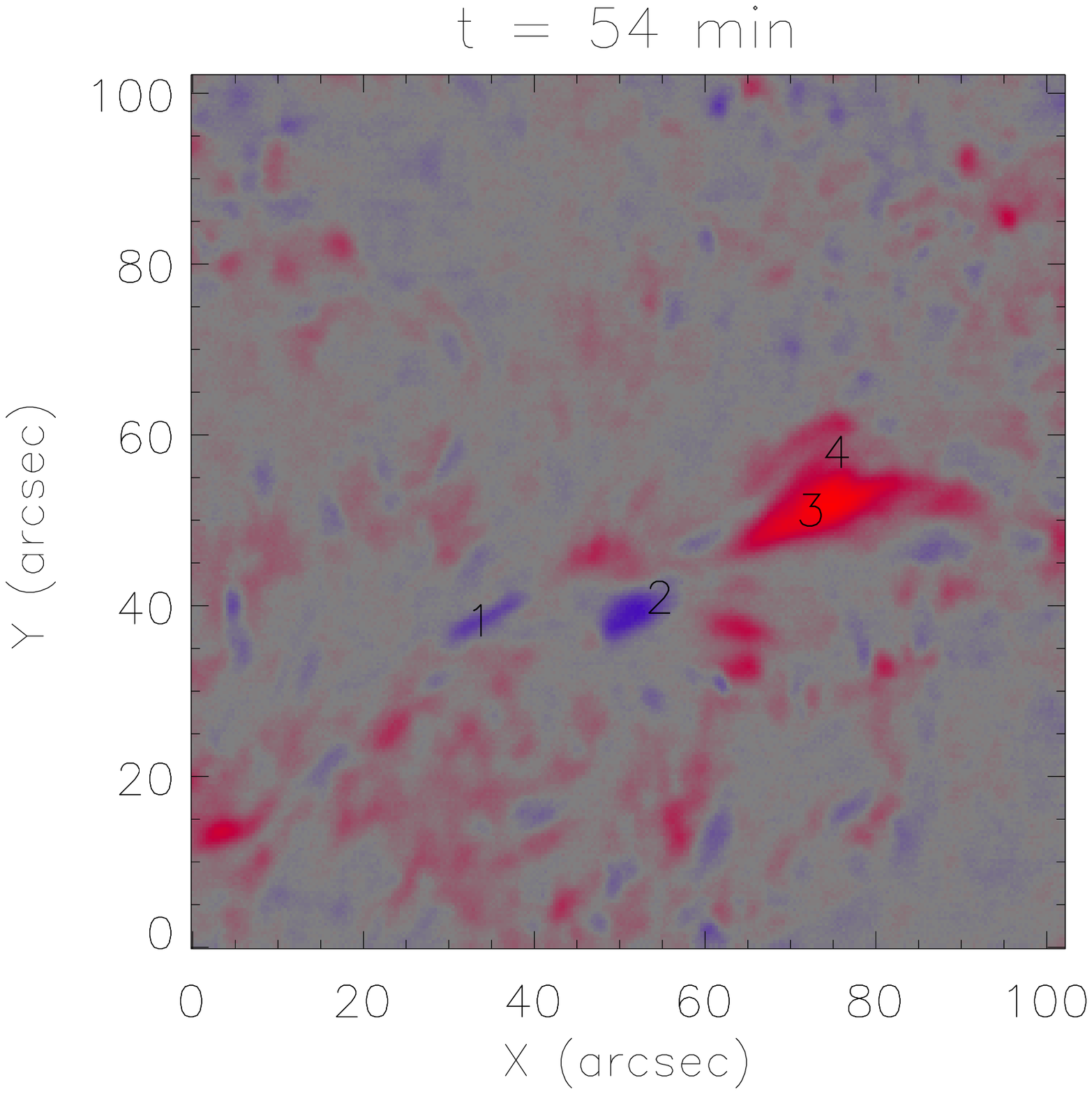}
   \caption{ Part of the velocity sequence in H$\alpha$ from  t=38 min to t=54 min.
   Colours correspond to upward velocities in blue (up to 15 km/s) and to downward velocities in red (up to 20 km/s).}
              \label{seqv1}
\end{figure*}

\begin{figure*}
   \centering
   \includegraphics[width=4cm,angle=90]{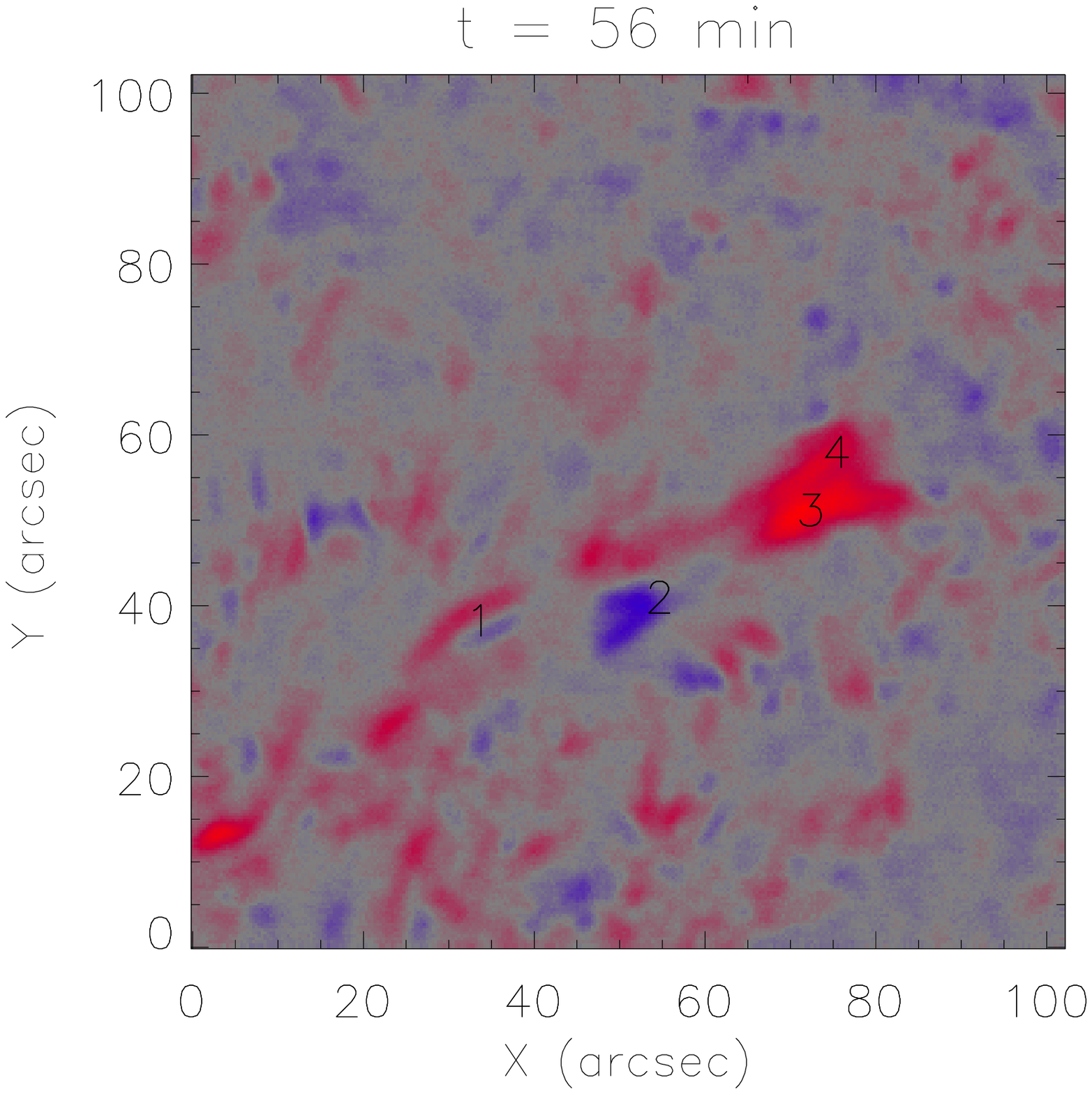}
   \includegraphics[width=4cm,angle=90]{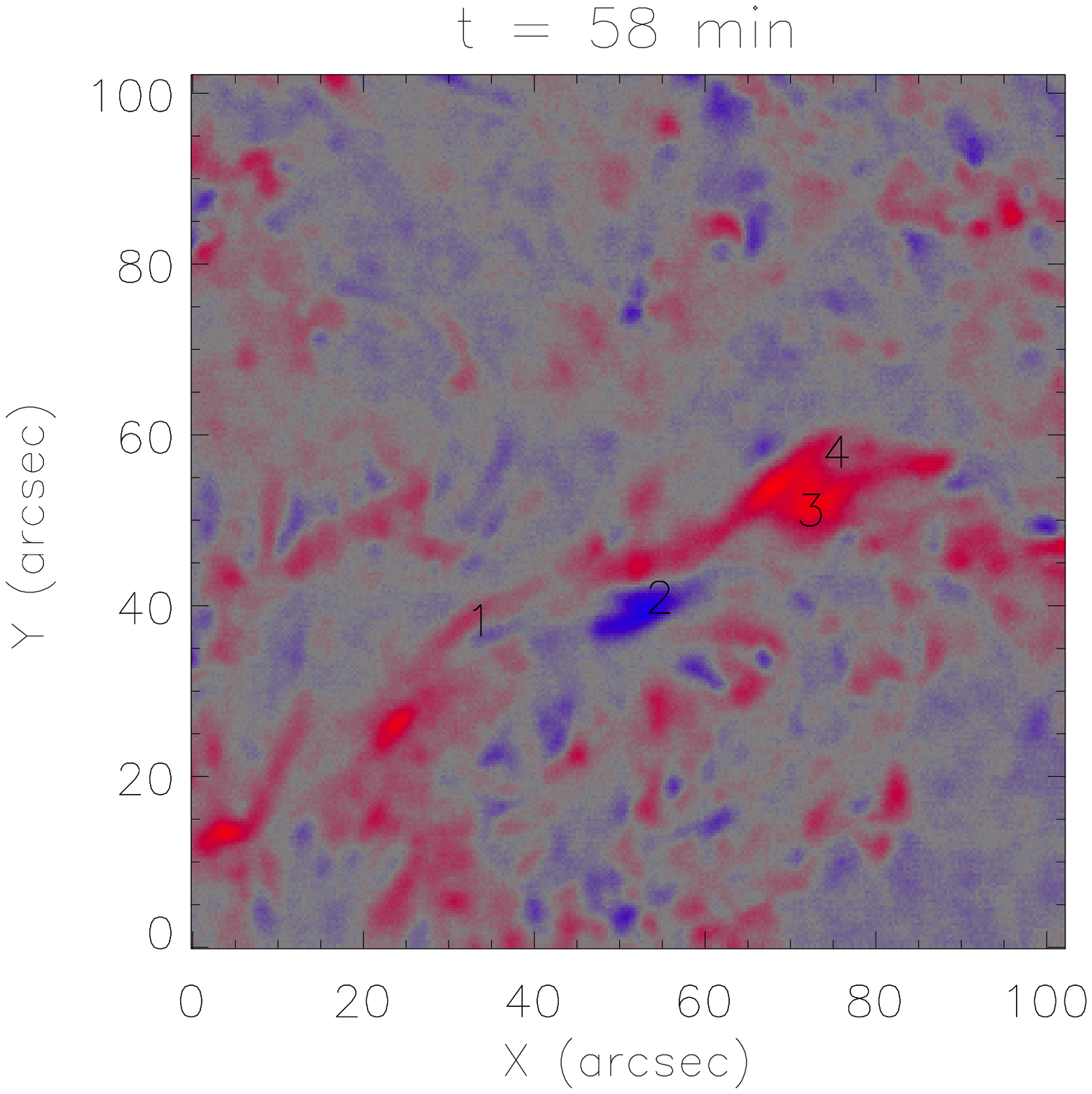}
   \includegraphics[width=4cm,angle=90]{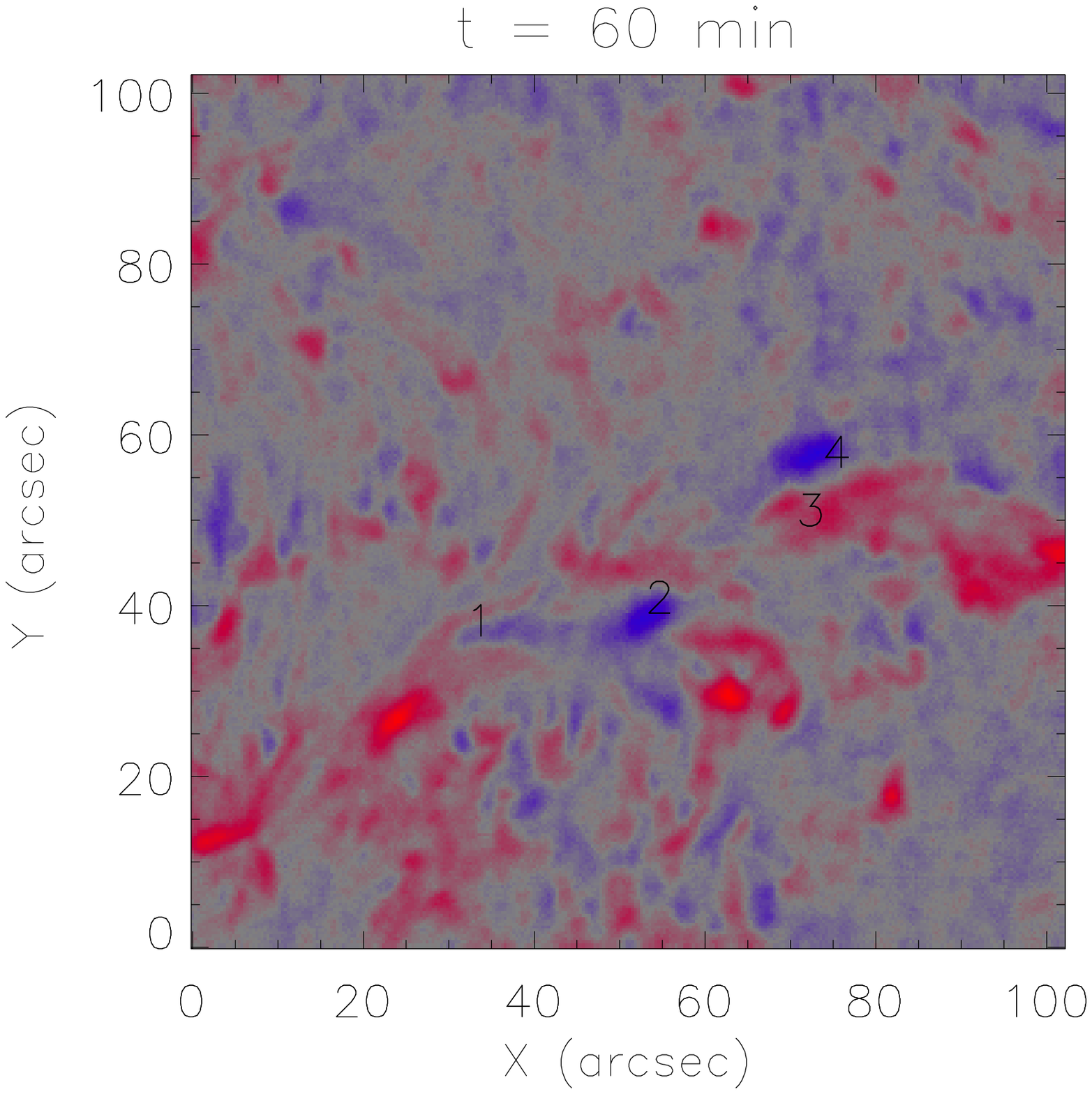}
   \includegraphics[width=4cm,angle=90]{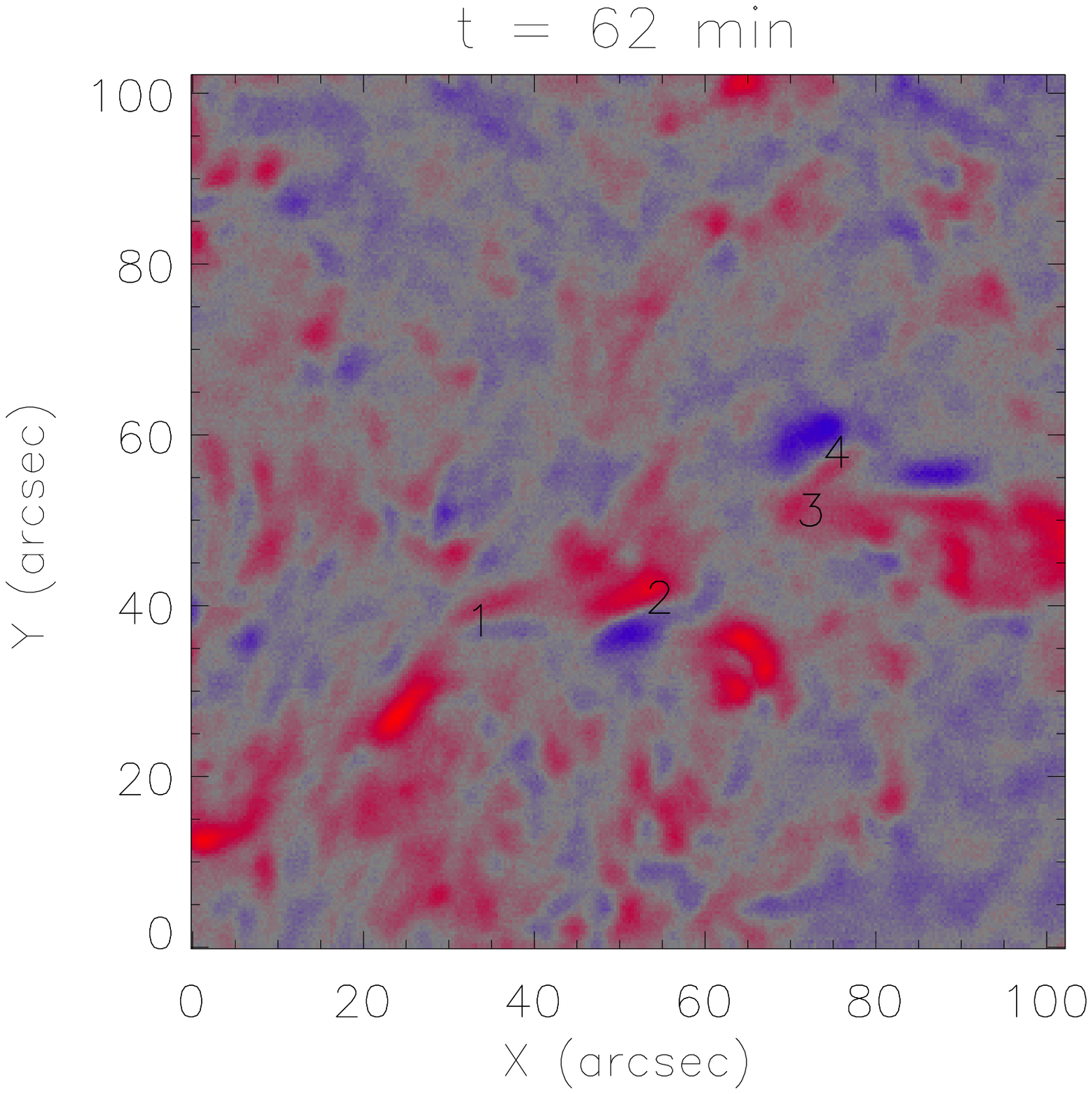}
   \includegraphics[width=4cm,angle=90]{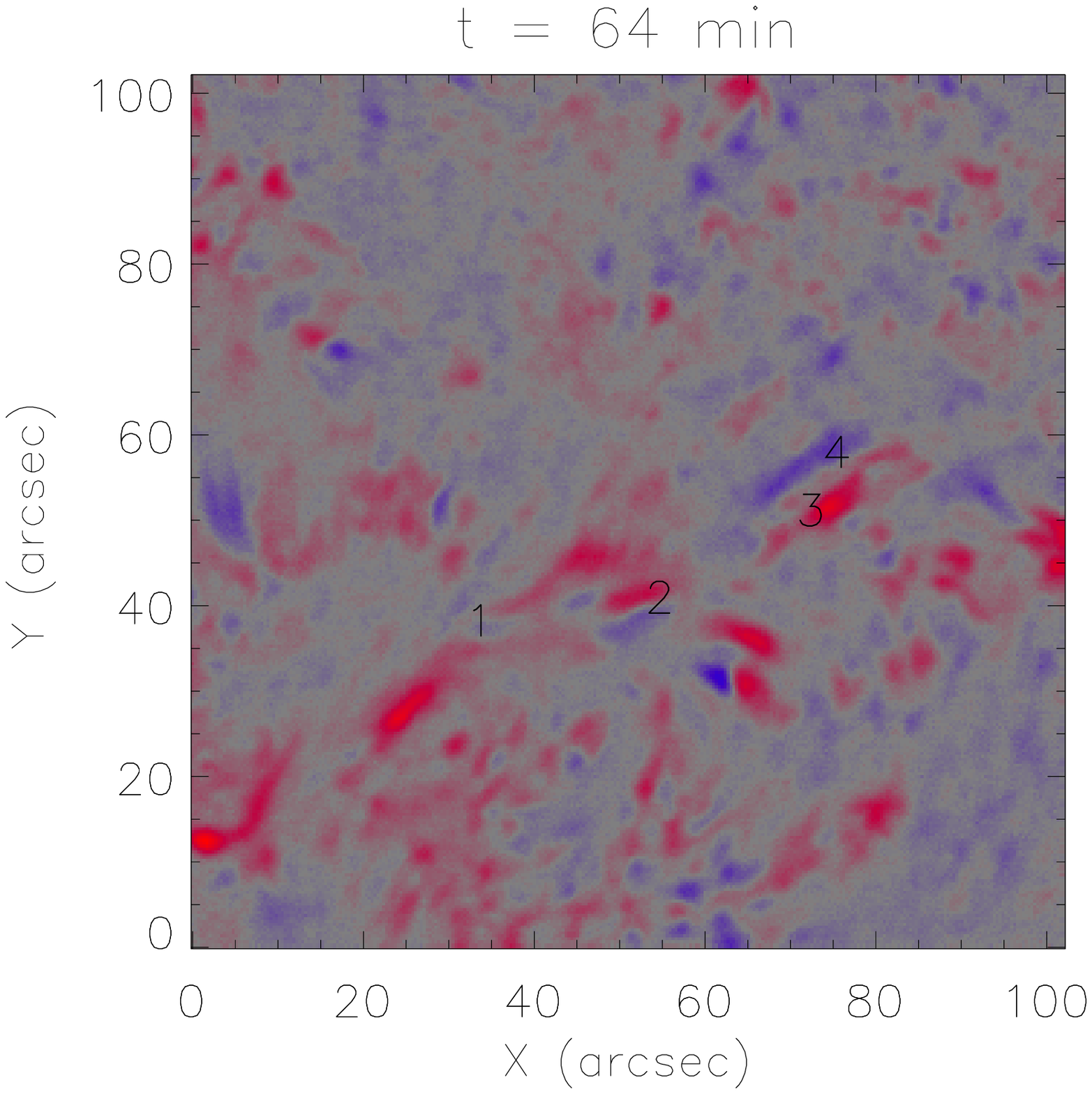}
   \includegraphics[width=4cm,angle=90]{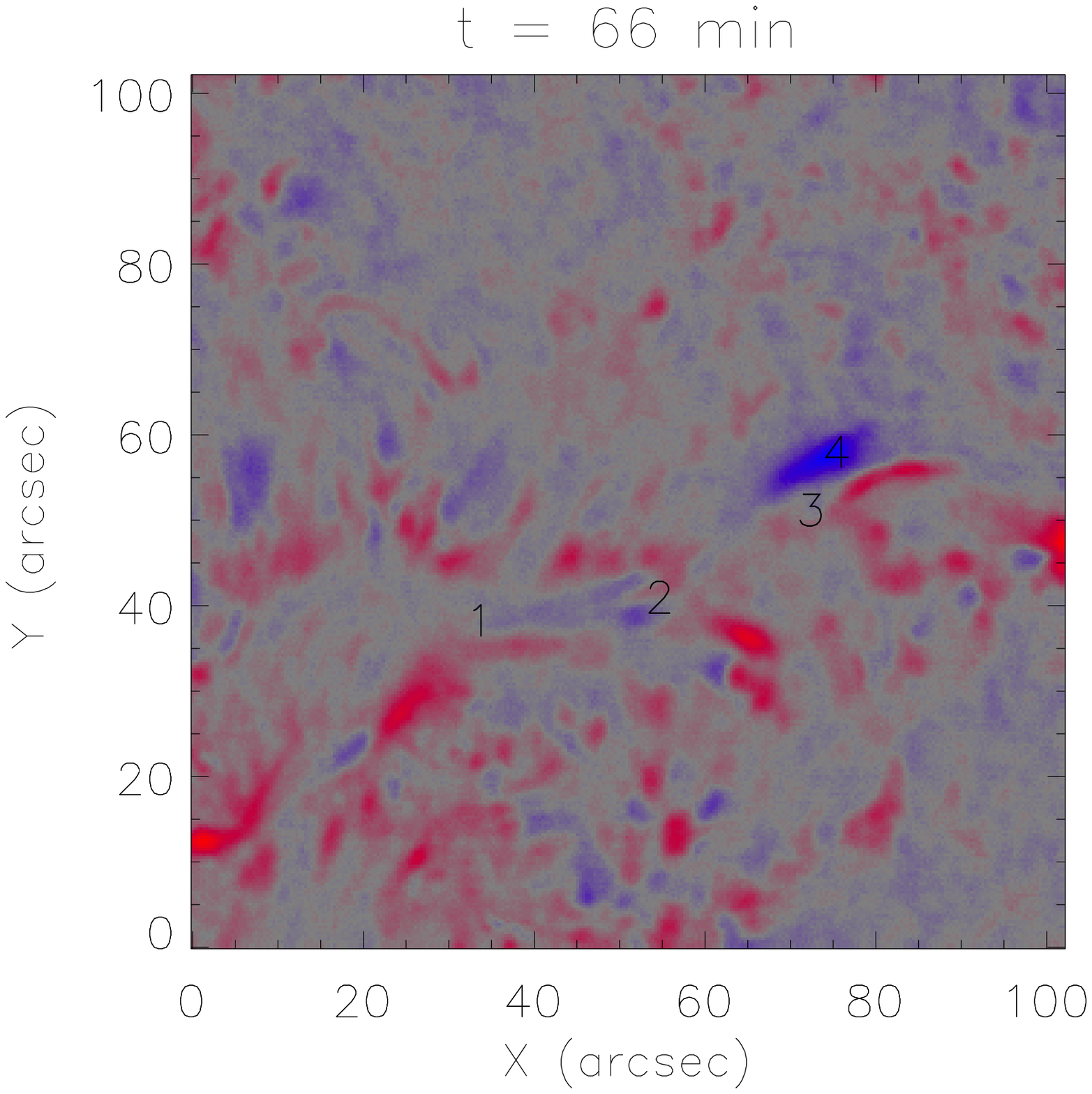}
   \includegraphics[width=4cm,angle=90]{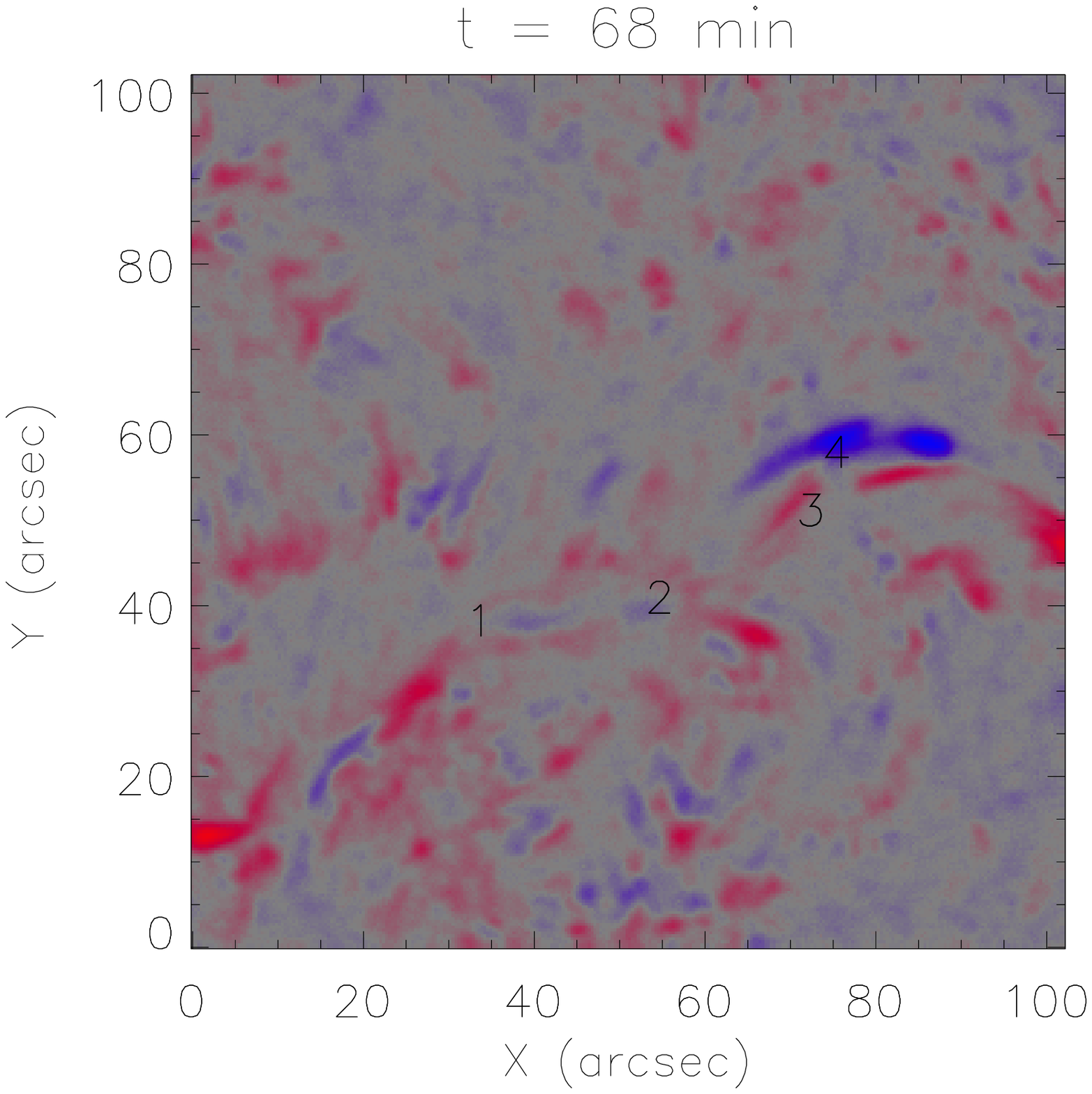}
   \includegraphics[width=4cm,angle=90]{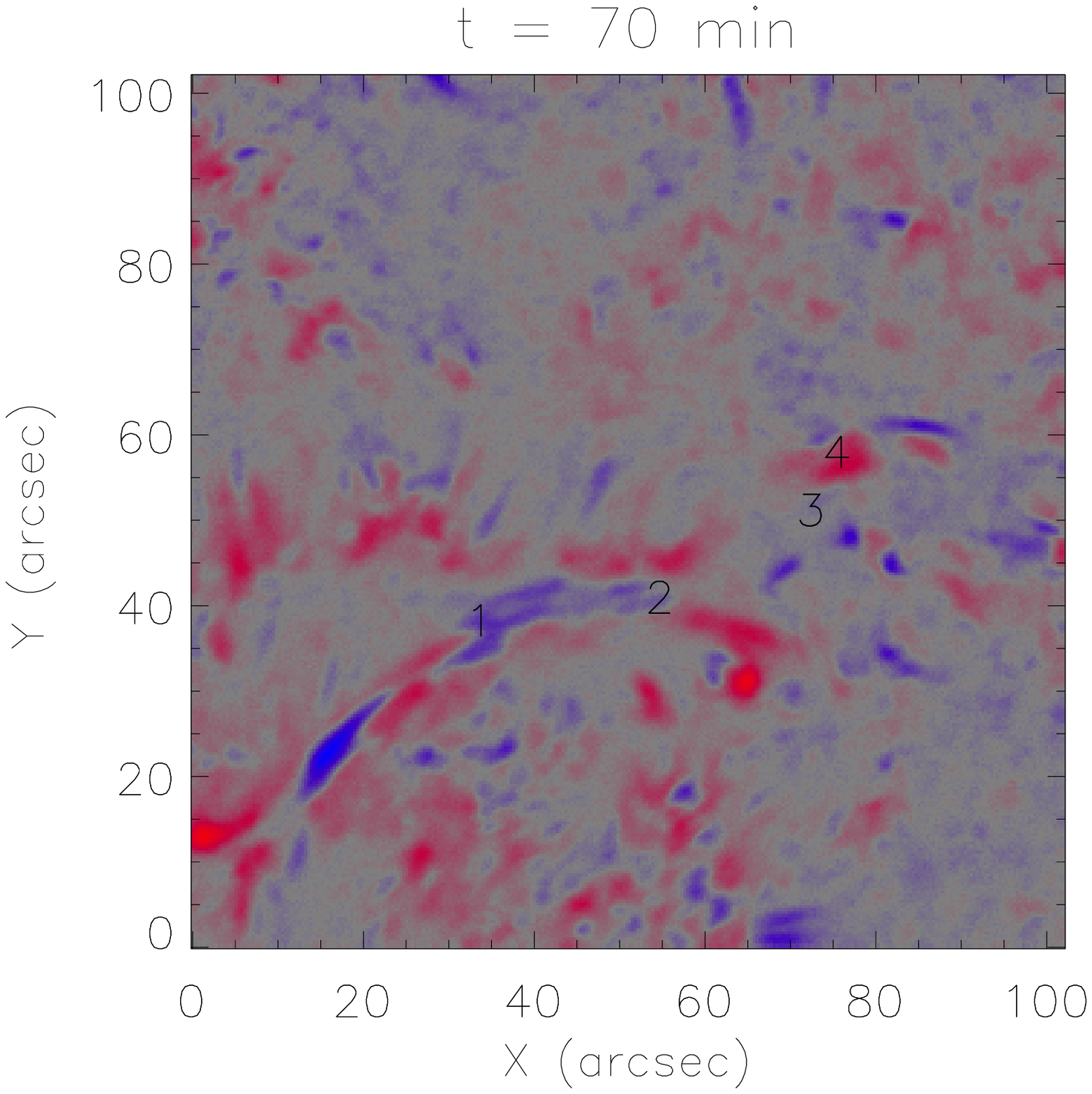}
   \includegraphics[width=4cm,angle=90]{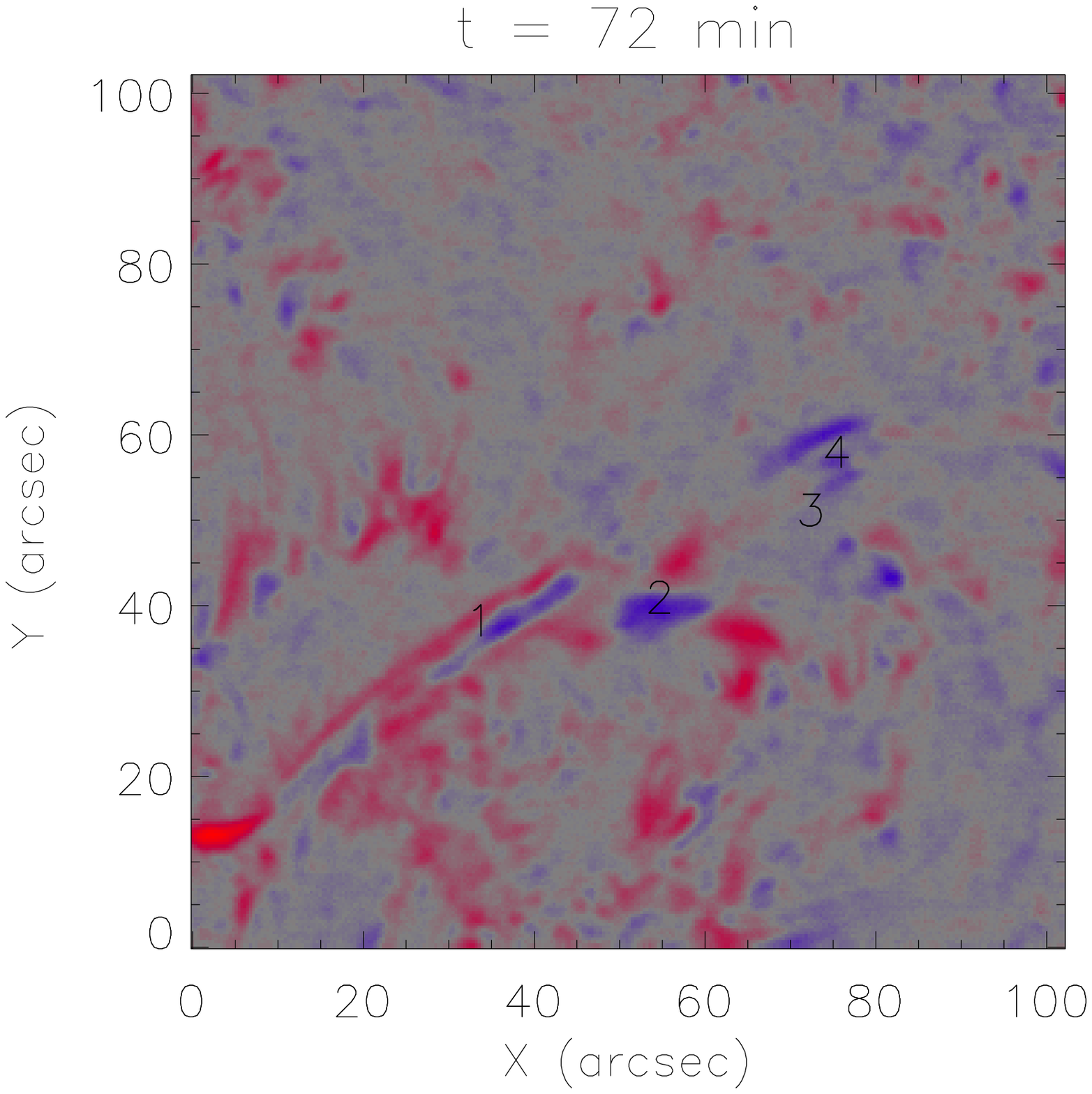}
   \caption{ Part of the velocity sequence in H$\alpha$ from  t=56 min to t=72 min.
   Colours correspond to upward velocities in blue (up to 15 km/s) and to downward velocities in red (up to 20 km/s).}
              \label{seqv2}
\end{figure*}

Figures \ref{seqi1} and \ref{seqi2} show the H$\alpha$ summed wings intensity sequence, and Figures \ref{seqv1} 
and \ref{seqv2}
show the velocity sequence, from $t$= 38 to $t$= 72 minutes after the 
beginning of the observations; the eruption occurs beween $t$= 38 and 50 minutes. A large part of the filament erupts as the filament
moves towards the Northwest, with an apparent proper motion of up to 25 km/s, as measured between $t$= 42 minutes and $t$= 54 minutes \citep{Baudin_98}. 
The velocity sequence shows some vertical motions as the two parallel blue and red areas clearly reveal: 
parts of the filament are moving upward (blue areas), whilst parts of the filament are moving downward (red areas).
The velocity ranges from 6 to 20 km/s for the zone going downward (red) and from 5 to 15 km/s for the ascending zone (blue).
A possible interpretation of the sequence is to assume rotationally coherent motions during approximately 10 minutes (from $t$= 40 to $t$= 50 minutes), 
in order to satisfy the mass conservation condition assuming no significant change of ionisation ratio (temperature) during this interval (see Section 4).
At the end of the eruption, a part of the filament fell down (only the red part of the velocity remained, figure \ref{seqv2}) 
while the signature of the filament in intensity was less extended (figure \ref{seqi2}): the filament lost a part of its mass 
during the eruption (disappearance). From $t$=70 minutes, the signature of a new filament is clearly visible in intensity and velocity, along 
the filament channel.\\
Two hours after the eruption, a large scale loop system was observed by Yohkoh in SXR moving outwardly, suggesting that large effects 
were produced in the hot surrounding corona.

%\clearpage

%Si pb de compilation ==> faire un \clearpage

\subsection{Spectral analysis of the signal in the filament}
\begin{figure*}
   \centering
\includegraphics[width=4.5cm,angle=90]{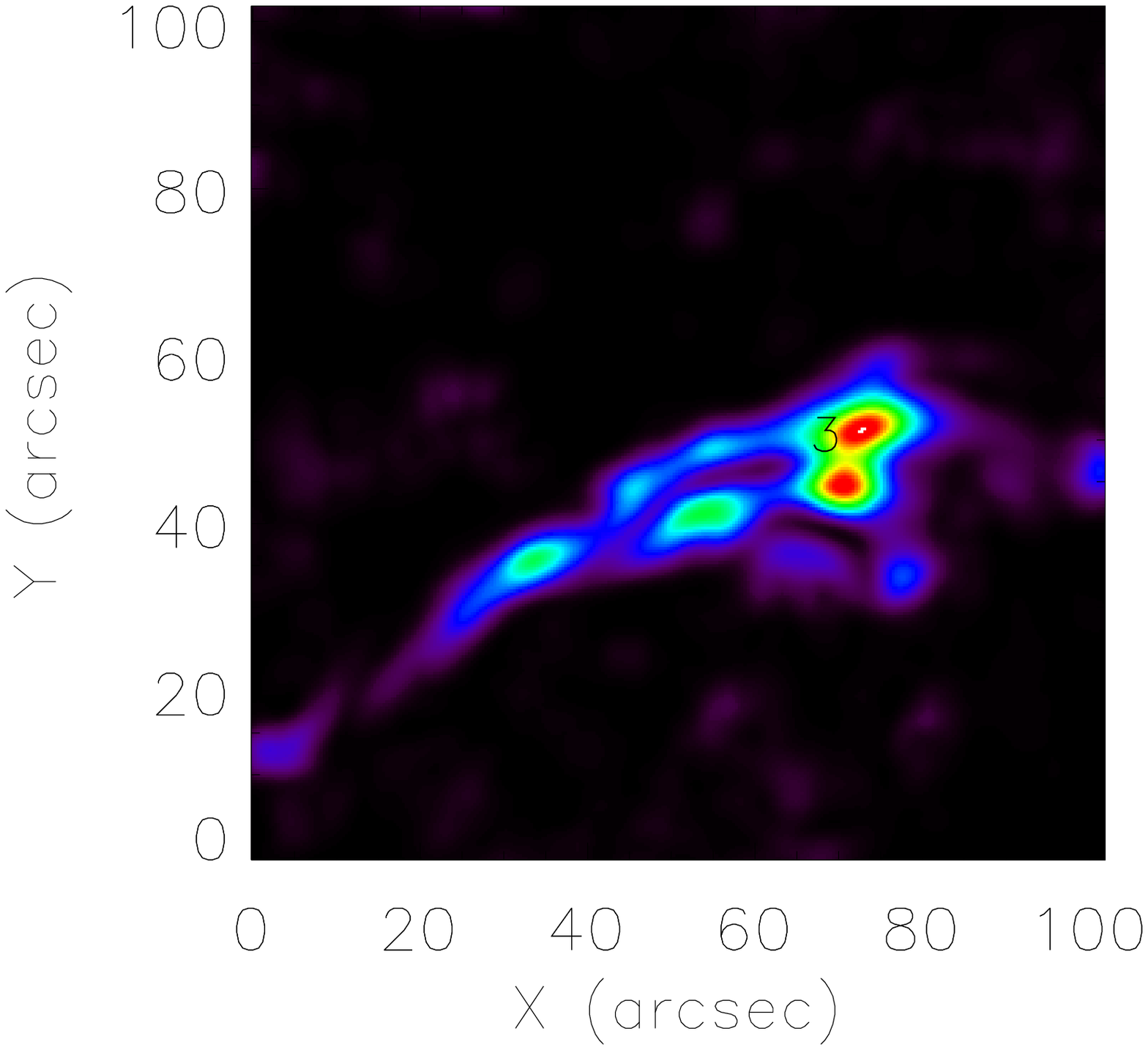}
\includegraphics[width=4.5cm,angle=90]{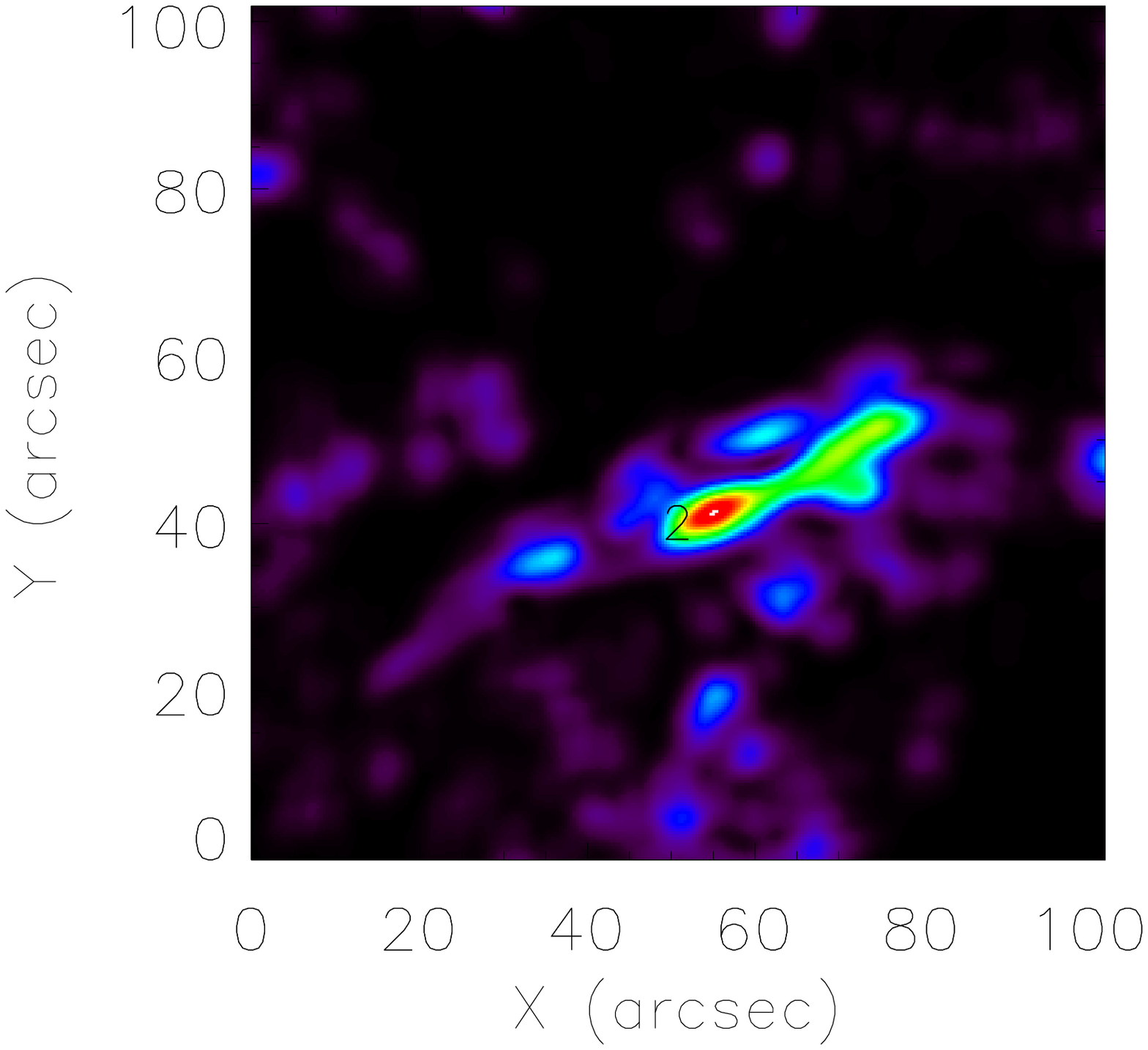}\\
\includegraphics[width=4.5cm,angle=90]{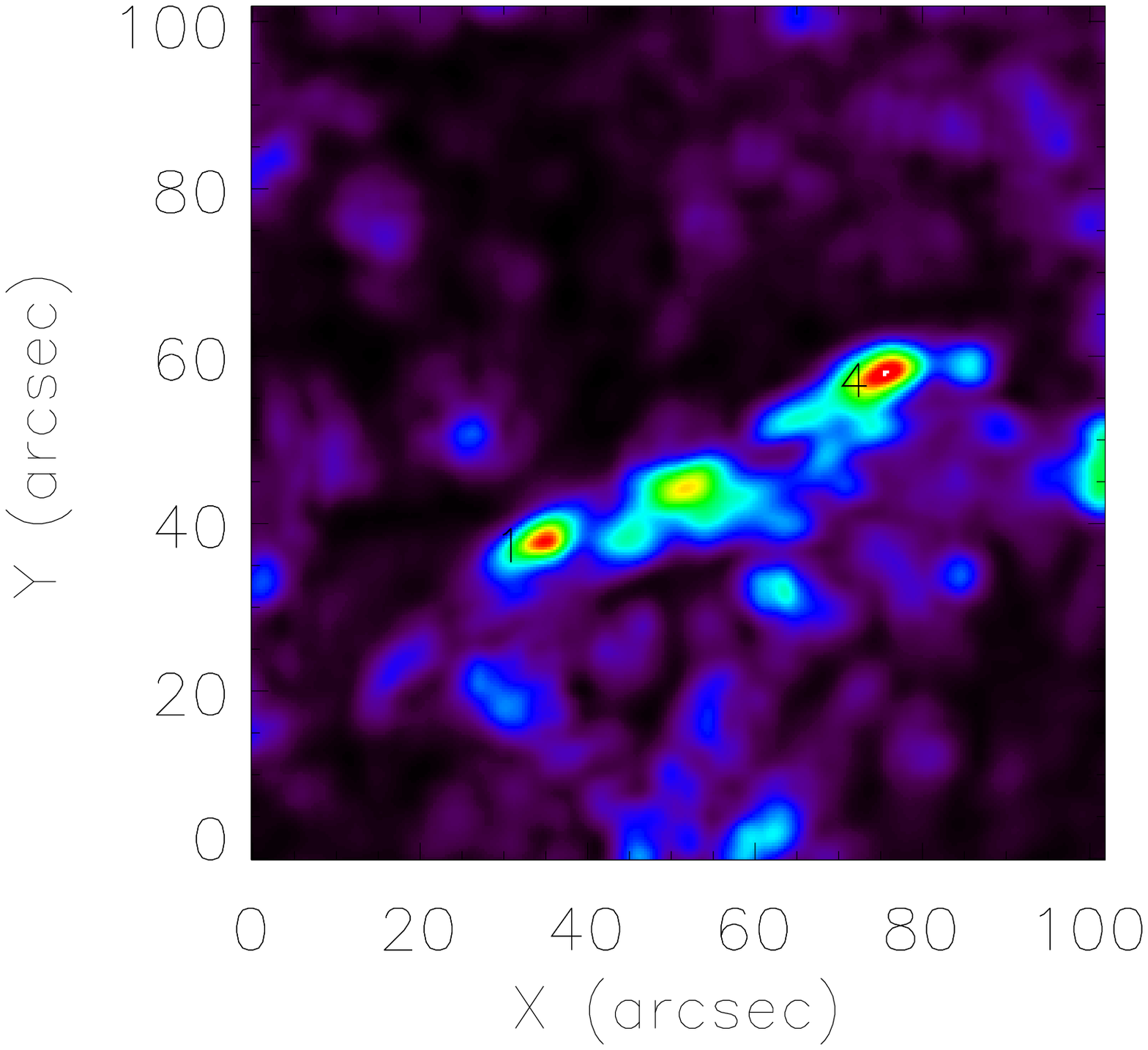}
\includegraphics[width=4.5cm,angle=90]{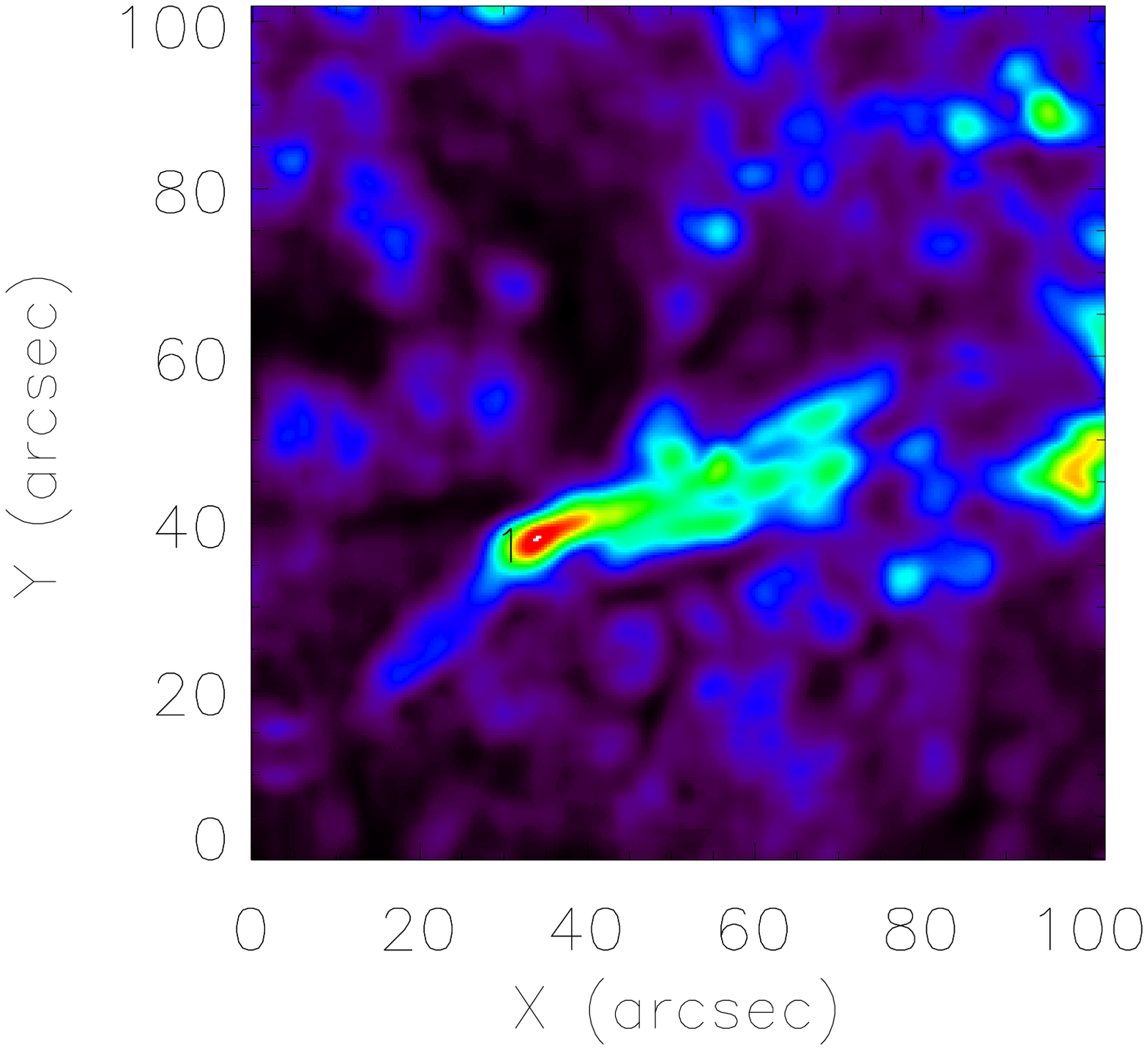}
  \caption{Power maps of the velocity sequence in H$\alpha$ from left to right and top to bottom, in the ranges
[15, 55 min], [8, 15 min], [4, 8 min] and [2, 4 min]. }
              \label{cartepuissance}
\end{figure*}

Power spectral maps are computed from the velocity signal, using the Fast Fourier Transform. The power maps 
lead us to retain four spatial zones with a high power, over several consecutive maps. These zones are indicated by 
the numbers 1, 2, 3 and 4 on Figures \ref{seqi1} and \ref{seqv1}. The zones correspond 
to the different positions of the filament whilst it is moving toward the Northwest: for zone 1 between $t$= 38 and $t$= 42 
minutes after the beginning of the observations, for zone 2 between $t$= 44 and $t$= 48 minutes, for zone 3, at $t$= 50 and 
52 minutes, and for zone 4, at $t$= 54 minutes. Figure \ref{cartepuissance} shows the 4 power maps averaged over 
the intervals [15, 55 min], [8, 15 min], [4 , 8 min] and [2, 4 min], corresponding to the intervals observed 
on the filament. Zone 1 is visible in the range [4, 8 min] and in the range [2, 4 min] (bottom left and right).
Zone 2 is visible in the range [8, 15 min] (top right). Zone 3 is visible in the range [15, 55 min] (top left).
Zone 4 is visible in the range [4, 8 min] (bottom left). \\

\begin{figure*}
   \centering
\includegraphics[width=4cm,angle=90]{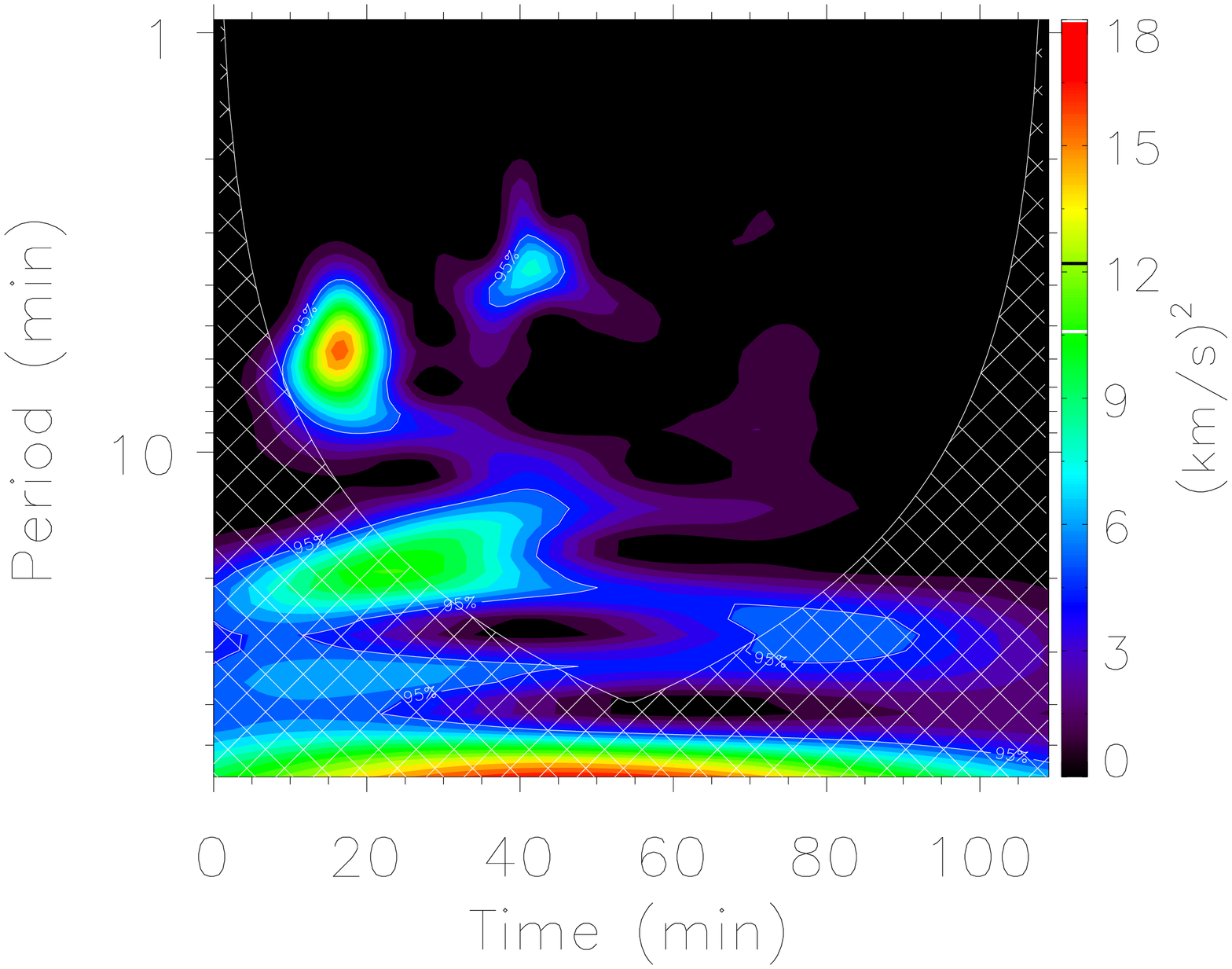}
\includegraphics[width=4cm,angle=90]{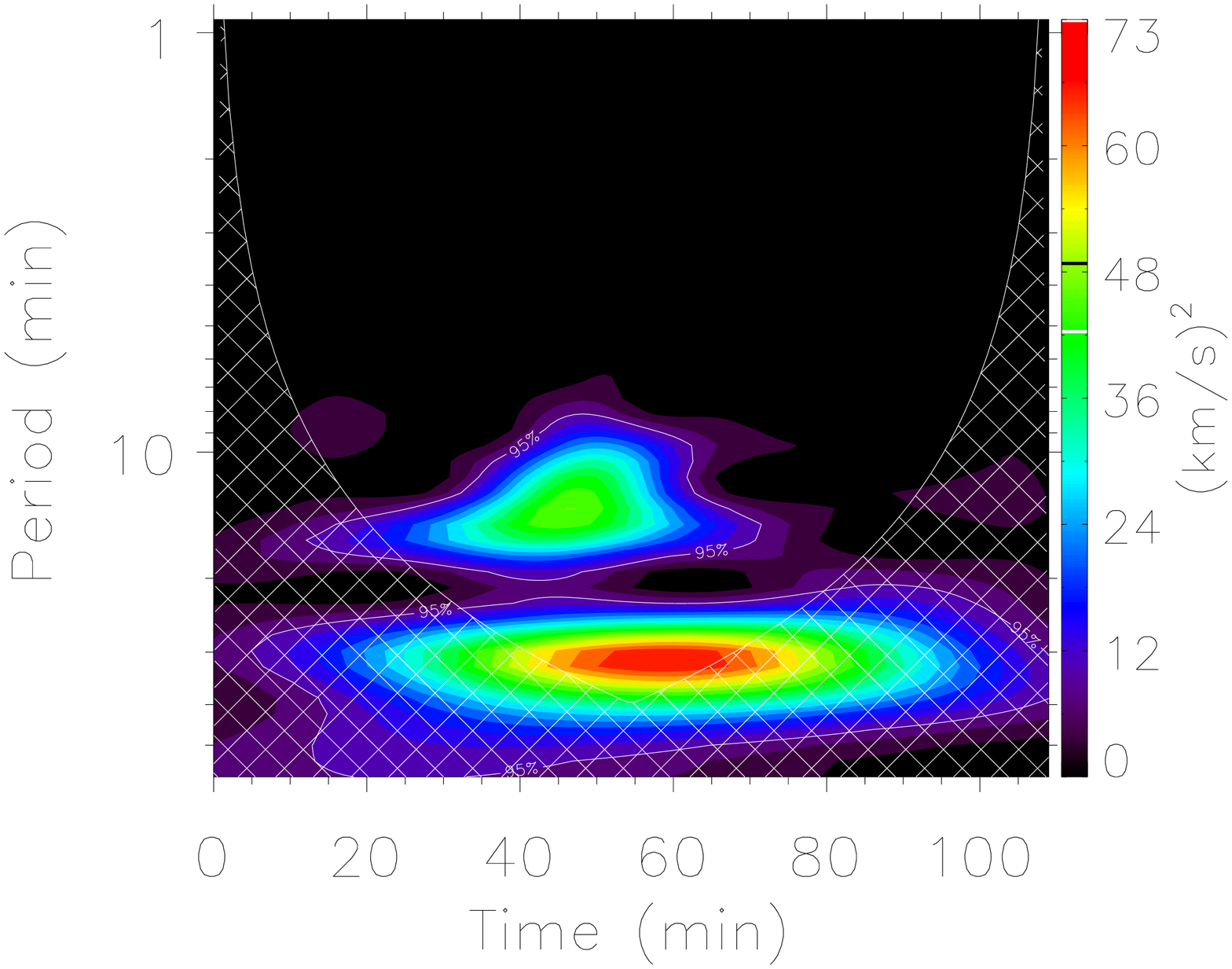}\\
\includegraphics[width=4cm,angle=90]{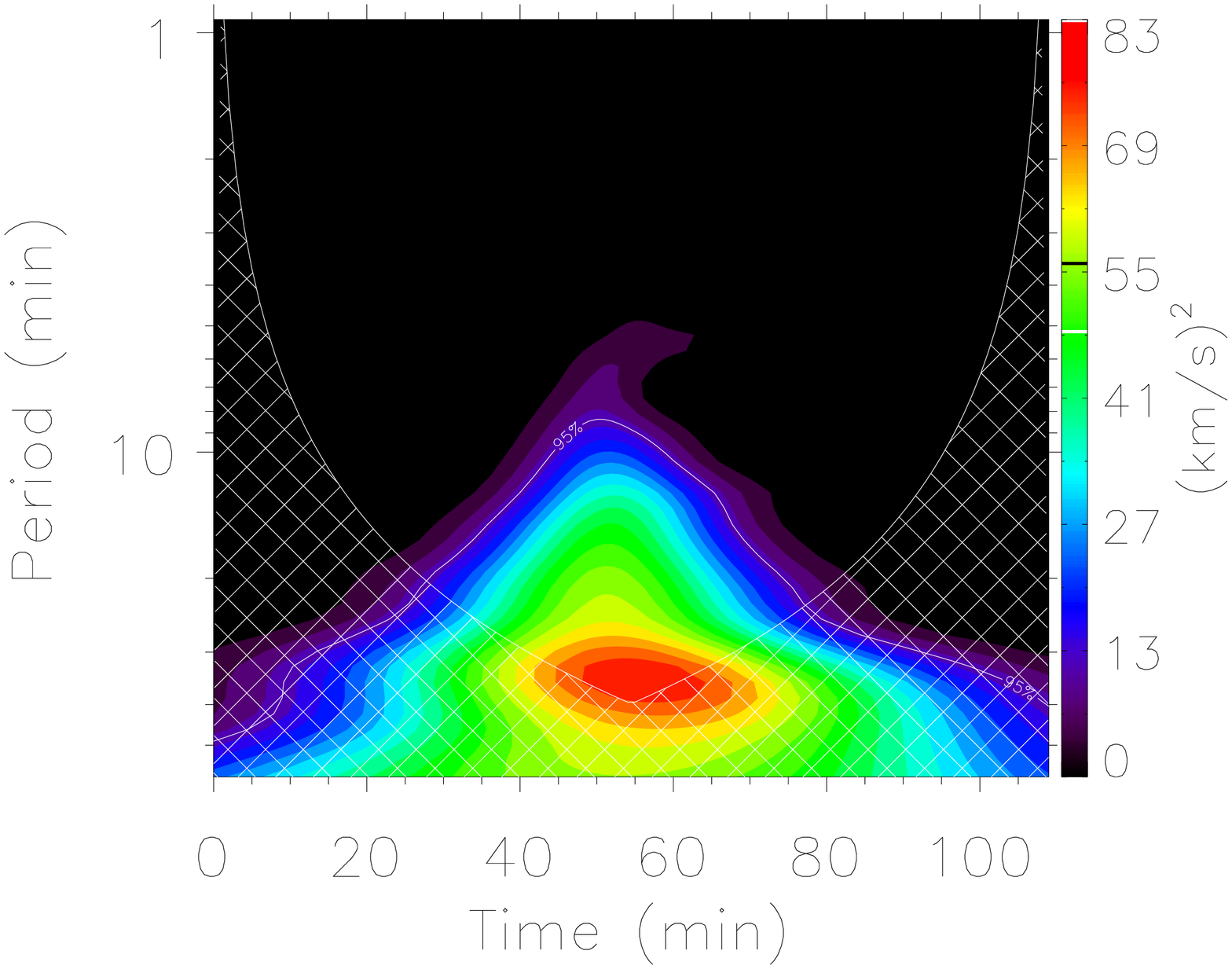}
\includegraphics[width=4cm,angle=90]{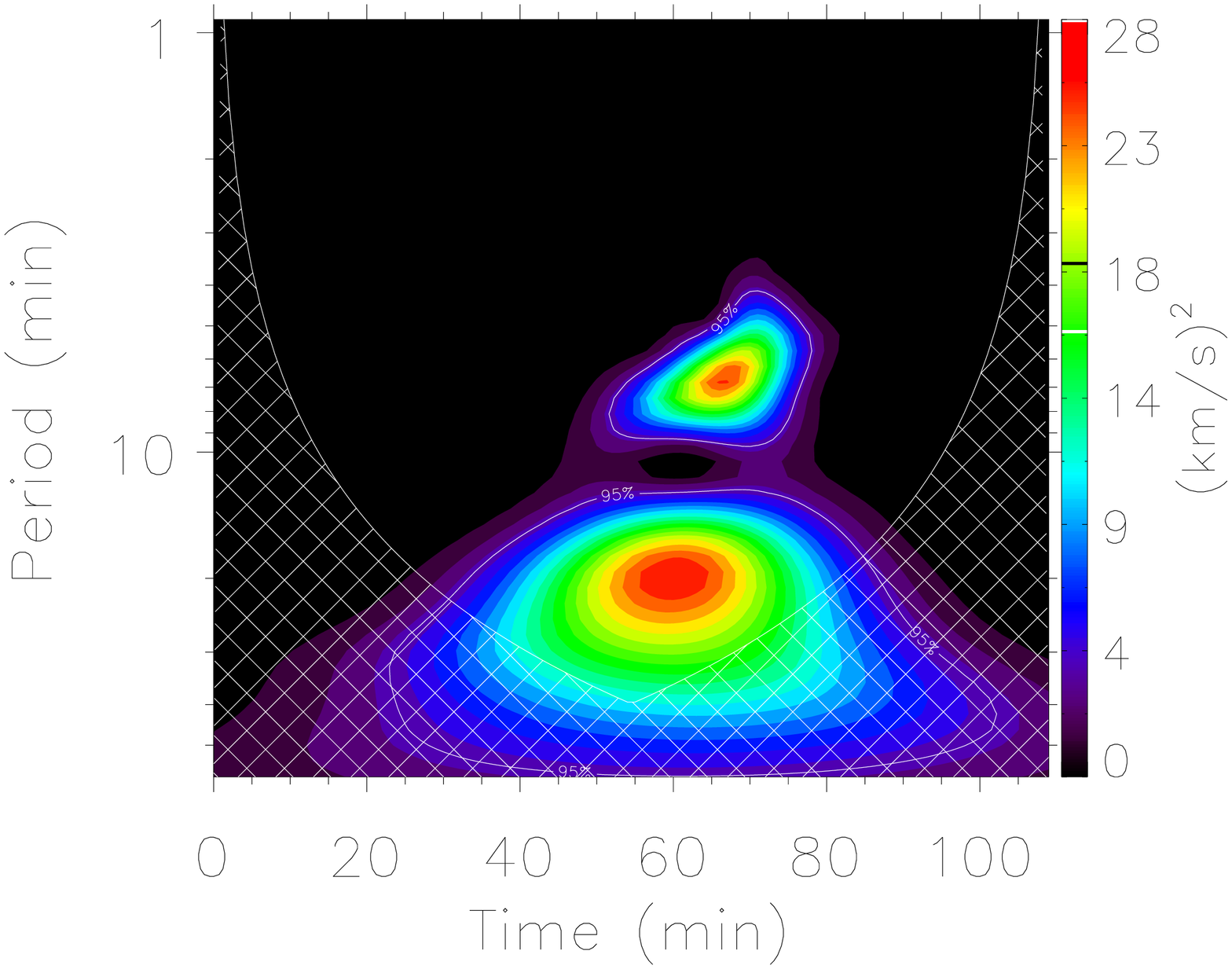}\\
   \caption{Wavelet analysis of the velocity in H$\alpha$ in zones 1 (top left), 2 (top right), 3 (bottom left) and 4 (bottom right); 
significant power is within the 95$\%$ confidence level contour and outside the hatched cones of influence.}
              \label{ondevsacpeak}
\end{figure*}

The wavelet analysis of the Doppler velocity for these 4 zones using the full 110 minute long sequence of the observations is displayed on Figure 
\ref{ondevsacpeak}: \\
-in zone 1, the period ranges from 3 to 10 minutes, as already shown in Figure \ref{cartepuissance}, at the beginning of the sequence\\
- zone 2 has a period of about 12 minutes and a period of 45 minutes centred on $t$= 55 minutes, during the rotational motion\\
- zone 3 has a weak period ranging from 8 to 40 minutes centred on $t$= 60 minutes, \\
-and zone 4 shows significant power near 7 and 20 minute periods centred on $t$= 65, while the filament was falling down.\\

\subsection{Summary of the analysis}
A 2D analysis of H$\alpha$ intensity observations in a filament during 110 minutes reveals a horizontal displacement of the filament 
at a speed of 25 km/s during 12 minutes, along with a vertical motion deduced from the Doppler velocity 
signal, followed by the eruption of a large part of the filament and the downward motion of the rest of the filament.
During 10 minutes, rotational motions inside the filament are detected before the occurence of the fast proper motion, 
starting at $t$= 40 minutes towards the Northwest.
Wavelet analysis of the velocity shows 4 zones with high power and different behaviors; the Doppler velocity in these zones exhibits strong 
variations (from a few km/s to more than 10 km/s) and periods ranging from 3 to 10 minutes; periods around 20 minutes and 45 minutes are also measured.

\section{Conclusion and discussion}
We presented the results of the analysis of two eruptive filaments exhibiting some similarities observed in completely different manners:
one observed in He~I and Mg~X with a spaceborne spectrometer, the other observed in H$\alpha$ wings with a groundbased imager. 
In both cases, the eruptive filaments show small amplitude oscillatory motions associated with an eruption and a partial disappearance:
the filaments lost a part of their mass, are heated and evaporate.

In He~I, the pseudo-period of the velocity oscillations is increasing from 20 to 50 minutes while the amplitude is decreasing: 
this damped velocity oscillation could be interpreted, as suggested by \cite{Vrsnak_90b}, as the signature of the eruption, due to the weakening of the 
restoring force 
giving rise to the increasing period. In this model dealing with the oscillatory motions in an eruptive filament, 
the authors explain that once the filament is in its eruptive phase, the filament jumps from an unstable position to a stable one at a higher altitude. 
In this stable position, if the damping factor (inverse of the damping time) is less than the frequency of the first oscillation mode, 
the filament will oscillate around its equilibrium position, as a damped oscillator. This behaviour corresponds to that described by our observations 
in Section 2. An increased heating of the top part of the filament is another possibility suggested by \cite{Filippov_02}.\\
We can link this increase of the period of the oscillating filament to the decreasing stiffness of an oscillating spring; this suggesting the weakening 
of the restoring force, which may also be related to the destabilisation and onset of the filament eruption which has modified the whole structure of the 
filament and thus the restoring forces \citep{Isobe_Tripathi_06}.\\
The value of $\tau/T$ = 1.9 obtained in our study may be compared to that obtained for a similar behaviour in a longitudinally damped oscillation 
of a filament observed in H$\alpha$ by \citet{Vrsnak_07}: $\tau/T$= 2.3.
They interpreted their observations in terms of poloidal magnetic flux injection by magnetic reconnection at one of the filament legs. 
They also pointed out an increase with time of the oscillation period and interpreted this behaviour in terms of nonlinear effects
in the framework of a harmonic oscillator model.\\

Results based on the velocity in He~I and Mg~X indicate that the corona in which the filament channel is embedded responds
coherently with the oscillating filament. The intensity signals in both lines are very closely in phase, showing nearly simultaneous variations.
Moreover, the peak of intensity seen in both lines appears after the beginning of the oscillations detected in velocity, which
indicates that the dynamical effects precede the thermal effects.\\

For the second case, the analysis of the velocity in H$\alpha$, shows a large power peak detected for zone 1 in the extreme eastern part of the filament 
rooted in the chromosphere: the small values of the period are close to the typical periods of the photospheric 
(5 minute) and chromospheric (3 minute) oscillations. These oscillations are detected before the motion of the 
filament toward the Northwest and before its eruption. We suggest that the filament is excited by the photospheric and 
chromospheric oscillations \citep{Bocchialini_01}, which are at the origin of the motion of the filament, with a
triggering of the eruption and the flare due to reconnections. This result can be compared to the 4 minute periods obtained in H$\alpha$ \citep{Schmieder_91} in 
chromospheric filamentary structures, interpreted either in terms of eigenmodes excited by the chromospheric oscillations, or 
in terms of Alfv\`en waves generated by the chromosphere and transmitted to the filamentary structures, as confirmed by the Hinode/SOT observations 
\citep{Okamoto_07}.\\

Indeed, these proper motions are
difficult to interpret as definite horizontal velocities. It seems rather that we are dealing with increasing turbulent proper motions with increasing ionisation 
ratio (temperature) leading to the disappearance of a large part of the filament after $t$= 54 minutes.
The analysis of the velocity, for both datasets, revealed the presence of oscillations with a period from 20 to 30 minutes in He~I, Mg~X, and H$\alpha$, 
cotemporal with the eruption of the filaments. These periods can be compared to the 20 minute period 
found by \citet{Anzer_09} who discussed the different oscillation periods for global modes of quiescent prominences. 
Anzer used simple 1D prominence configurations to describe the magnetohydrostatic equilibrium and their small amplitude oscillations. 
The restoring force along the $z$-direction, perpendicular to the solar surface, resulting from the magnetic tension of the stretched field lines, 
leads to periods of 20 minutes in this direction.
Our results can also be compared to those of \citet{Terradas_05} who showed that for a typical prominence temperature of 6000\,K, the oscillation 
period computed for a quiescent prominence is around 25 minutes and the damping time is around 19 minutes. The damping time increases with
temperature and depends on the radiation time. \citet{Ballester_06} points out in his review on solar corona seismology that other mechanisms are possible
such as wave leakage, resonant absorption and ion-neutral collisions. \citet{Soler_07} and \citet{Soler_09} demonstrate that 
radiative effects of the prominence plasma are responsible for the attenuation of slow modes, connected to intermediate period
(10 to 40 minutes) prominence oscillations. 
The simultaneous upflows and downflows detected in the filament observed in the wings of H$\alpha$ could be compared to the results obtained 
by \citet{Berger_10}: 
high spatial resolution observations, in H$\alpha$ with Hinode/SOT, of quiescent prominences reveal the presence of upflowing plumes. 
Their occurrrence combined with the plume dimensions indicate that ``the plumes are sources of upward mass flux into the prominence which partially offset 
the continual downward mass flux of heavy prominence material''.\\

These results on small amplitude oscillations of the intensity and of the Doppler velocity signals in eruptive filaments contribute to the statistical 
study needed to check the link between oscillations and eruption precursors.

\begin{acknowledgements}
The H$\alpha$ images are courtesy of the Kiepenheuer Institute and NSO/Sacramento Peak. 
EIT, CDS, and MDI data are courtesy of SOHO consortia. 
SOHO is a project of international cooperation between ESA and NASA. 
The Yohkoh/SXT images are from the JAXA consortium.
The authors thank the anonymous referee for helpful comments and suggestions and L.M.R. Lock for his help.
\end{acknowledgements}

\bibliographystyle{aa}
\bibliography{bibliography}

\end{document}